\documentclass[11pt,
              a4paper,
              index=totoc,
              headsepline,
              footsepline,
              BCOR=12mm,
              DIV=13]{scrbook}

\usepackage{graphicx}
\usepackage[
bookmarksopen,
bookmarksdepth=2,
breaklinks=true
]{hyperref}
\usepackage{tabularx}
\usepackage{multirow}
\usepackage{tikz}
\usepackage{pgfplots}
\usepackage{adjustbox}
\usepackage[graphicx]{realboxes}
\usepackage{caption}
\usepackage{xcolor}
\usepackage{algorithm}
\usepackage[noend]{algpseudocode}
\usepackage{mathtools}
\usepackage{amssymb}
\DeclareCaptionFormat{citation}{%
   \ifx\captioncitation\relax\relax\else
     \captioncitation\par
   \fi
   #1#2#3\par}

\let\captioncitation\relax
\newcounter{notecounter}
\newcommand{\enotesoff}{\long\gdef\enote##1##2{}}

\DeclareOldFontCommand{\bf}{\normalfont\bfseries}{\mathbf}

\pdfoutput=1

\enotesoff

\graphicspath{ {./} }

\def\university{Technische Universit{\"a}t M{\"u}nchen}
\def\universityLogo{tum_logo}
\def\program{Computational Science and Engineering \\(International Master's Program)}
\def\programLogo{cse_logo}
\def\doctype{Master's Thesis}

\def\title{Lifelong Neural Topic Learning in Contextualized Autoregressive Topic Models of Language via Informative Transfers}
\def\author{Yatin Chaudhary}
\def\examinerOne{Prof.\ Dr.-Ing\ Thomas A. Runkler, TUM, Siemens AG}

\def\primaryAdvisor{Pankaj Gupta, LMU, Siemens AG}

\def\date{March 25th, 2019}

\def\keywords{{keyword1}, {keyword2}, {keyword3}}

\def\metaTitle{\title}
\def\metaAuthor{\author}
\def\metaSubject{\doctype\ -\ \university}
\def\metaKeywords{\keywords}

\def\footertext{}

\usepackage{palatino}

\usepackage{scrpage2}
\usepackage{scrhack}
\ifoot[\footertext]{\footertext} 

\usepackage{ifthen}

\usepackage{rotating}

\usepackage{hyperref}

\usepackage[toc, xindy]{glossaries}

\usepackage{makeidx}

\usepackage{graphicx}
\graphicspath{{figures/}}

\usepackage{float}

\usepackage{subcaption}

\usepackage{amsmath}

\usepackage{amsfonts}

\usepackage{tabu}

\definecolor{Pantone300C}{HTML}{0065BD} 
\definecolor{Pantone301}{HTML}{005293}  
\definecolor{Pantone540}{HTML}{003359}  
\definecolor{DarkGray}{HTML}{333333}    
\definecolor{MediumGray}{HTML}{808080}  
\definecolor{LightGray}{HTML}{CCCCC6}   
\definecolor{Pantone7527}{HTML}{DAD7CB} 
\definecolor{Pantone158}{HTML}{E37222}  
\definecolor{Pantone383}{HTML}{A2AD00}  
\definecolor{Pantone283}{HTML}{98C6EA}  
\definecolor{Pantone542}{HTML}{64A0C8}  

\def\colorLinks{Pantone300C}
\def\colorUrl{Pantone542}
\def\colorCitations{Pantone158}

\hypersetup{
  linkcolor=\colorLinks,%
  urlcolor=\colorUrl,%
  citecolor=\colorCitations
}

\usepackage[ocgcolorlinks]{ocgx2}[2017/03/30]

\hypersetup{
  pdftitle={\metaTitle},%
  pdfauthor={\metaAuthor},%
  pdfkeywords={\metaKeywords},%
  pdfsubject={\metaSubject}
}

\usepackage{hyperxmp}

\usepackage{xspace}

\newcommand{\clearemptydoublepage}{%
  \ifthenelse{\boolean{@twoside}}{\newpage{\pagestyle{empty}\cleardoublepage}}%
  {\clearpage}}

\newglossaryentry{computer}
{
  name=computer,
  description={is a programmable machine that receives input,
               stores and manipulates data, and provides
               output in a useful format}
}

\newglossaryentry{poc}
{
  name={proof of concept},
  description={}
  }
\newglossaryentry{ui}
{
  name={user interface},
  description={}
  }
\newglossaryentry{ai}
{
  name={arithmetic intensity},
  description={a measure of floating-point operations (FLOPs)
              \hyphenation{per-formed} performed by a \hyphenation{gi-ven} given code or code section relative
              to the amount of memory accesses (Bytes) that are required
               to support those operations\cite{AI}}
  }

\newglossaryentry{speed-up}
{
  name={speed-up},
  description={the factor of temporal acceleration a program
  exhibits when additional computational resources are dedicated to it's execution.}
}

\newglossaryentry{directive pragmas}
{
  name={directive pragma},
  description={a computer programming language construct that specifies how a compiler
  should process input data} 
}

\newglossaryentry{rc}{
name={race condition},
description={A race condition or race hazard is the behavior of an electronic,
 software, or other system where the output is dependent on the sequence or
 timing of other uncontrollable events. It becomes a bug when events do not
 happen in the order the programmer intended. The term originates with the idea
 of two signals racing each other to influence the output first.}
}
\newglossaryentry{dd}{
name={data dependencies},
description={}
}
\newglossaryentry{sisd}{
name={single instruction single data},
description={}
}
\newglossaryentry{simt}{
name={single instruction multiple threads},
description={}
}

\newglossaryentry{simd}{
name={single instruction multiple data},
description={}
}
\newglossaryentry{gp}{
name={Gaussian Plane},
description={The two dimensional plane of complex numbers.}
}
\newglossaryentry{CURAND}{
name={CURAND},
description={
The CURAND library provides facilities that focus on the simple and efficient
generation of high-quality pseudorandom and quasirandom numbers.\cite{cuRAND}
}
}

\newacronym[longplural={partial differential equations}]{PDE}{PDE}{partial differential equations}
\newacronym{mpi}{MPI}{Message Passing Interface}

\newacronym[longplural={Random Walks on Spheres}]{RWoS}{RWoS}{Random Walk on Spheres}

\newacronym[longplural={graphical processing units}]{GPU}{GPU}{graphical processing unit}

\newacronym[longplural={central processing units}]{CPU}{CPU}{central processing unit}
\newacronym{hpc}{HPC}{high performance computing}

\newacronym[longplural={arithmetic logic units}]{ALU}{ALU}{arithmetic logic unit}

\newacronym[longplural={streaming multi-processors}]{SM}{SM}{streaming multi-processor}

\newacronym[longplural={boundary value problems}]{BVP}{BVP}{boundary value problem}
\newacronym[longplural={general purpose graphical processing units}]{GPGPU}{GPGPU}{general purpose graphical processing units}
\newacronym{CUDA}{CUDA}{compute unified device architecture}
\newacronym{RAM}{RAM}{random access memory}
\newacronym{SRAM}{SRAM}{static random access memory}
\newacronym{DRAM}{DRAM}{dynamic random access memory}
\newacronym{I/O}{I/O}{input/output}
\newacronym{PTX}{PTX}{Parallel Thread eXecution}
\newacronym{jit}{JIT}{just in time}

\makeglossary

\begin{document}

 \frontmatter



\def\bcorcor{0.15cm}
\addtolength{\hoffset}{\bcorcor}

\thispagestyle{empty}

\vspace{4cm}
\begin{center}
	\includegraphics[width=4cm]{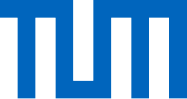}\\
	\vspace{5mm}
	\huge \program \\
	\vspace{0.5cm}
	\large \university
\end{center}

\vspace{20mm}
\begin{center}
	{\Large \doctype}\\
	\vspace{20mm}
	{\huge \textbf \title}\\
	\vspace{15mm}
	{\LARGE  \author}\\
	\vspace{\fill}
	\includegraphics[width=4cm]{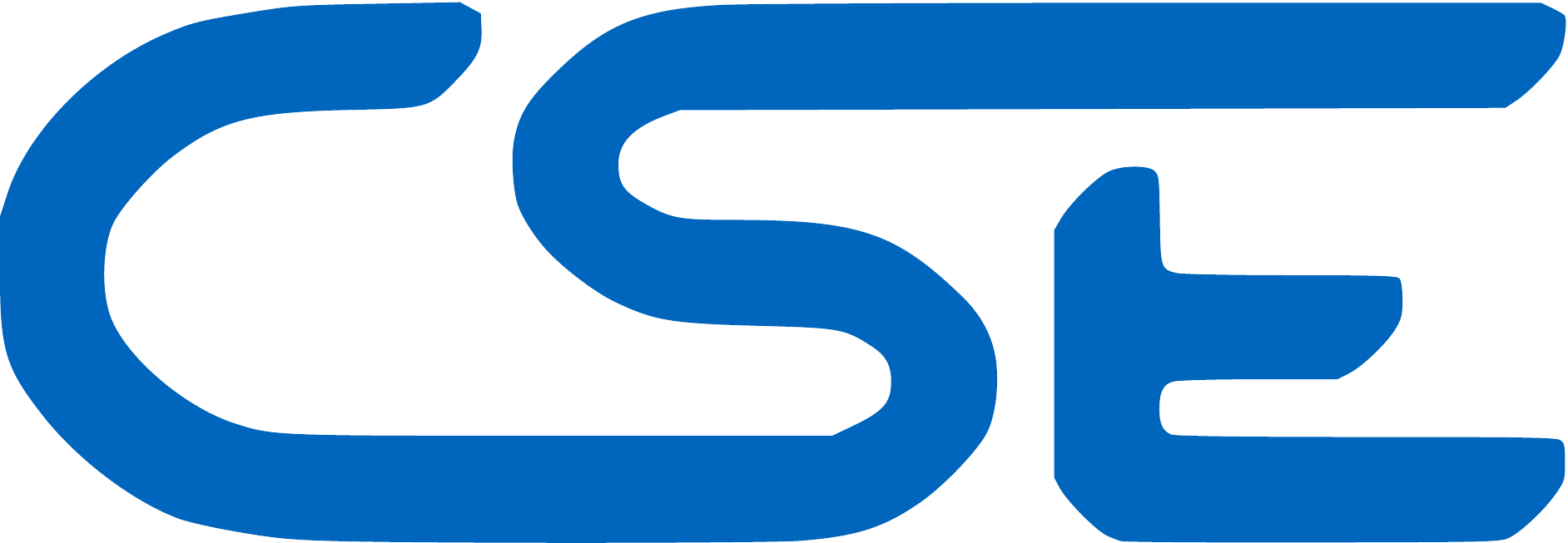}
\end{center}

 \clearemptydoublepage



\def\bcorcor{0.15cm}
\addtolength{\hoffset}{\bcorcor}

\thispagestyle{empty}

\vspace{4cm}
\begin{center}
    \includegraphics[width=4cm]{\universityLogo}\\
    \vspace{5mm}
    \huge \program \\
    \vspace{0.5cm}
    \large \university
\end{center}

\vspace{10mm}
\begin{center}
    {\Large \doctype}\\
    \vspace{10mm}
    {\LARGE \title}\\
    \vspace{10mm}

    \begin{tabular}{ll}
      \Large Author:                        & \Large \author \\[2mm]
      \Large 1\textsuperscript{st} examiner:& \Large \examinerOne\\[2mm]
      \Large Primary advisor:             & \Large \primaryAdvisor \\[2mm]
      \Large Submission Date:               & \Large \date
    \end{tabular}

    \vspace{\fill}
    \includegraphics[width=4cm]{\programLogo}
\end{center}

\addtolength{\hoffset}{\bcorcor}

\clearemptydoublepage

\thispagestyle{empty}
\vspace*{0.8\textheight}
\noindent
I hereby declare that this thesis is entirely the result of my own work except where otherwise indicated. I have only used the resources given in the list of references.

\vspace{15mm}
\noindent
\date \hspace{5cm} \author
\newpage

\clearemptydoublepage
\phantomsection
\addcontentsline{toc}{chapter}{Acknowledgements}

\vspace*{2cm}

\begin{center}
{\Large \textbf{Acknowledgments}}
\end{center}

\enote{PG}{please remove vspace from the whole thesis}

\enote{PG}{rearrange/edit some figure to beautify them. cite figure in caption.}

\enote{PG}{Missing reference: cite: iDocNADe in DocNADE, 'Keyword Learning for Classifying Requirements in Tender Documents' in RBMs, 'Identifying Patients with Diabetes using Discriminative Restricted Boltzmann Machines' in RBMs, 'Bi-directional recurrent neural network with ranking loss for spoken language understanding' in RNN, 'Deep Learning Methods for the Extraction of Relations in Natural Language Text' in RNN, 'Table Filling Multi-Task Recurrent Neural Network for Joint Entity and Relation Extraction' in RNN, 'Neural Architectures for Relation Extraction Within and Across Sentence Boundaries in Natural Language Text' RNN, 'LISA: Explaining Recurrent Neural Network Judgments via Layer-wIse Semantic Accumulation and Example to Pattern Transformation' in RNN,
'Deep Temporal-Recurrent-Replicated-Softmax for Topical Trends over Time' in RSM variants, 
'Replicated Siamese LSTM in Ticketing System for Similarity Learning and Retrieval in Asymmetric Texts' in LSTM and composite models}


It would be safe to say that this Master's thesis would not be possible without the support of many people. First and foremost, I would like to thank my thesis supervisor Prof. Dr.-Ing Thomas A. Runkler, for providing me with this opportunity to work under his guidance. You have greatly helped me in structuring this thesis' topic and providing me with the holistic guidance from time to time. I would like to thank you for the proposal discussion and review meetings which defined my project with more impact. Thank you also for reading the thesis draft and providing valuable review comments.


I would also like to thank Prof. Runkler to introduce me to Pankaj Gupta, a PhD student at LMU and Siemens AG. Pankaj, you have been a great advisor and I absolutely enjoyed working with you. I would also like to thank you for carefully imparting your technical knowledge to me as an advisor and a friend. I really liked your never-ending pragmatic approaches to solve complex problems. Technical sessions with you provided me with lot of insight and in-depth knowledge. I would like to thank you for allowing me to contribute in your research at Siemens AG. It was truly challenging and really exciting for me at the same time. I thank you for keeping me motivated this whole time and for advising on how to finish the experiments in due time. 
Thank you for reading so many versions of this thesis report. Your indispensable review comments helped in refining this thesis report. I admire your ability to see the nuances in everything. I feel extremely honored being your Master's thesis student and hope that we continue our collaboration in future too. Pankaj, you are an amazing personality, very helpful, caring, friendly and open by heart. You have always considered me as a friend which made my time more productive and fun. I would also like to thank Dr. Florian B\"uttner for your advice and discussions during our collaborative work for the conference papers. Thank you for being my technical and non-technical advisor for all the time we had together.


I thank you Dr. Ulli Waltinger and Dr. Bernt Andrassy to provide me with the computational resources at Siemens AG. Once again I would like to thank Pankaj at Siemens AG to help me with the required infrastructure, especially GPUs, for my thesis work. Without you all, it wouldn't have been possible to finish the countless experiments within the due time.




I would like to thank God to bless me with such a talented, loving and caring friend Sunita Pateer. You are a very special person to me. Without your constant motivation, love and support, this thesis would have been very difficult. I thank my parents for constantly keeping a check on my physical and mental well being and for supporting me with love as they always do. I would also like to thank Soumya Dubey for supporting me as an elder sister and as a friend. Soumya and Pankaj, you both have always provided constant support to me.

I acknowledge, with gratitude, the financial assistance from Siemens AG, especially Machine Intelligence (MIC) department, to support my thesis work. Thank you all my colleagues, including Subburam Rajaram at Siemens AG, for supporting me.


It is the outcome of the friendly and supportive environment at Siemens AG that I was fortunate enough to find so many great people to work with. I really enjoyed my collaborations with you all.

 
 \clearpage
\phantomsection

\begin{center}
\vspace*{11cm}
\textit{"Computers are incredibly fast, accurate and stupid; humans are incredibly slow, inaccurate and brilliant; together they are powerful beyond imagination."}
\end{center}
\par
\hspace*{7cm}
\textit{-Albert Einstein}

\clearemptydoublepage
\phantomsection
\addcontentsline{toc}{chapter}{TL;DR}

\vspace*{2cm}
\begin{center}
{\Large \textbf{TL;DR}}
\end{center}


Topic models such as LDA, DocNADE, iDocNADEe have been popular in document analysis. However, the traditional topic models have several limitations including: (1) Bag-of-words (BoW) assumption, where they ignore word ordering, (2) Data sparsity, where the application of topic models is challenging due to limited word co-occurrences, leading to incoherent topics and (3) No Continuous Learning framework for topic learning in lifelong fashion, exploiting historical knowledge (or latent topics) and minimizing catastrophic forgetting.   

This thesis focuses on addressing the above challenges within neural topic modeling framework. We propose: (1) Contextualized topic model that combines a topic and a language model and introduces linguistic structures (such as word ordering, syntactic and semantic features, etc.) in topic modeling, (2) A novel lifelong learning mechanism into neural topic modeling framework to demonstrate continuous learning in sequential document collections and minimizing catastrophic forgetting. Additionally, we perform a selective data augmentation to alleviate the need for complete historical corpora during data hallucination or replay.


\clearemptydoublepage
\phantomsection
\addcontentsline{toc}{chapter}{Abstract}

\vspace*{2cm}
\begin{center}
{\Large \textbf{Abstract}}
\end{center}



Availability of huge amount of unstructured text data demands the development of smart text mining techniques and algorithms. Topic modeling is one such technique in which the underlying semantic structures, i.e., topics, are extracted, from a large corpus of documents, in a co-occurrence pattern. While, the Natural Language Processing (NLP) community, in recent years, has developed state-of-the-art Topic Models (TMs) like Latent Dirichlet Allocation (LDA) \cite{DBLP:blei_LDA}, Replicated Softmax (RSM) \cite{DBLP:ruslan_RSM}, Document Neural Autoregressive Distribution Estimator (DocNADE) \cite{DBLP:Lauly_DocNADE} and Document Informed Neural Autoregressive Distribution Estimator (iDocNADE/iDocNADEe) \cite{DBLP:iDocNADE}, they have their own drawbacks. First drawback is the missing context information in Topic Models, i.e., loss of sequential (word order) information, after summarization into bag-of-words (BoW) representation and hence semantics of the context around the words is permanently lost. Second drawback comes from the issue of data sparsity, as Topic Models tend to generate less coherent and less meaningful latent document representations for datasets with low word co-occurence statistics i.e., small corpus or short text documents. Another motivation, rather than a drawback, is the idea to develop a topic model which can successively learn from each newly available dataset, in a sequential fashion, and retain the past knowledge, i.e., lifelong learning, to become an ever increasing knowledge base of latent topic and word representations.

In this thesis work, we have developed (1) a Contextualized Topic Model and (2) a Lifelong Learning Topic Model, which do not have the afore-mentioned drawbacks and learn informative latent topic and word representations which is shown with higher performance on TM evaluation metrics. In order to achieve these goals, we focus on three tasks. First, the missing language structure (word order) can be incorporated in Topic Model by combining an LSTM based language model (LSTM-LM) with TM to become a Contextualized Topic Model. While, TM learns latent topic and word representations from the entire document in a co-occurrence pattern, the LSTM-LM learns latent word representations in the language modeling framework. Second, the missing contextual knowledge, in sparse datasets, can be transferred via pre-trained distributed word embeddings (GloVe), trained on very large corpus (Wikipedia) in the form of a static prior knowledge to augment the knowledge of the local context around a word. Third, to sub-divide the task of lifelong learning into elementary tasks and study the effect of knowledge transfer from source datasets and knowledge retention of source datasets during topic modeling of dataset under consideration (target dataset). We call this Topic Model as Lifelong Learning Topic Model as datasets are modeled, one after another, in a sequential manner. Document Neural Autoregressive Distribution Estimator (DocNADE) \cite{DBLP:Lauly_DocNADE} has been used as the base Topic Model (TM) in all of the three tasks.

Chapter ~\ref{chapter:background} discusses past works on topic models (TMs) and addresses the drawbacks in the existing methods, and gives an introduction about the Document Neural Autoregressive Distribution Estimator (DocNADE). It explains how language models, like recurrent neural networks (RNNs) and long short-term memory (LSTM) networks, learn the underlying language structure semantics of text data. It also discusses the difference between language modeling and topic modeling. The last section of chapter ~\ref{chapter:background}, briefly, discusses some recent works related to the composition of language models and topic models, i.e., composite models, and the problems they target.


Further, chapters ~\ref{chapter:ctx_docnade} and ~\ref{chapter:lifelong_learning} focus on the architecture and design of contextualized topic model and lifelong learning topic model respectively. This is the core chapter of our thesis work, where we emphasize on two different areas. First, we explore the contextualized topic model architecture. We explain its weight sharing concept, attention parameter and demonstrate the integration of LSTM-LM into TM. Second, we explore the lifelong learning topic model architecture. We explain its knowledge transfer concept, knowledge retention concept, individual components of its final loss term and the attention parameters. All this exploration contribute to the major outcome of this thesis work.


Chapters ~\ref{chapter:results_ctx_docnade} and ~\ref{chapter:results_lifelong_learning} give a detailed overview of the experiments performed and the evaluations done in this thesis work across a wide variety of long and short text datasets under different settings of hyper-parameters. It explains the different evaluations metrics used in this thesis work, i.e., Information Retrieval (IR), Perplexity (PPL), Topic Coherence (COH) and Classification (F1), to quantify the improvements. 



In conclusion, we observe that the introduction of language structure in DocNADE Topic Model helps in learning the better latent topic and word representations which can be observed from the significant improvement in Topic Coherence (COH), Information Retrieval (IR) and Perplexity-per-word (PPL) evaluation metrics. For short text datasets, the introduction of pre-trained distributed word embeddings reports further improvement in latent topic and word representations. For the Lifelong Learning Topic Model, we observe that there is an inverse relationship between the degree of retention of past learning (from source datasets) and degree of learning on the target dataset. However, we observe that both retention and transfer of previous knowledge results in learning the better latent topic and word representations on the target dataset. Our work related to the contextualized topic model has been accepted at the \textit{Seventh International Conference on Learning Representations} (ICLR-2019), New Orleans.

 \tableofcontents

 \mainmatter


\addtolength{\evensidemargin}{-12mm}

%
%

\chapter{Introduction}
\label{chapter:Introduction}

\section{Overview}
\enote{PG}{remove red color, use italics instead}

To extract a sensible knowledge or get a meaningful representation from text data is one of the major tasks in the field of Natural Language Understanding (NLU). As more and more information is becoming available every day, that too mostly in the form of unstructured text data, it is becoming difficult to access the relevant and desired information. In recent years, the Natural Language Processing (NLP) community has developed some powerful techniques which can be used to mine through the data and get sensible information. There are a number of levels at which we can extract meaningful information - from words to sentences to paragraphs to documents. 


Topic modeling is one such technique, at document level, which can extract meaningful information in the form of latent semantic topical structures from a collection of documents. Contrasting from rule-based text mining techniques, it is an unsupervised technique used for finding a collection of words, i.e., a topic, such that the words exists in a repeating pattern of co-occurrence in a corpus of documents. For example, consider two topics mentioned below: 

\begin{itemize}
    \item \textbf{T1:} \{\texttt{atheism, sin, Jesus, god, bible}\}
    \item \textbf{T2:} \{\texttt{farm, crops, wheat, barley, tractor}\}
\end{itemize}

 The words present in \textbf{T1} would, mostly, co-occur \textit{in a similar context} in different documents across the document corpus and they together constitute a topic related to \textit{Religion}. Similarly, the words present in \textbf{T2} follow a high co-occurrence statistics and together constitute a topic related to \textit{Farming}. 


Probabilistic topic models, such as LDA  (Blei et al., 2003) \cite{DBLP:blei_LDA}, Replicated Softmax (RSM) (Salakhutdinov \& Hinton, 2009) \cite{DBLP:ruslan_RSM} and neural topic models like Document Neural Autoregressive Distribution Estimator (DocNADE) (Larochelle \& Lauly, 2012 \cite{DBLP:larochelle_DocNADE}; Zheng et al., 2016 \cite{DBLP:Zheng_deep_DocNADE}; Lauly et al., 2017 \cite{DBLP:Lauly_DocNADE}) are often used to extract meaningful topics from a document corpus. Subsequently, they learn latent document representations that can be used to perform NLP tasks such as information retrieval (IR), document classification or text summarization. In other words, topic models (TMs) are very useful for the purpose for document clustering, organizing large blocks of textual data, information retrieval from unstructured text and feature selection. 

\section{Problem statements}

While there are many advantages of topic models as mentioned before, there are some drawbacks of these probabilistic and neural topic models which deter their performance. We have mentioned below three such specific challenges.


\begin{enumerate}
    \item \textbf{Missing language structure information in topic models (TMs)}: Probabilistic topic models ignore word order by summarizing a given context as a \textit{bag-of-words} representation. Therefore, the semantics of the words in the context is lost. Consider the two sentences: 
    
    \enote{PG}{why all uppercase?}
    
    \begin{itemize}
        \item \texttt{Bear falls into market territory}
        \item \texttt{Market falls into bear territory}
    \end{itemize}

    The \textit{bag-of-words} representation of these two sentences would be the same (i.e., same uni-gram statistics) but they refer to different semantics. While, the word \textit{bear} in the first sentence is a proper noun \& subject and relates to the \textit{animal} topic; it is an object in the second sentence and relates to the \textit{stock market trading} topic.
    
    \item \textbf{Data sparsity}: Sparse datasets such as a corpus of short text documents, a corpus of few documents or both, generally, do not contain enough word co-occurrences across the document corpus for efficient application of topic models. Therefore, the application of topic models (TMs) can be challenging because the generated topics would not be coherent enough to be considered as a good representation of the entire document corpus.
    
    \item \textbf{Lifelong learning}: In the settings of online learning, where, topic models (TMs) learn latent topic representations progressively from each dataset that becomes available, TMs tend to forget the learning, i.e., \textit{catastrophic forgetting}, from previous datasets, while learning latent topic representations on the current dataset. Similarly, in the settings of transfer learning, TMs adapt the transferred knowledge according to new datasets, thus forgetting learning from previous datasets.
\end{enumerate}


The first two challenges of topic models (TMs) are more of their limitations which, ultimately and unfortunately, affect the latent topic representations of the document corpus and would have a significant impact on the performance of evaluation metrics like clustering, information retrieval (IR), perplexity per word (PPL), topic coherence (COH) etc. However, the third challenge is a way of operation, which, if successfully implemented, would enable the topic models to progressively learn and maintain topic and word knowledge information in a sequential modeling of text datasets.

\section{Proposals}

For the problems of topic models (TMs) stated above, we have briefly explained our proposed ideas below:

\begin{enumerate}
    \item To handle the above-mentioned problem of \textit{missing context in topic models (TMs)}, we focus on incorporating the language structure information from  language models (LMs) such as Long Short-Term Memory (LSTM) network into topic models (TMs). We have tried to keep the applicability of the model as general as possible. To quantify the improvements we mainly focus on information retrieval (IR), perplexity (PPL) and topic coherence (COH) as the evaluation metrics.
    
    \item To handle the problem of \textit{data sparsity} in topic models (TMs), we focus on the transfer of contextual knowledge via distributional priors i.e., pre-trained distributed word embeddings. Distributed word embedding representations learned on huge datasets, like Wikipedia, using language models (LMs) encode the important information about the different types of context in which a particular word appear.
    
    \item For integration of \textit{lifelong learning} in topic modeling process, we focus our attention on three different aspects of \textit{lifelong learning}: 
    
    \begin{enumerate}
        \item Selection of those documents from previous datasets which have a good domain overlap with the current dataset and use them in co-training with the new dataset to enhance the word co-occurrence statistics. This would result in more coherent topic representations.
        
        \item Retention of knowledge from previous datasets by minimizing the deviation of new model parameters from the parameters learned on previous datasets.
        
        \item Explicit transfer of learning from previous datasets via word embeddings transfer with attention. Accumulation and transfer of the contextual information of words from previous datasets in the form of word embeddings would result in the more meaningful topic representations for the new dataset.
    \end{enumerate}
    
    Successful integration of these aspects in topic models (TMs) would result in the improved latent topic and word representations on the new dataset while minimizing the \textit{catastrophic forgetting} of knowledge from previous datasets.
\end{enumerate}

\chapter{Background Theory}
\label{chapter:background}

\section{Topic Modeling}
\enote{PG}{cite all figure you used from. Put the citation in the caption}

\subsection{Restricted Boltzmann Machine (RBM)}

\begin{figure}[t]
\includegraphics[]{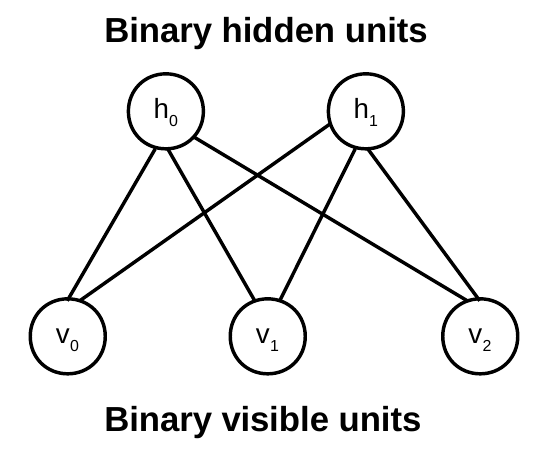}
\centering
\caption{Restricted Boltzmann Machine (RBM) model with bi-partite connections between binary visible and hidden units.}
\label{fig:rbm}
\end{figure}

Modeling of multidimensional data, like text data, by estimating its probability distribution is one of the most common problems in natural language processing (NLP) research. One core challenge that comes in the way of distribution estimation is the \textit{curse of dimensionality}. This problem had confounded NLP researchers for a long time. But, significant efforts have been made in the past to overcome this problem and achieve very efficient and fast algorithms for probability distribution estimation of text corpora/discrete multidimensional data. One example for such type of model is Restricted Boltzmann Machine (RBM) (Hinton 2002) \cite{DBLP:Hinton_2002}, (Gupta 2015) \cite{gupta2015keyword} \cite{gupta2015identifying} shown in fig. ~\ref{fig:rbm}. It transforms binary input data into latent feature space, in an unsupervised fashion, and then regenerate the input data to learn probability distribution of input data. RBM is an undirected probabilistic graphical model which associate probability to a particular configuration of binary visible ($\textbf{v} \in \{0,1\}^{D}$) and hidden/latent layer ($\textbf{h} \in \{0,1\}^H$) in terms of a joint energy function, where $D$ is the size of the input vector and $H$ is the size of the latent vector. The joint energy function is described in eq. (~\ref{eq:rbm_energy}) and its associated probability is described in eq. (~\ref{eq:rbm_prob}), where $\textbf{W} \in \mathbb{R}^{D \times H}$, $\textbf{b} \in \mathbb{R}^D$ and $\textbf{c} \in \mathbb{R}^H$ are the model parameters.

\begin{equation}
    E(\textbf{v},\textbf{h}) = - \textbf{b}^T - \textbf{c}^T - \textbf{v}^T\textbf{W}\textbf{h}
    \label{eq:rbm_energy}
\end{equation}

\begin{equation}
    P(\textbf{v},\textbf{h}) = \frac{1}{Z} \exp (-E(\textbf{v},\textbf{h}))
    \label{eq:rbm_prob}
\end{equation}

\begin{equation}
    Z = \sum_{\textbf{v}}\sum_{\textbf{h}} \exp (-E(\textbf{v},\textbf{h}))
    \label{eq:rbm_partition}
\end{equation}


One drawback of RBM is that the exact inference of the probability term, in eq. (~\ref{eq:rbm_prob}), is intractable. The normalization constant ($Z$), in eq. (~\ref{eq:rbm_partition}), is called as \textit{partition function} which is responsible for making the sum of the probabilities of all possible configurations of visible (\textbf{v}) and hidden units (\textbf{h}) go to unity, as seen in eq. (~\ref{eq:rbm_prob}). However, summation over all possible binary configurations of visible (\textbf{v}) and hidden (\textbf{h}) units is exponential in number, hence computationally intractable. Again, when the log-likelihood of a set of input data is calculated, the partition function ($Z$) makes it intractable. So, RBM uses approximation algorithms to compute the inexact log-likelihood. As a result, it would be impractical to compute the gradient of an inexact log-likelihood function of input data. Therefore, for these particular reasons, training of RBM is done efficiently using Contrastive Divergence (CD) \cite{DBLP:Hinton_2002} or Persistent Contrastive Divergence (PCD) \cite{DBLP:Tieleman_PCD} and samples from RBM  can be generated by using a Markov Chain Monte Carlo (MCMC) algorithm to convergence, with Gibbs sampling as the transition operator using conditional probability eq. (~\ref{eq:rbm_cond_h_v}), for generating hidden state given visible state, and eq. (~\ref{eq:rbm_cond_v_h}) for generating visible state given hidden state.

\begin{equation}
    p(\textbf{h}|\textbf{v}) = \prod_{j=1}^H p(h_j|\textbf{v}) = \prod_{j=1}^H \mbox{sigmoid}(c_j + \textbf{v}^T W_{:,j})
    \label{eq:rbm_cond_h_v}
\end{equation}

\begin{equation}
    p(\textbf{v}|\textbf{h}) = \prod_{i=1}^D p(v_i|\textbf{h}) = \prod_{i=1}^D \mbox{sigmoid}(b_i + W_{i,:} \textbf{h})
    \label{eq:rbm_cond_v_h}
\end{equation}

\subsection{Latent Dirichlet Allocation (LDA)}

Topic modeling is the \enote{PG}{task} task of identifying multiple sets of words called topics which can best describe a given document corpus. These topics are usually called the latent information and helps in efficient processing of a large collection of documents. Latent Dirichlet Allocation (LDA) (Blei et al., 2003) \cite{DBLP:blei_LDA} is a generative probabilistic graphical model that represents an item (document) of a collection as a finite composite over a fixed latent set of topics and each topic is characterized by a distribution over words. This composite representation of documents over latent topics is very useful for basic text processing tasks like classification, information retrieval, sentiment analysis, summarization etc. while preserving essential statistical relationships. This type of topic modeling is identical to the probabilistic latent semantic analysis (pLSA) \cite{DBLP:Hofmann_pLSA}, except with the assumption that LDA has a sparse Dirichlet prior as topic distribution. The sparse Dirichlet priors encode the intuition that documents cover only a small set of topics and that topics use only a small set of words frequently. In practice, this results in a precise topic distribution for documents and better disambiguation of words. LDA is a generalization of the pLSA model, which is equivalent to LDA under a uniform Dirichlet prior distribution.


The generative process of a document (\textbf{w}) in LDA is described below:
\begin{enumerate}
\item Choose the number of words the document is going to have according to Poisson distribution i.e., \{$N \sim Poisson(\xi)$\}
\item Choose the probability mixture of topics the document is going to have according to Dirichlet distribution (over a fixed set of $K$ topics) i.e., \{$\theta \sim Dirichlet(\alpha)$\}
\item Now, generate all the words in document by:
\begin{enumerate}
\item Choose a topic from a multinomial distribution over the chosen set of topics $\theta$ i.e., \{$z_n \sim Multinomial(\theta)$\}
\item Choose a word from multinomial distribution of topic ($z_n$) over words using parameter $\beta$ i.e., \{$w_n \sim p(w_n|z_n, \beta)$\}
\end{enumerate}
\end{enumerate}

\begin{figure}[t]
\includegraphics[scale=0.5]{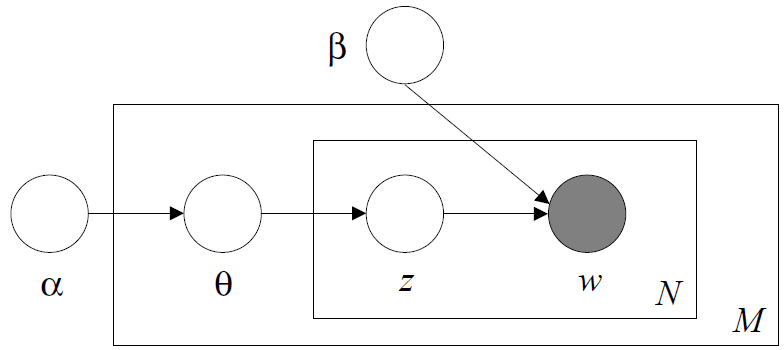}
\centering
\caption{Graphical model representation \cite{DBLP:blei_LDA} of Latent Dirichlet Allocation (LDA)}
\label{fig:lda}
\end{figure}


The graphical representation of LDA is shown in fig. ~\ref{fig:lda}. Let's say we are going to generate M documents with each document having N words. Now, $\alpha$ is a dirichlet distribution parameter (a vector) which samples the probability distribution $\theta$, of a fixed mixture of K topics, M times for each document. Using multinomial distribution, from each $\theta$, a topic ($z$) is sampled N times for N words in each document. Finally, using the topic $z$ and $\beta$ which is the probability distribution vector for topic $z$, a word is sampled N times. It is important to note that the joint posterior probability of $\theta$ and $z$, i.e., $p(\theta,z|\textbf{w},\alpha,\beta)$,  is intractable to compute for exact inference. Therefore, variational inference algorithm is used for inference and learning.

\subsection{Replicated Softmax (RSM)}

Probabilistic graphical models using Restricted Boltzmann Machines (RBMs) have been developed in the past for the task of topic modeling. While RBM has been able to produce good distributed latent representations on the input data and has performed well in tasks like information retrieval and clustering, it has been unable to properly deal with documents of different lengths, which makes learning very unstable and hard. For undirected models, like RBM, marginalizing over latent variables is generally an intractable operation, which makes modeling far more difficult. 


Replicated Softmax (RSM) (Salakhutdinov \& Hinton, 2009) \cite{DBLP:ruslan_RSM} model, and its variant RNN-RSM (Gupta et al., 2018) \cite{gupta2018deeptemporal}, is a combination of different sized Restricted Boltzmann Machines (RBMs). RSM models word count data by creating a seperate RBM for each document with as many softmax units, in visible layer, as there are words in the document ($D$). Further, it ignores the word order in the document to share the same set of weights, among all the softmax units, connecting them to binary latent units. Another way of putting it is instead of $D$ softmax units a multinomial unit can be used which can be sampled as many times as there are words in the document ($D$). Therefore, it has a better way of dealing with documents of different lengths.

\begin{figure}[t]
\includegraphics[width=\textwidth]{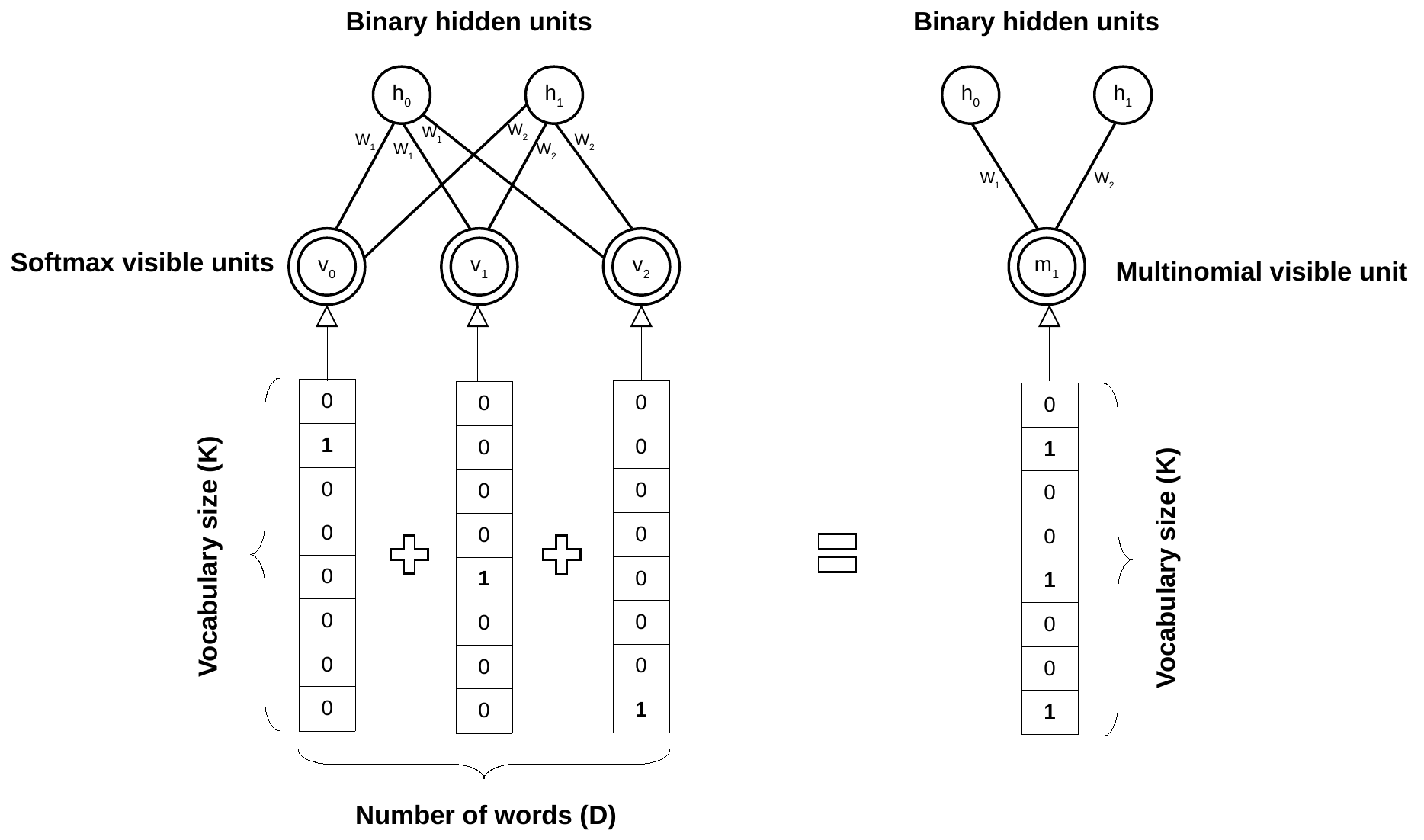}
\centering
\caption{Replicated Softmax (RSM) model with an illustration of how multiple softmax visible units can be combined into a multinomial visible unit for topic modeling. The connection weights between hidden units and softmax visible units are shared.}
\label{fig:rsm}
\end{figure}


Consider modeling a word count matrix $\textbf{V} \in \mathbb{R}^{K \times D}$ of a document where, $K$ is the vocabulary size and $D$ is number of words in the document. Each column of $\textbf{V}$ represents each word of the document in a \textit{one-hot encoding}. To model this observed binary matrix \textbf{V}, $D$ different RBMs with the same latent binary vector ($\textbf{h} \in \{0,1\}^H$), can be used and the combined energy function of the configuration \{\textbf{V},\textbf{h}\} is given in eq. (~\ref{eq:rsm_energy}), where $\{W,a,b\}$ are the model parameters and $H$ is the size of the latent unit. The probability of this visible binary matrix \textbf{V} as per the model is given in eq. (~\ref{eq:rsm_prob}), where, $Z$ is again the partition function described in eq. (~\ref{eq:rsm_partition}). Now, if we ignore the order of the words and share the weights connecting the visible softmax units to the binary hidden units, the same combined energy function (~\ref{eq:rsm_energy}) of the configuration \{\textbf{V},\textbf{h}\} can be described as in eq. (~\ref{eq:rsm_energy_reduced}), where $\hat{v}^k = \sum_{i=1}^D v_i^k$ denotes the count for the $k^{th}$ word in the document. The weights can now be shared by the whole family of different-sized RBM’s that are created for documents of different lengths, fig. (~\ref{fig:rsm}). Also, the bias term $a$ is scaled up by the document length $D$, in eq. (~\ref{eq:rsm_energy_reduced}), to allow the hidden topic units to efficiently deal with the documents of different lengths.

\begin{equation}
    E(\textbf{V},\textbf{h}) = -\sum_{i=1}^{D}\sum_{j=1}^{F}\sum_{k=1}^{K}W_{ij}^{k}h_jv_i^k -\sum_{i=1}^{D}\sum_{k=1}^{K}v_i^kb_i^k -\sum_{j=1}^{F}h_ja_j
    \label{eq:rsm_energy}
\end{equation}

\begin{equation}
    P(\textbf{V}) = \frac{1}{Z}\sum_{h} \exp (-E(\textbf{V},\textbf{h}))
    \label{eq:rsm_prob}
\end{equation}

\begin{equation}
    Z = \sum_{\textbf{V}}\sum_{\textbf{h}} \exp (-E(\textbf{V},\textbf{h}))
    \label{eq:rsm_partition}
\end{equation}

\begin{equation}
    E(\textbf{V},\textbf{h}) = -\sum_{j=1}^{F}\sum_{k=1}^{K}W_{j}^{k}h_j\hat{v}^k -\sum_{k=1}^{K}\hat{v}^kb^k -D\sum_{j=1}^{F}h_ja_j
    \label{eq:rsm_energy_reduced}
\end{equation}

Notice that the RSM model still has the partition function in probability equation, which makes its way to the \textit{log-likelihood} of the model, ultimately, making the exact inference intractable. Therefore, in RSM, the minimization of negative log-likelihood is done using Contrastive Divergence (CD) \cite{DBLP:Hinton_2002} algorithm.

\subsection{Neural Autoregressive Distribution Estimator (NADE)}

\begin{figure}[t]
\includegraphics[width=0.8\textwidth]{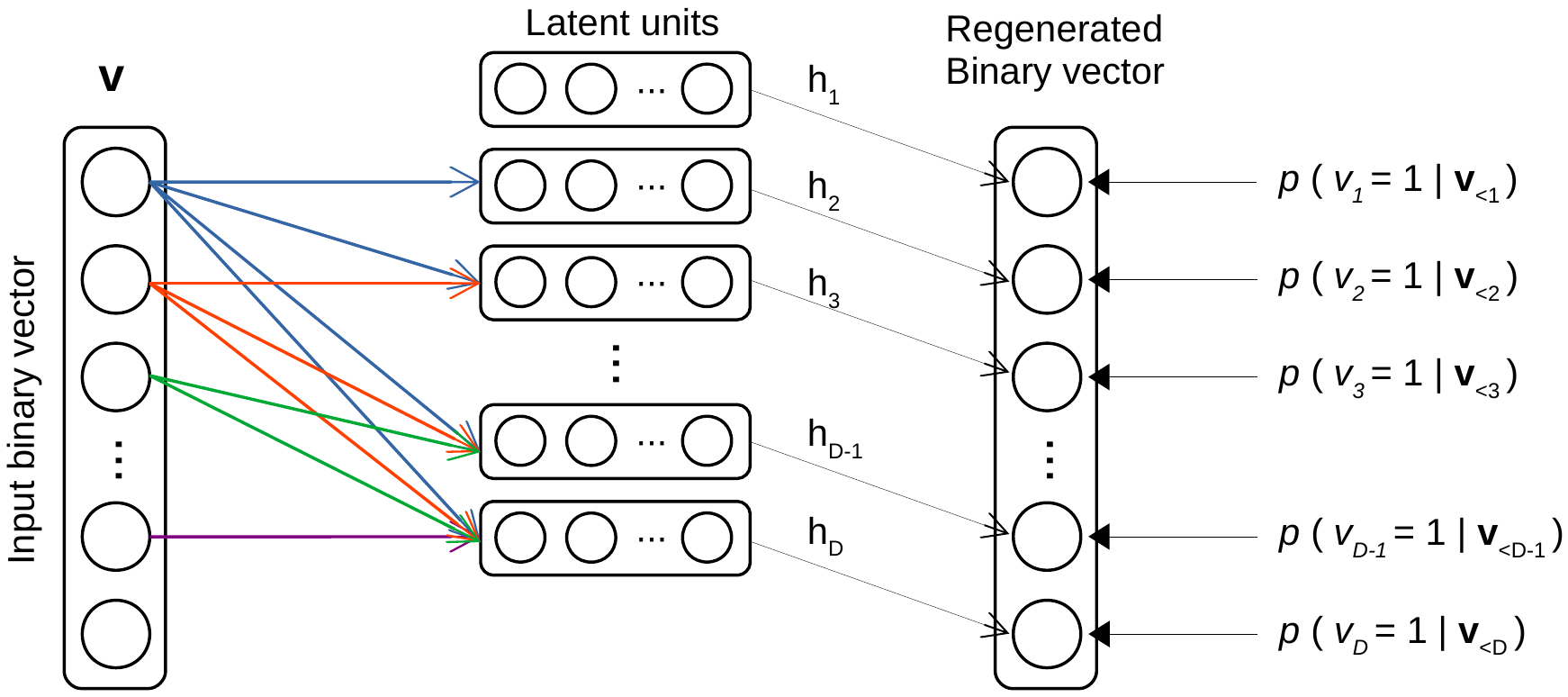}
\centering
\caption{Neural Estimator Distribution Estimator (NADE) \enote{PG}{use v to represent input, not x, as in equations} model with autoregressive connections between binary input and hidden units. Connections shown with the same color share same weights. Connections between hidden units and regenerated output are linear in nature.}
\label{fig:nade}
\end{figure}

We have already seen that the calculation of the exact probability of a binary observable in RBM is computationally intractable for moderately large models. The consequence of this is that the RBM cannot be trained using normal gradient based methods like Stochastic Gradient Descent method (SGD) but using Contrastive Divergence (CD) \cite{DBLP:Hinton_2002}, which makes an approximation of the log-likelihood of the model. Also, it approximates the probability of unseen test samples which makes it hard to understand how good the learned model distribution fits the actual data. To tackle this problem Larochelle \& Murray (2011) \cite{DBLP:Larochelle_NADE} introduced a feed forward neural network called Neural Autoregressive Distribution Estimator (NADE) which is inspired by RBM but is asymmetrical in structure. The connections between binary input units ($\textbf{v} \in \{0,1\}^D$) and hidden units ($\textbf{h} \in \mathbb{R}^H$) is autoregressive in nature as seen in fig. ~\ref{fig:nade}, where $D$ is the size of the input/output layers and $H$ is the size of the hidden layer. This autoregressive framework is employed to model the exact probability distribution of high dimensional binary input variables. In contrast to RBM, computation of the probability of binary observable ($\textbf{v}$) under NADE model is tractable and efficient.


Consider modeling of a $D$ dimensional binary vector $\textbf{v} = [v_1, ... , v_{i-1}, v_i, ... , v_D]^T$. NADE computes the exact probability of vector $\textbf{v}$, using chain rule, as mentioned in eq. ~\ref{eq:nade_prob}, where, $\textbf{v}_{<i}$ consists of all the binary units before $v_i$ i.e., $[v_1, ... , v_{i-1}]$, and $v_i$ is the $i_{th}$ binary unit in vector $\textbf{v}$. However, the probability conditional $p(v_i|\textbf{v}_{<i})$ can be converted into a more suitable form as in eq. ~\ref{eq:nade_conditional}.

\begin{equation}
    p(\textbf{v}) = \prod_{i=1}^{D}p(v_i|\textbf{v}_{<i})
    \label{eq:nade_prob}
\end{equation}

\begin{equation}
    p(v_i|\textbf{v}_{<i}) = \sum_{\textbf{v}_{>i}}\sum_{\textbf{h}}p(v_i,\textbf{v}_{>i},\textbf{h}|\textbf{v}_{<i})
    \label{eq:nade_conditional}
\end{equation}

It is already established that the conditional $p(v_i,\textbf{v}_{>i},\textbf{h}|\textbf{v}_{<i})$ is intractable to compute in RBM. Therefore, this conditional should be approximated with another distribution \,\,\,\,\, $q(v_i,\textbf{v}_{>i},\textbf{h}|\textbf{v}_{<i})$ in such a way that $q(v_i|\textbf{v}_{<i})$ is easy to compute and acts as an approximation of the actual conditional $p(v_i|\textbf{v}_{<i})$. Larochelle \& Murray used \textit{mean-field distribution}, in which a factorial decomposition is assumed to compute the approximation $q(v_i,\textbf{v}_{>i},\textbf{h}|\textbf{v}_{<i})$. They further proceed by minimizing the KL divergence between $p(v_i,\textbf{v}_{>i},\textbf{h}|\textbf{v}_{<i})$ and \,\,\,\, $q(v_i,\textbf{v}_{>i},\textbf{h}|\textbf{v}_{<i})$.


However, this mean-field approximation computation is quite slow with convergence taking upto many iterations. The calculation is also impractical for large dimensions ($D$) of binary visible vector ($\textbf{v}$) because of the repetition of the computation for all units of the binary visible vector. To keep the things simple, NADE model uses only one iteration of mean-field approximation, which transforms to a feed forward neural network with a single hidden layer with tied, autoregressive, connections between visible and hidden units. So, the equations for computation of hidden units and output units can be described as in equations (~\ref{eq:nade_hidden_conditional}) and (~\ref{eq:nade_prob_conditional}). These autoregressive connections also helps in speeding up the computation of all conditionals in eq. (~\ref{eq:nade_prob}). Also, NADE model is different from RBM as it has different matrices for encoding and decoding as seen in fig. (~\ref{fig:nade}). Finally, the training is done by minimizing the average negative log-likelihood ($NLL$) of the training dataset using any gradient based methods, eq. (~\ref{eq:nade_nll}).

\begin{figure}[h]
\includegraphics[scale=0.78]{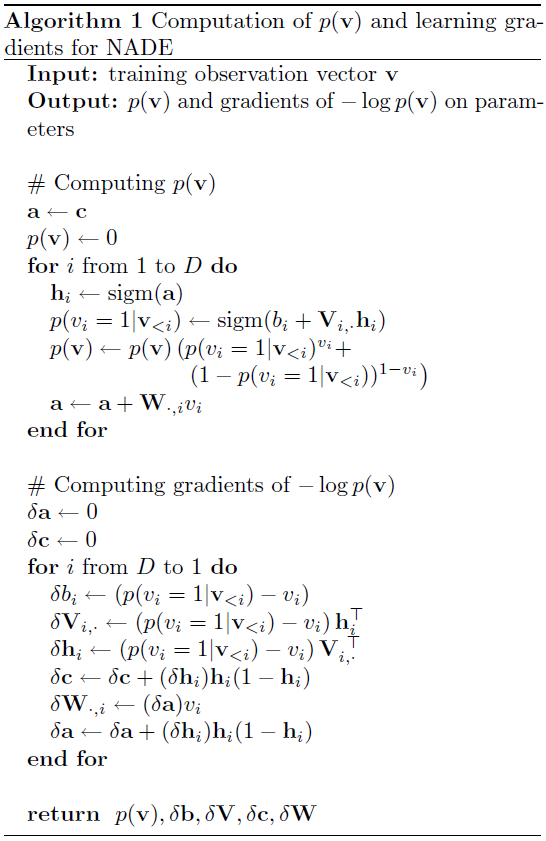}
\centering
\caption{Algorithm \cite{DBLP:Larochelle_NADE} for computation of $p(\textbf{v})$ and gradients in NADE model}
\label{fig:nade_algorithm}
\end{figure}

\begin{equation}
    NLL = \frac{1}{N}\sum_{n=1}^{N}-\log p(\textbf{v}^n) = \frac{1}{N}\sum_{n=1}^{N}\sum_{i=1}^{D}-\log p(v_i^n|\textbf{v}_{<i}^n)
    \label{eq:nade_nll}
\end{equation}

\begin{equation}
    \textbf{h}_i(\textbf{v}_{<i}) = g(\textbf{c} + \sum_{k<i} \textbf{W}_{:,k} v_{k})
    \label{eq:nade_hidden_conditional}
\end{equation}

\begin{equation}
    p(v_i=1|\textbf{v}_{<i}) = \mbox{sigmoid}(b_i + \textbf{V}_{i,:} \cdot \textbf{h}_i(\textbf{v}_{<i}))
    \label{eq:nade_prob_conditional}
\end{equation}

where, 
\begin{itemize}
    \item g() is any activation function and $N$ is the total number of documents in dataset
    \item $\textbf{W} \in \mathbb{R}^{D \times H}$ is the encoding matrix connecting visible layer to the hidden layer
    \item $\textbf{V} \in \mathbb{R}^{H \times D}$ is the decoding matrix connecting the hidden layer to the output layer
    \item $\textbf{c} \in \mathbb{R}^H$ and $\textbf{b} \in \mathbb{R}^D$ are the encoding and decoding biases respectively
\end{itemize}

NADE is also closely related to autoencoders, i.e. neural networks trained to reproduce the input at the output. An alternative view of NADE is as an autoencoder that has been wired such that its output can be used to assign probabilities to input observations in a valid way. Algorithm for computation of $p(\textbf{v})$ and gradients in NADE model is detailed in fig ~\ref{fig:nade_algorithm}.


In conclusion, the NADE model transforms an intractable distribution from RBM into a tractable and efficient distribution. In NADE the probabilities of input binary variables can be computed accurately and training can be done using stochastic gradient descent (SGD) or any other gradient based methods without making any approximation.

\subsection{Document Neural Autoregressive Distribution Estimator (DocNADE)}

We have already seen that, for the task of topic modeling, RBM is capable of modeling binary vectors observations, but it is incapable of modeling word count vectors i.e., RBM cannot model \textit{bag-of-words} representations of documents. For that task we have Replicated Softmax (RSM) model, which has softmax
\enote{PG}{use $\times$ for multiplication sign}
observable units and can model the binary observable matrix using separate RBMs for each word. RSM can handle variable length documents by sharing visible to hidden weights among $D$ softmax observable units or single multinomial observable unit sampled $D$ times.

\begin{figure}[t]
\center
\includegraphics[scale=0.75]{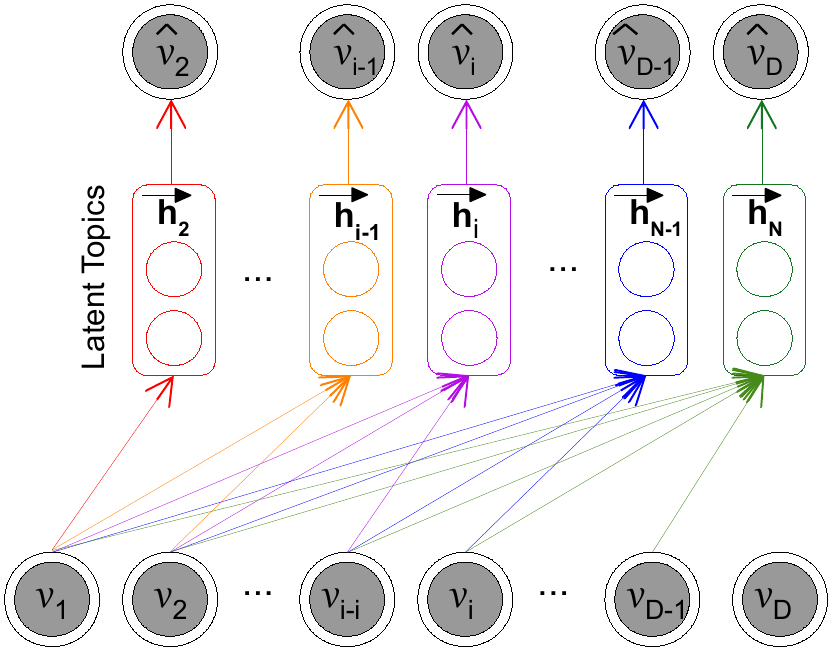}
\caption{DocNADE \cite{DBLP:Lauly_DocNADE} \enote{PG}{cite. show weight sharing. put in caption the significance of colors. why arrow on the hidden vector. it is uni-directional only. what is $\hat{v}$. Put here in caption} topic model with autoregressive connections between input softmax units and latent units. The arrow ($\rightarrow$) inside the hidden units ($\textbf{h}_i$) denotes that each hidden unit ($\textbf{h}_i$) takes into account the previous words only ($v_1, ..., v_{i-1}$) for regeneration probability $\hat{v}_i$ of the word $v_i$. Also, connections between a particular softmax visible ($v_i$) and its corresponding hidden units ($\textbf{h}_{i+1}, ..., \textbf{h}_D$) share weights.}
\label{fig:docnade_background}
\end{figure}

Similarly, Document Neural Autoregressive Distribution Estimator (DocNADE) (Laroc\-helle \& Lauly, 2012) \cite{DBLP:Lauly_DocNADE} (Gupta et al., 2019) \cite{DBLP:iDocNADE} is an extension of NADE model which can learn meaningful representations of texts from a collection of documents in an unsupervised fashion. Similar to NADE, it has a feed forward neural network architecture and learns the probability distribution of the \textit{bag-of-words} representation of documents. It has the same autoregressive connections between the softmax visible units and hidden layers as in NADE as seen in fig. ~\ref{fig:docnade_background}. The hidden layers, in DocNADE, capture the latent semantic structures of the documents called as topics, hence topic modeling. DocNADE generate topics based on the co-occurence statistics of words in different documents.

\begin{figure}[t]
\includegraphics[scale=0.5]{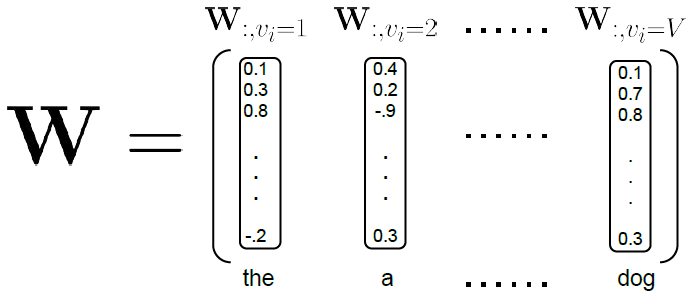}
\centering
\caption{Word representation matrix of DocNADE with each column as vector representation of words in the vocabulary V \cite{DBLP:Lauly_DocNADE}}
\label{fig:docnade_W}
\end{figure}

\begin{figure}[t]
\includegraphics[scale=0.5]{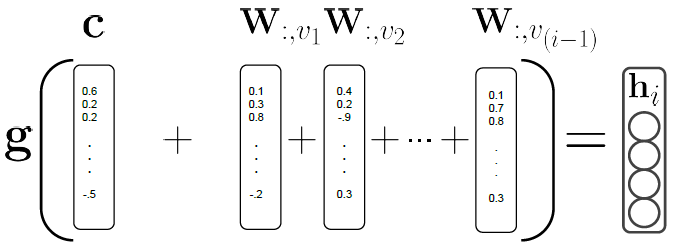}
\centering
\caption{Computation of hidden layer \cite{DBLP:Lauly_DocNADE}, \textbf{g}(\,) can be any activation function}
\label{fig:docnade_hidden}
\end{figure}

The \textit{bag-of-words} representation of a document of arbitrary size $D$ can be written as $\textbf{v} = [v_1, v_2, v_3, ... , v_D]$, where $v_i \in \{1,2, ... , V\}$ is the index of the $i^{th}$ word in the fixed vocabulary of size $V$. DocNADE uses a word representation matrix $\textbf{W} \in \mathbb{R}^{H\times V}$, where $H$ is the dimension of hidden layer and each column $\textbf{W}_{:,v_i}$ of the matrix is a vector representation of the word $v_i$ in the vocabulary $V$ as seen in fig. ~\ref{fig:docnade_W}. The hidden layer representation can be calculated using eq. (~\ref{eq:docnade_hidden}), where the embedding of the $i^{th}$ word in the document is the column with index $v_i$ in the matrix \textbf{W} and $\textbf{c} \in \mathbb{R}^H$ is the input bias. Figure ~\ref{fig:docnade_hidden} illustrates the hidden layer computation as mentioned in eq. (~\ref{eq:docnade_hidden}), where g() is any activation function.

\begin{equation}
    \textbf{h}_i(\textbf{v}_{<i}) = g(\textbf{c} + \sum_{k<i} \textbf{W}_{:,v_k})
    \label{eq:docnade_hidden}
\end{equation}

\begin{figure}[t]
\includegraphics[scale=0.25]{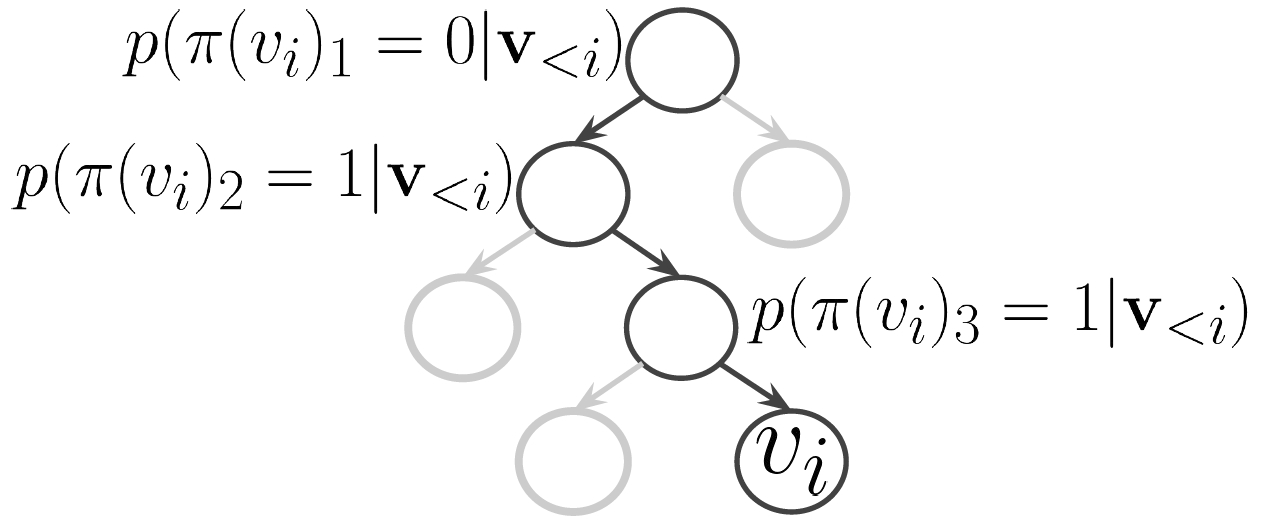}
\centering
\caption{The highlighted nodes indicate the path from main node to word $v_i$ in probabilistic binary tree \cite{DBLP:Lauly_DocNADE}}
\label{fig:docnade_prob_tree}
\end{figure}

Using hidden layer representation, total probability of a document of size $D$ can be calculated, in an autoregressive manner like NADE, as mentioned in eq. (~\ref{eq:docnade_prob}). To compute each conditional $p(v_i|\textbf{v}_{<i})$ in eq. (~\ref{eq:docnade_prob}), a simple softmax layer can be used with a shared weight matrix $\textbf{V} \in \mathbb{R}^{H \times V}$ and bias $\textbf{b} \in \mathbb{R}^V$ with each hidden layer $\textbf{h}_i$. But, as the vocabulary size $V$ of a document corpus is generally a very high number and a softmax layer scales linearly with vocabulary size $V$ i.e., computational complexity of $O(V)$. Therefore, it is prohibitive to use a linear softmax layer, instead DocNADE uses a probabilistic binary tree softmax to reduce the complexity to $O(\log V)$, for a balanced tree. This probabilistic tree approach has been used in probabilistic language models in the past (Morin and Bengio, 2005 \cite{DBLP:Morin}; Mnih and Hinton, 2009 \cite{DBLP:Mnih}). Each of the word $v_i$ in the vocabulary is present at the leaves of this binary tree and the probability of a word can be calculated by multiplication of all node probabilities in the path, $\textbf{l}(v_i) = [l(v_i)_1, l(v_i)_2, ...]$ to reach the word $v_i$ in the tree as seen in fig. ~\ref{fig:docnade_prob_tree} and eq. (~\ref{eq:docnade_prob_tree_vi}), where $\Pi(v_i) = [\pi(v_i)_1, \pi(v_i)_2, ...]$ is the sequence of left/right decisions at every node in the path $\textbf{l}(v_i)$. The value of $\pi(v_i)_m$ would be 0 if $v_i$ is in the left sub-tree or 1 if it is in the right subtree. The probability of a node $\pi(v_i)_{m}$, in path $\textbf{l}(v_i)$, is calculated with logistic classifier using hidden layer $\textbf{h}_i(\textbf{v}_{<i})$ and weights $\textbf{V}_{l(v_i)_{m,:}}$ as shown in eq. (~\ref{eq:docnade_prob_tree_vi_node}). The matrix $\textbf{V}^{V\times H}$ store the weights for each logistic classifier in its rows. Then, the total probability of the tree path $\textbf{l}(v_i)$ can be calculated, using eq. (~\ref{eq:docnade_prob_tree_vi_node}), as shown in eq. (~\ref{eq:docnade_prob_tree_vi}).

\begin{equation}
    p(\textbf{v}) = \prod_{i=1}^{D}p(v_i|\textbf{v}_{<i})
    \label{eq:docnade_prob}
\end{equation}

\begin{equation}
    p(v_i = w|\textbf{v}_{<i}) = \prod_{m=1}^{\Pi(v_i)} p(\pi(v_i)_m = \pi(w)_m|\textbf{v}_{<i})
    \label{eq:docnade_prob_tree_vi}
\end{equation}

\begin{equation}
    p(\pi(v_i)_m = 1|\textbf{v}_{<i}) = \mbox{sigmoid}(\textbf{b}_{l(v_i)_m} + \textbf{V}_{l(v_i)_m,:} \cdot \textbf{h}_i(\textbf{v}_{<i}))
    \label{eq:docnade_prob_tree_vi_node}
\end{equation}

\begin{figure}[t]
\includegraphics[scale=0.65]{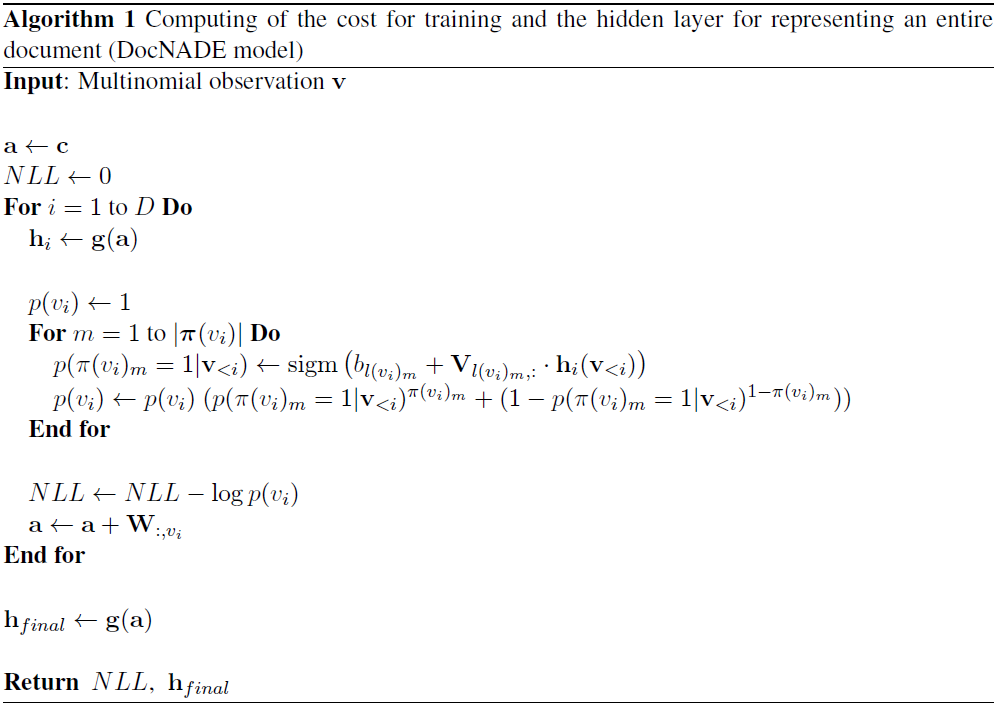}
\centering
\caption{Algorithm \cite{DBLP:Lauly_DocNADE} for computing hidden layer and negative log-likelihood loss for DocNADE topic model. Here, $|\pi(v_i)|$ denotes the length of $\Pi(v_i)$ vector of node probabilities to reach the word $v_i$ is the probabilistic binary tree and g() is an activation function. \enote{PG}{ + cite this}}
\label{fig:docnade_algorithm}
\end{figure}

Since, there are $\log(V)$ nodes in any path from main node to a leaf in a balanced binary tree, the computational complexity of regeneration of a word is $O(\log(V)H)$, instead of $O(VH)$ in case of a linear softmax layer. Therefore, for a document of $D$ words the complexity of regeneration is $O(\log(V)HD)$. One thing to note here is that the word order in a document is lost when converted to \textit{bag-of-words} representation $\textbf{v}$. But, DocNADE still learns good representations of documents by training on different permutations of the same \textit{bag-of-words} vector. With this procedure, DocNADE learns to predict a new word at a random position in a document while preserving overall semantics of the document which helps in identifying the intruder words in the document. Algorithm for computing hidden layer ($\textbf{h}$) and negative log-likelihood in DocNADE model is detailed in fig ~\ref{fig:docnade_algorithm}.


Therefore, DocNADE is a simple, powerful feed-forward neural network architecture which has a fast and efficient way to train on documents which have lost the information about the ordering of words.

\subsection{Latent Dirichlet Allocation with Product of Experts (ProdLDA)}

Generally, probabilistic topic models, like LDA, have the disadvantage of the intractable inference of the posterior distribution. To circumvent this problem, mean field approximation is used to do an efficient inference of the the posterior distribution. Therefore, the objective becomes to minimize the Kullback-Liebler (KL) divergence between the approximate variational posterior and the true posterior distributions. For LDA, there exist a closed form coordinate descent equations for this optimization problem because dirichlet and multinomial distribution are conjugate in nature. But, if the model changes, then it depends on the author's ability to derive the closed form equations for that model. In this way mean field approximation is not flexible in nature.


So, there is a need of a black-box inference method to circumvent this issue of deriving the optimization equations. Autoenconding Variational Bayes (AEVB) is one of the black-box inference methods generally used instead. In AEVB, an \textit{inference network} is used to calculate variational parameters using data as input and a Monte Carlo estimation, also known as ``reparameterization trick'', is used to calculate the expectation of variational posterior distribution. However, using AEVB and applying this ``reparameterization trick'' in topic models has some practical challenges like choice of reparameterization function and component collapsing. A new method called Autoencoding Variational Inference for Topic Models (AVITM) (Srivastava \& Sutton, 2018) \cite{DBLP:ProdLDA} propose to solve these problems.


LDA \cite{DBLP:blei_LDA} assume the document distribution as a mixture of multinomial distributions. The problem with this assumption is that this mixed distribution can never make predictions that are sharper than the individual components of the mixture. ProdLDA model offers to alleviate this problem by replacing the mixture distribution by product of experts, which by design is able to have sharper predictions than individual components. Additionally, the ProdLDA model is trained using AVITM as black-box inference method. 


In summary, ProdLDA trained with AVITM black-box method has following advantages:

\begin{enumerate}

\item It has better topic coherence than LDA.
\item Training with AVITM is very fast and efficient than standard mean-field because AVITM requires only one forward pass, through the neural network, on new data.
\item AVITM is a black-box, so no new mathematical derivations to handle changes in existing model.

\end{enumerate}

\subsection{Sparse Contextual Hidden and Observed Language Autoencoder\,\,\, (SCHOLAR)}

Document metadata like date, source, timestamp, rating and author is generally ignored during document topic modeling. Inspired by SLDA (Wang \& Grimson, 2007) \cite{DBLP:Blei_SLDA} and SAGE (Eisenstein et al., 2011) \cite{DBLP:Eisenstein_SAGE}, SCHOLAR (Card et al., 2017) \cite{DBLP:Card_SCHOLAR} incorporates this metadata information into topic modeling by making variations in LDA generative process on the lines of ProdLDA model. These changes in generative process of LDA enforce changes in the inference mechanism. Therefore, SCHOLAR uses variational inference framework as a black-box mechanism, like AVITM \cite{DBLP:ProdLDA}, to circumvent the problem of deriving new optimization equations. Therefore, it is safe to say that SCHOLAR topic model is equivalent to ProdLDA topic model with metadata information using similar autoencoding variational inference as black-box inference tool. In conclusion, SCHOLAR produces more coherent topics at the cost of worse perplexity-per-word than LDA topic model.

\section{Neural Language Modeling}

\subsection{Recurrent Neural Network (RNN)}

The sequence in which the particular words are written to make a meaningful sentence is very important to understand the syntax and semantics of the language used. As you read the sentence word by word you accumulate the meaning from the understanding of the previous words i.e., previous context. While, the sequence of words in a sentence is important, but equally important is the persistence of the knowledge from previous context to establish long-term dependency between two sentences or two paragraphs. If a machine can understand the natural language then it can perform tasks like sentiment analysis, language translation, answering the questions etc. However, if a machine has to understand the meaning of a sentence then it has to understand the syntax of the language used and also the semantics accumulated from previous context. 



\begin{figure}[t]
\includegraphics[]{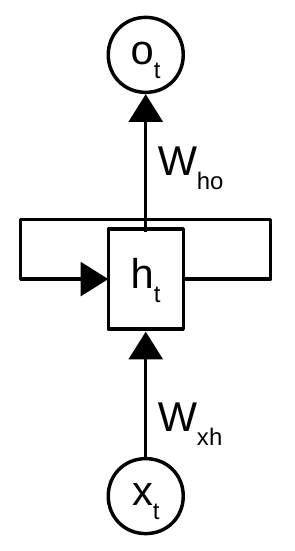}
\centering
\caption{Recurrent neural network with loop}
\label{fig:rnn_rolled}
\end{figure}

\begin{figure}[t]
\includegraphics[scale=0.95]{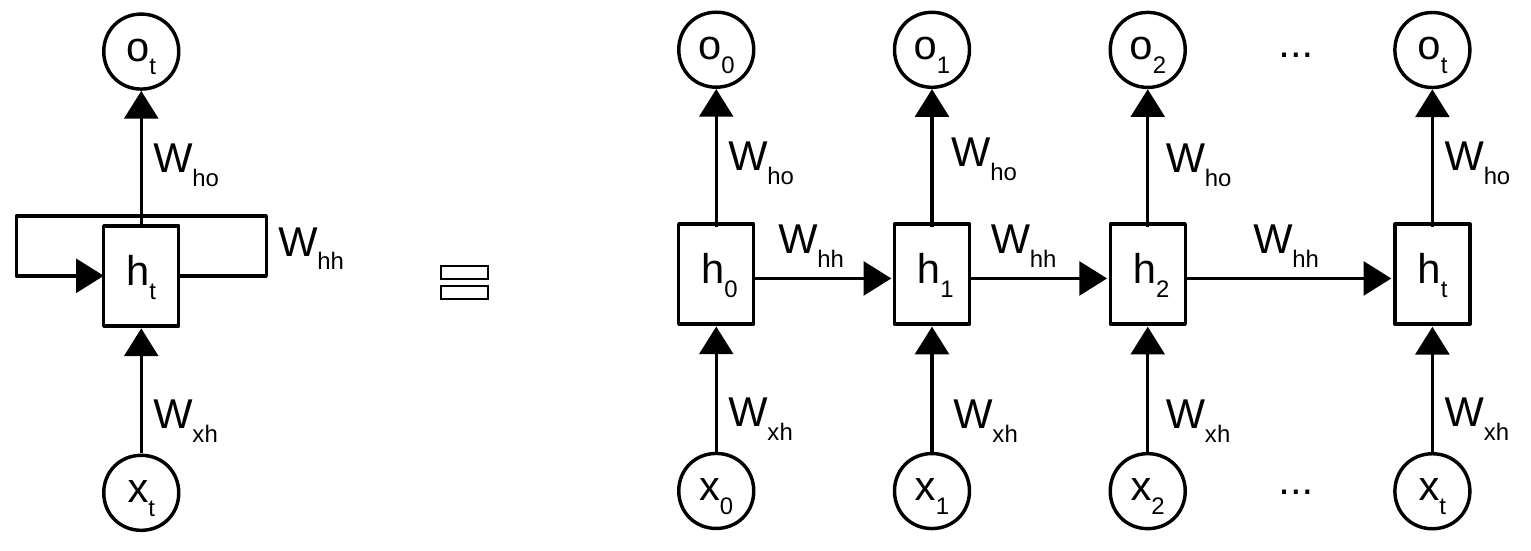}
\centering
\caption{Recurrent neural network with unfolding in time. Information from one time step ($h_1$) is transferred to the next time step ($h_2$) using $W_{hh}$ parameter.
\enote{PG}{say unfolding in time}}
\label{fig:rnn_unrolled}
\end{figure}

However, there is special type of neural network called Recurrent neural network (RNN) (Gupta et al., 2015) \cite{gupta2015masterthesis} (Gupta et al., 2016) \cite{gupta2016table} (Vu et al., 2016) \cite{DBLP:ThangVu_ranking} (Rajaram et al., 2018) \cite{subburam_thesis} (Gupta \& Sch\"utze, 2018) \cite{gupta2018lisa} which can process input data word by word in a given sequence and accumulates the information for a long period. As seen in fig. ~\ref{fig:rnn_rolled} RNN has a loop in its architecture which can be thought of as the transfer of accumulated information. This loop represents the temporal dimension of RNN where at each point in time ($t$) an input ($x_t$) from the sequence is given and the loop provides the information about the previous context [$x_1, ... , x_{t-1}$]. Upon unrolling the loop, as illustrated in fig. ~\ref{fig:rnn_unrolled}, RNN can be seen as the multiple copies of the same network, each passing some information to the successor. 

For language modeling using RNN, the input is given as a sequence of words ($x_t$) at different time steps ($t$) and calculate the generative probability of next words from hidden units ($h_t$) using parameters {$W_{xh}, W_{hh}, W_{ho}$}, shared across all time steps. Consider modeling a sequence of vector {$x_1, ... , x_n$}, we first calculate the hidden representation ($h_t$) according to eq. (~\ref{eq:rnn_hidden}) and then calculate output vector ($o_t$) and then, using a softmax layer at the output ($o_t$) we can calculate the prediction probabilities of all the words $p(x|h_t)$ in vocabulary ($V$) as per eq. (~\ref{eq:rnn_output}). Finally, we can calculate the total generative probability of the input sentence as mentioned in eq. (~\ref{eq:rnn_prob}).

\enote{PG}{$\hat{x}_t = p(x_t)$. correct it}

\begin{equation}
    h_t = g(W_{hh}\cdot h_{t-1} + W_{xh}\cdot x_t)
    \label{eq:rnn_hidden}
\end{equation}

Where, $W_{xh}$ is the parameter connecting input to hidden layer, $W_{hh}$ is the parameter responsible for accumulation of information by connecting hidden layer of one time step to the hidden layer of next time step, and $W_{ho}$ is the parameter connecting hidden layer to output. Function $g(\,)$ can be any activation function.

\begin{equation}
    p(x|h_t) = \mbox{softmax}(o_t); o_t = W_{ho}\cdot h_t
    \label{eq:rnn_output}
\end{equation}

\begin{equation}
    p(\textbf{x}) = \prod_{t=1}^n p(x = x_t|h_{t-1})
    \label{eq:rnn_prob}
\end{equation}

\begin{figure}[t]
\includegraphics[scale=0.95]{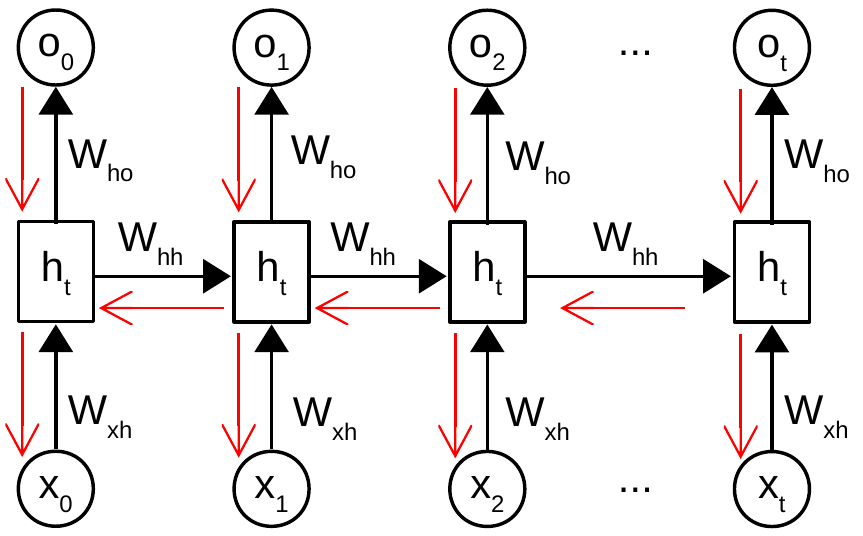}
\centering
\caption{Black arrows indicate the flow of information during forward pass (forward in time) in RNN and red arrows indicates the backward flow of final loss through time (BPTT) to calculate the gradients of weight parameters.}
\label{fig:rnn_unrolled_gradient}
\end{figure}

In the past RNNs have proven to be very successful in tasks like language modeling, machine translation and sentiment analysis. Traditional neural networks optimize their parameters using backpropagation algorithm. In backpropagation algorithm, the final loss moves backwards, from the output, through the weight parameters to the inputs in a way that assigns an error term to every weight parameter. Using that error term, the derivative of the final loss with respect to the weight parameters can be calculated which is essential for gradient based optimization methods. But, RNNs use an extension of backpropagation algorithm called backpropagation through time (BPTT) \cite{DBLP:Mozer_BPTT}, in which the final loss flows back in temporal dimension as demonstrated in fig. ~\ref{fig:rnn_unrolled_gradient} with gradients shown by red arrows.

\subsection{Long Short-Term Memory (LSTM)}

Sometimes, prediction of a new word require information about the recent past context, so RNN can learn that gap and can perform well in word prediction probability task. Let's consider the sentence ``Berlin is the capital of Germany'', here to predict the word ``Germany'' correctly the model can understand the context from ``Berlin'' and ``capital of'' that next word is going to be ``Germany''. Like this, RNNs can learn to use the past context where the gap is small i.e., short-term dependency. Now consider the new sentence ``I grew up in India. I studied ... I speak fluent \textit{Hindi}.'' the words ``speaks'' \& ``fluent'' suggests that the next word is the name of a language but doesn't narrow it down to ``Hindi''. But, if we look further back into the context then the word ``India'' suggests that language should be ``Hindi''. In theory, RNN should be able to learn these long-term dependencies but, in practice, it seems difficult for RNN to learn long-term dependencies. This problem of RNNs is due to Vanishing gradient problem. Vanishing gradient problem in RNNs was first explored by Hochreiter (1991) \cite{hochreiter1991}.


Now when we do backpropagation through time (BPTT) i.e moving backward in the network and calculating gradients of final loss with respect to the weights, the gradients tends to get smaller and smaller as we keep on moving backward in the Network, it is therefore called as the vanishing gradient problem. This means that the neurons in the Earlier layers learn very slowly as compared to the neurons in the later layers in the network hierarchy. The Earlier layers in the network are slowest to train. Earlier layers in the Network are important because they are responsible to learn and detecting the abstract patterns and are actually the building blocks of the whole network. Obviously, if the earlier layers give improper and inaccurate results, then how can we expect the next layers and the complete network to perform nicely and produce accurate results. The maximum value of the gradient of \textit{sigmoid} activation function is 0.25, which multiplies with itself in multiple layers during backpropagation algorithm and hence vanishes the gradients. That is why \textit{sigmoid} and \textit{tanh} activation functions are rarely used in RNN.


\begin{figure}[!htbp]
    \begin{minipage}[c]{0.48\textwidth}
    \includegraphics[width=\linewidth]{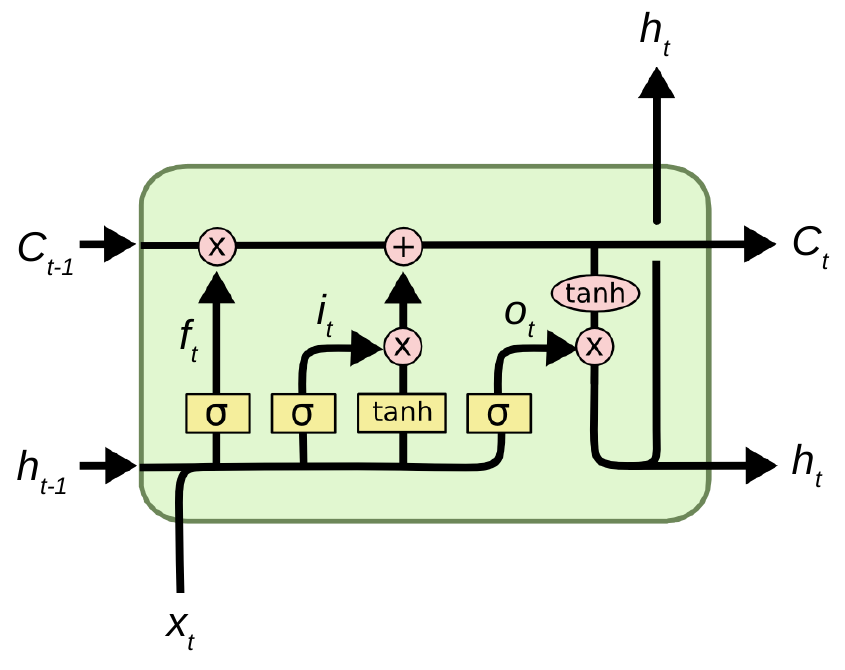}
    \end{minipage}%
    \hspace{5mm}
    \begin{minipage}[c]{0.42\textwidth}
    $f_t = \mbox{sigmoid}(W_{hf}\cdot h_{t-1} + W_{xf}\cdot x_t + b_f)$
    
    $i_t = \mbox{sigmoid}(W_{hi}\cdot h_{t-1} + W_{xi}\cdot x_t + b_i)$
    
    $\hat C_t = \mbox{tanh}(W_{hc}\cdot h_{t-1} + W_{xc}\cdot x_t + b_c)$
    
    $C_t = f_t \ast C_{t-1} + i_t \ast \hat C_t$
    
    $o_t = \mbox{sigmoid}(W_{ho}\cdot h_{t-1} + W_{xo}\cdot x_t + b_o)$
    
    $h_t = \mbox{tanh}(C_t) \ast o_t$
    \end{minipage}
    \caption{On the left, the gated internal structure of LSTM \cite{colah_LSTM} unit. On the right, the equations used to calculate the outputs of those gates using inputs, hidden units and weight parameters.}
    \label{fig:lstm_with_equations}
\end{figure}

\enote{PG}{where is the caption of LSTM cell unit}

\enote{PG}{figure and equation references in the text missing}

Long Short Term Memory (Hochreiter \& Schmidhuber, 1997) \cite{DBLP:Hochreiter_LSTM} (Gupta et al., 2018) \cite{gupta2018replicated} network, usually just called as ``LSTM'', is a special type of RNN, capable of learning long-term dependencies. LSTM is explicitly designed to avoid the long-term dependency problem of the conventional RNN. Remembering information for long periods of time is practically the default behavior of LSTM network, not something it struggle to learn. Instead of having a non-linearity repetition structure in RNN, LSTM has a 4 gated repetition structure as shown in fig. ~\ref{fig:lstm_with_equations}(left). In addition to transmission of previous hidden state to next hidden state, LSTM also transmits cell state $C_t$. Let's look at the equations of the LSTM cell and their description.

\enote{PG}{missing reference to equation}

First gate is called the forget gate ($f_t$) which decides what information in cell state ($C_t$) is not useful and has to be forgotten as can be seen in the (first) eq. in fig.  ~\ref{fig:lstm_with_equations}(right). The output of forget gate, between 0 and 1, gets multiplied in previous cell state ($C_{t-1}$) to forget the less important information and keep the more important ones. Second gate is called the input gate consisting of two parts, first is the input gate importance factor ($i_t$) with values between 0 and 1 and second is the new input information ($\hat C_t$). Therefore, the new important information is calculated using factor $i_t \ast \hat C_t$ and added in cell state ($C_t$) as seen in (second, third and fourth) equations in fig.  ~\ref{fig:lstm_with_equations}(right). The fourth and final gate is called the output gate ($o_t$) which decides how much information from new cell state ($C_t$) gets transmitted to output of the cell as can be seen in (fifth) eq. in fig.  ~\ref{fig:lstm_with_equations}(right). In a gist, forget gate ($f_t$) removes less important information, input gate ($i_t$) adds the new important information and output gate ($o_t$) decides how much of the new cell state gets transmitted through the output of the cell.


Now, it is evident that LSTM can learn to remove less important information or add more important information in the network at any time step and hence has a mechanism to remember log-term dependencies and not remember recent irrelevant information. To conclude, LSTM is a special type of RNN which is designed in a way to avoid the vanishing gradient problem and remember long-term dependencies, using gated architecture, which is very important for generative language modeling tasks.

\section{Composite Modeling}

\subsection{Topically Driven Neural Language Model (TDLM)}

Topically driven language model (TDLM) (Lau et al., 2017) \cite{DBLP:Lau_TDLM} propose to include the global semantic knowledge into the language model to increase the generative likelihood probability as opposed to language model itself. TDLM utilizes convolution based topic model and LSTM based language model (LSTM-LM). Given a document as input, TDLM uses the convolution network to generate a document vector from word embeddings of the words present in a document. After that, a document specific topic vector is calculated by association of the document vector with the topic matrix using an attention mechanism to compute a weighted mean of topic vectors. This document specific topic vector is then incorporated in the LSTM-LM for the prediction of succeeding words. While, the LSTM-LM performs langauge modeling at the sentence level (local syntax and semantics), the global document semantics is provided by the topic model using full document context except the sentence used by LSTM-LM, to prevent the TDLM model from memorizing the next word via the topic model.

\begin{figure}[h]
  \centering
  \includegraphics[width=0.95\textwidth]{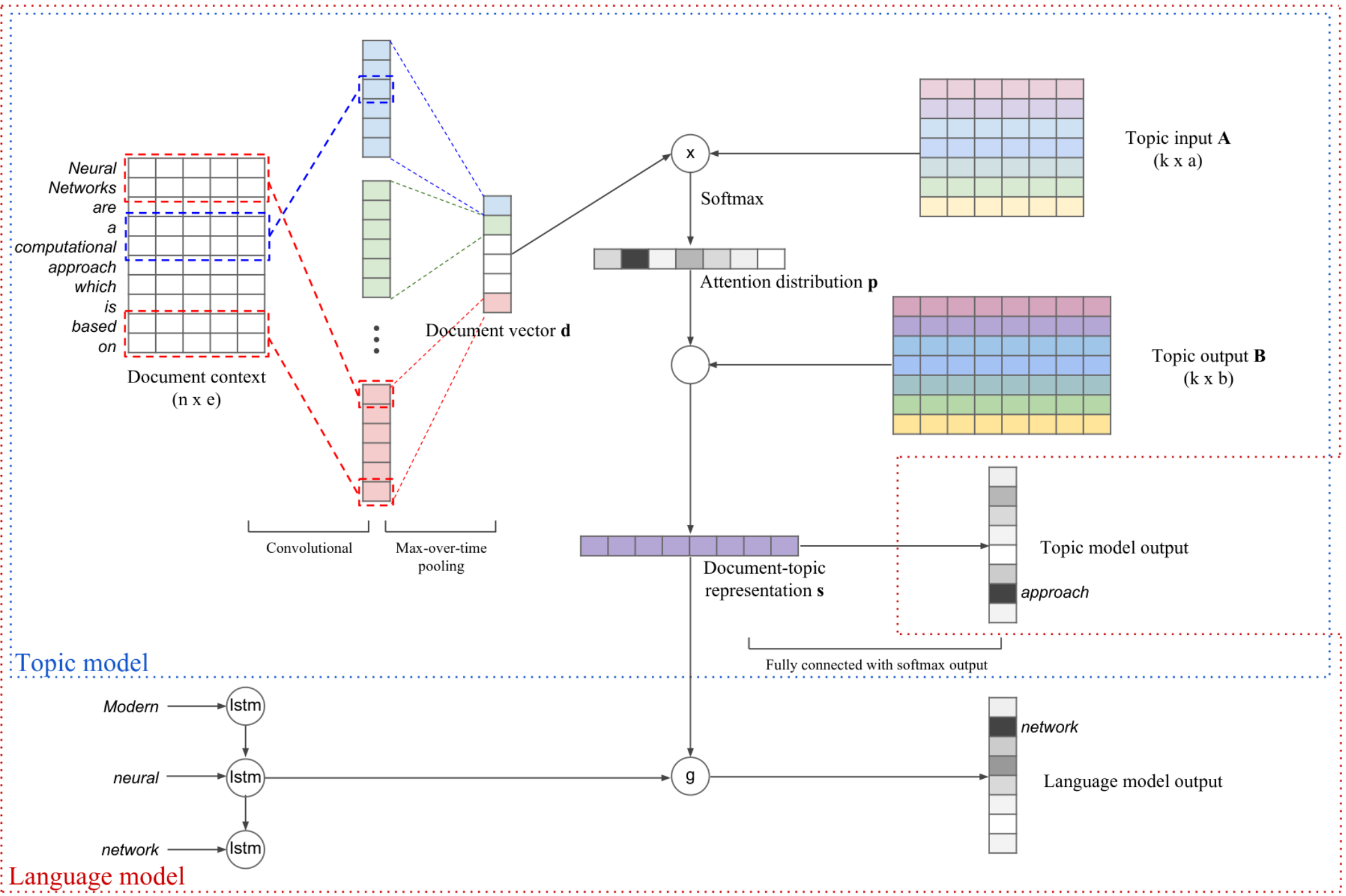}
  \caption{Overall architecture of TDLM model \cite{DBLP:Lau_TDLM}. It clearly illustrates how the global semantic knowledge in the form of latent topic information is incorporated into the language modeling side of TDLM model.
  \enote{PG}{move source in caption, cite paper instead. Do it for all other figures.}}
\end{figure}

As, TDLM model is fully neural network based, the whole model is trained, end-to-end, using stochastic gradient descent (SGD). TDLM resolves the problem of missing context in sentence level language model by including document topic information vector using convolution neural network. However it can be argued that instead of using sentences, why not use full document in language model? But, there will again be the problem of long-term dependencies and vanishing gradient in using full document as input. Refer to the original paper (Lau et al., 2017) \cite{DBLP:Lau_TDLM} for more detailed description of TDLM model.

\subsection{A Recurrent Neural Network with Long-Range Semantic Dependency (TopicRNN)}

TopicRNN (Dieng et al., 2017) \cite{DBLP:Dieng_TopicRNN} model, similar to TDLM, proposes a generative composite model by inclusion of long-term semantic dependencies from topic model into language model to improve the word prediction probabilities and sentence generation of language model. It combine latent dirichlet allocation (LDA) \cite{DBLP:blei_LDA} topic model and RNN based language model into a composite model called TopicRNN. TopicRNN uses Gaussian distribution instead of Dirichlet distribution in LDA topic model to allow for more flexibility during sequence prediction.

\begin{figure}[h]
  \centering
  \includegraphics[width=\textwidth]{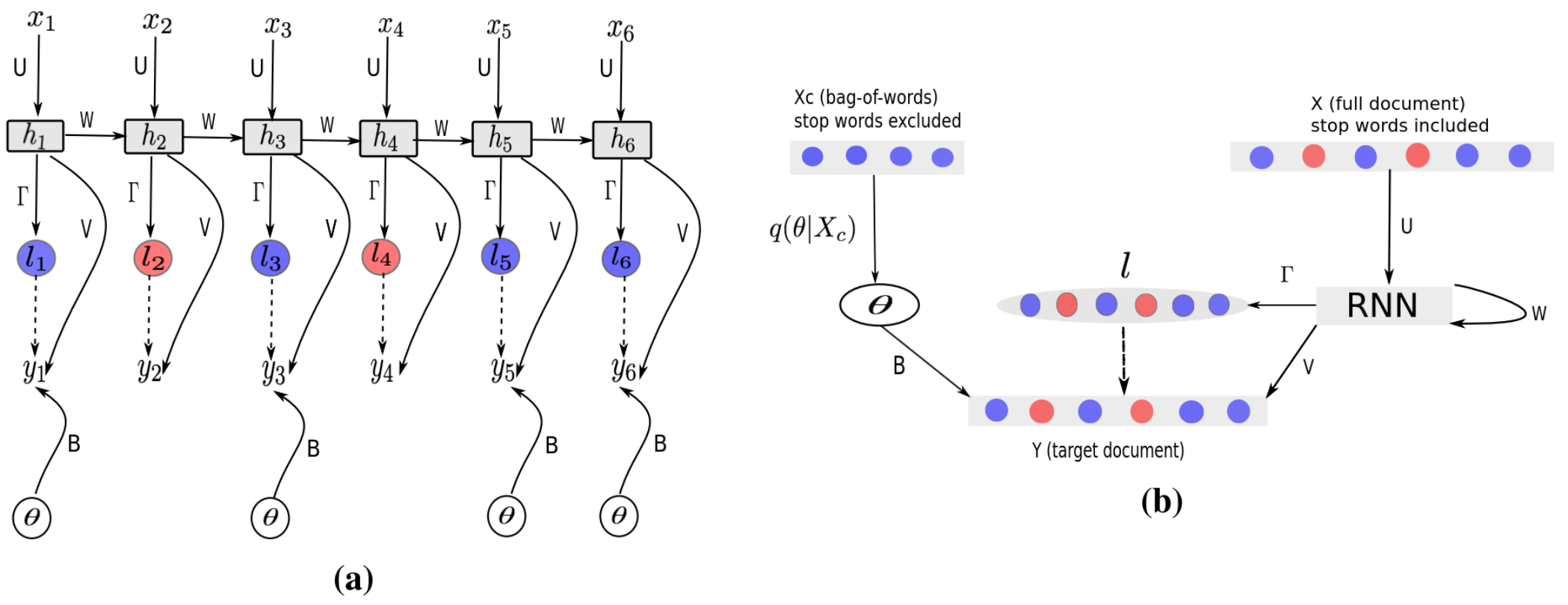}
  \caption{Overall architecture of TopicRNN model \cite{DBLP:Dieng_TopicRNN} with a visual representation of combination of topic vector ($\theta$) and RNN hidden states (\textit{h}) for joint prediction of words in the sequence}
  \label{fig:topicrnn}
\end{figure}

\enote{PG}{use LSTM-LM or LSTMLM throughout}

For a given \textit{bag-of-words} representation of a document, a topic vector ($\theta$) is derived. Using the topic vector $\theta$, RNN inputs $\textit{x}_i$, RNN hidden states $\textit{h}_i$ and the stop word indicator vectors $\textit{l}_i$ in a combined fashion the next word is predicted as illustrated in fig. ~\ref{fig:topicrnn}. The stop word indicator vector \textit{l} is introduced to automatically handle the stop words in language model as they are difficult to handle in topic model. Because of the use of Gaussian distribution based probabilistic topic model, the exact inference of the input data likelihood is intractable. Therefore, the training of TopicRNN is done end-to-end in an unsupervised fashion by maximizing the evidence lower bound (ELBO) of the variational inference of the objective function. TopicRNN addresses the problem of long-term dependencies in RNN based language model by incorporating global document semantics, in the form of a topic vector $\theta$, during word prediction. Refer to the original paper (Dieng et al., 2017) \cite{DBLP:Dieng_TopicRNN} for more detailed explanation of model architecture and inference mechanism.

\subsection{Topic Compositional Neural Language Model (TCNLM)}

Topic compositional neural language model (TCNLM) (Wang et al., 2018) \cite{DBLP:wang_TCNLM} follows a similar idea, as TopicRNN \cite{DBLP:Dieng_TopicRNN} and TDLM \cite{DBLP:Lau_TDLM}, of incorporating the global document semantic information from topic model into the local semantic information of language model in order to improve the word prediction probabilities and generation of meaningful sentences. They use neural topic model (NTM) based on variational autoencoder framework and LSTM based language model (LSTM-LM) to combine together into the TCNLM composite model. 

\begin{figure}[h]
  \centering
  \includegraphics[width=\textwidth]{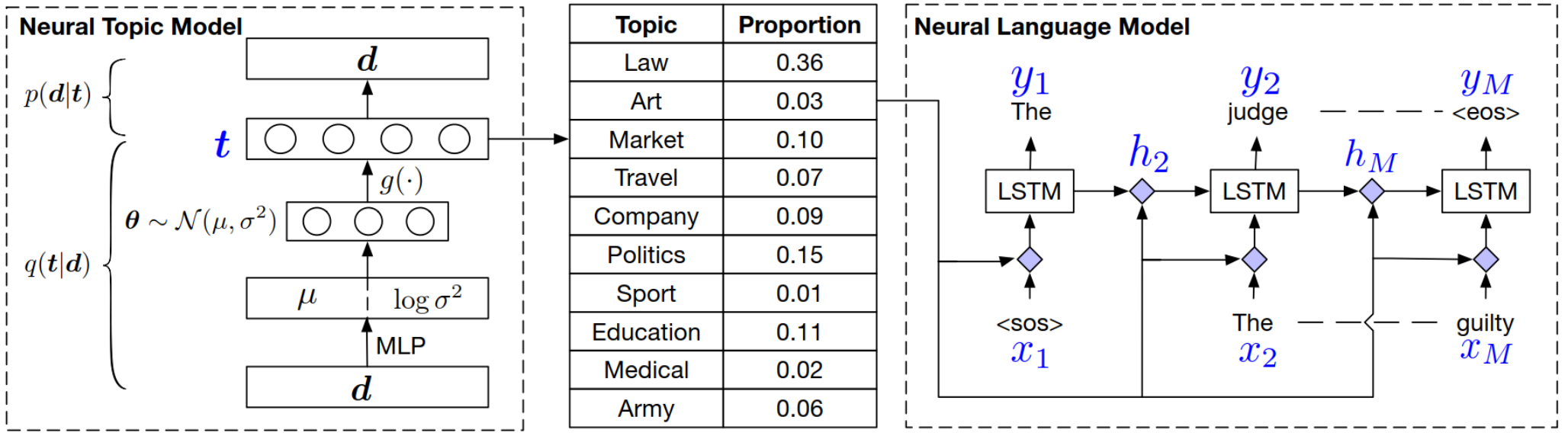}
  \caption{Overall architecture of TCNLM model \cite{DBLP:wang_TCNLM} with an illustration of the composition of the variational autoencoder based topic model and Mixture-of-Experts (MoE) based language model.}
  \label{fig:tcnlm}
\end{figure}

Given a \textit{bag-of-words} representation (\textbf{d}) of a document, TCNLM model uses the variational autoencoder structure in NTM to map \textbf{d} onto a topic vector \textbf{t}, which in turn is used to regenerate \textbf{d} as illustrated in fig. ~\ref{fig:tcnlm} (left). On the language modeling side, TCNLM uses Mixture-of-Experts (MoE) approach by keeping a separate list of LSTM weight matrices for each of the topics. For a given topic vector \textbf{t}, it combines the different LSTM weight matrices of all the topics according to the topic probabilities in \textbf{t}, then use the combined weight matrix for language modeling. Exact inference, in TCNLM, is intractable because of the variational autoencoder framework used on the topic modeling side. Therefore, the training of TCNLM model is done end-to-end by maximizing the evidence lower bound (ELBO) of the variational inference of the objective function. Refer to the original paper (Wang et al., 2018) \cite{DBLP:wang_TCNLM} for more detailed explanation of model architecture and inference framework.

All of the composite models mentioned above, ie., TDLM, TopicRNN and TCNLM, address the limitations of the long-term dependency and the missing context in LSTM/RNN based language models by incorporating the global semantic and contextual information via topic models. However, it is important to note that, our proposed models share the idea of a composite model architecture with TDLM, TopicRNN and TCNLM, but we are entirely focused on addressing the limitations of topic models using syntactic and local contextual information from language model, while TDLM, TopicRNN and TCNLM are focused on improving language model using global semantic knowledge from topic models.

%

\chapter{Contextualized topic model}
\label{chapter:ctx_docnade}

To summarize the modeling processes of topic models and language models, we have listed down some ideas we have already discussed in previous chapters:

\begin{enumerate}
    \item Topic models while very good at extracting latent semantic features, at the document level, does not take into account the order of words which can be a problem in some cases.
    \item Language models are very good at learning about the syntax and semantics of the language, mostly at the sentence level, by taking into account the actual order of the words.
    \item In case of sparse datasets, i.e., a small number of documents or the document length is very small, the topic models do not capture the latent features efficiently because of the low word co-occurrence statistics.
\end{enumerate}

To tackle the problems of \textit{missing language structure information} and \textit{data sparsity} in topic models, we describe two contributions below. We have used DocNADE as the base topic model in all of our contributions.


\section{Missing language structure information in DocNADE}

In the recent past, an LSTM based language model (LSTM-LM) called ELMo (Peters et al., 2018) \cite{DBLP:Peters_ELMo} has shown the capabilities of language models to capture different language concepts in a layer-wise fashion i.e., the lowest layer captures language syntax and the topmost layer capture language semantics. However, generally, in LSTM-LMs to regenerate a word, the model does not look beyond its current sentence. Therefore, the word occurrences and langauge semantics are modeled in a \textit{fine granularity} and do not capture semantics at the document level.


To mitigate this, recent studies such as TDLM (Lau et al., 2017) \cite{DBLP:Lau_TDLM}, Topic-RNN (Dieng et al., 2016) \cite{DBLP:Dieng_TopicRNN} and TCNLM (Wang et al., 2018) \cite{DBLP:wang_TCNLM} have integrated the merits of latent topic models (TMs) and neural language models (LMs). These composite models have focused on improving LMs with global semantics using latent topical information from TMs.


In contrast, DocNADE (Larochelle \& Lauly, 2012) \cite{DBLP:Lauly_DocNADE} learns word occurrences across the whole document i.e., \textit{coarse granularity} 
(in the sense that the regeneration probability of a word in a document equally depends on all the other words in the previous context of that word in the document)
However, since DocNADE is based on the \textit{bag-of-words} assumption, all of the language structure of the document is ignored.


To tackle this problem of missing language structure in topic models we incorporate language structure information using LSTM based language model (LSTM-LM), thereby accounting for word order (semantics) and language concepts (syntax).

\begin{figure}[t]
\center
\includegraphics[width=\textwidth]{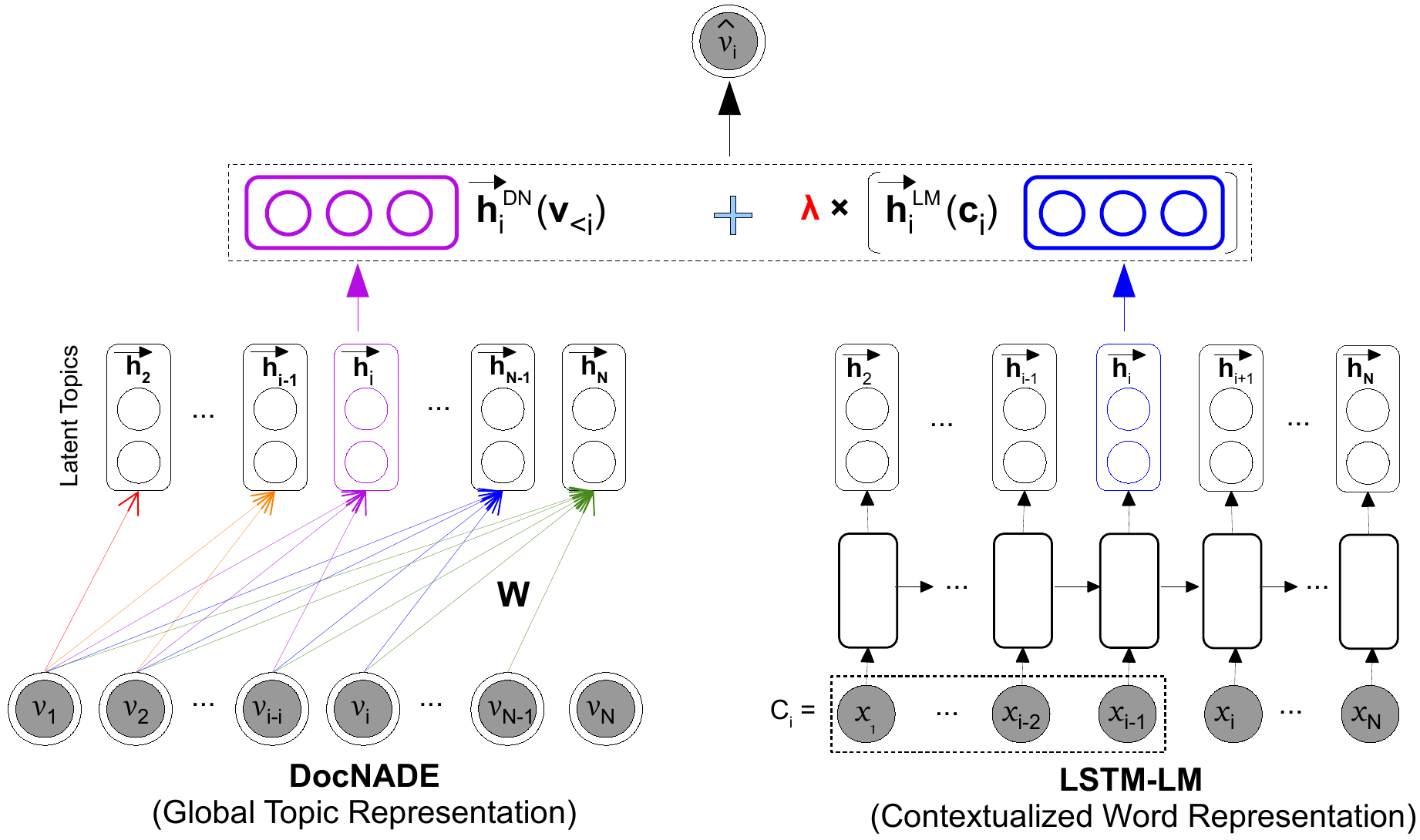}
\caption{ctx-DocNADE topic model. On left is the DocNADE topic model which provide \textit{global topic representations} ($h_i^{DN}$) of a document. In DocNADE, the connections between a particular softmax visible ($v_i$) and corresponding hidden units ($h_{i+1}, ..., h_N$) share the same weight parameter. On right is the LSTM-LM language model which provide the \textit{contextualized word representations} ($h_i^{LM}$) of every word in a sequential manner. The arrow ($\rightarrow$) inside hidden units ($h_i$) of both models denotes that the previous context of a word ($v_i$) is taken to calculate the regenerative probability ($\hat{v}_i$). Both hidden representation are combined using a $\lambda$ parameter (in red) to calculate ($\hat{v}_i$). The word representation matrix \textbf{W} is shared between DocNADE and LSTM-LM.
\enote{PG}{what does the arrow above means on hidden. remove LSTM from boxes. where is lambda in the fig. the weight equation not correct. what is hat over v.}}
\label{fig:ctxdocnade}
\end{figure}

This allows for the combined use of global context, i.e., \textit{coarse granularity}, from DocNADE model, without word order information, and local context, i.e., \textit{fine granularity}, from LSTM-LM, with consideration of word order information as seen in Fig ~\ref{fig:ctxdocnade}, for better prediction probabilities of words in DocNADE topic model.


The proposed topic model is named as \textit{contextualized-Document Neural Autoregressive Distribution Estimator} (ctx-DocNADE). ctx-DocNADE helps in learning complementary semantics by combining language and latent topic learning in a unified neural autoregressive framework.

\begin{figure}[t]
\center
\includegraphics[scale=0.75]{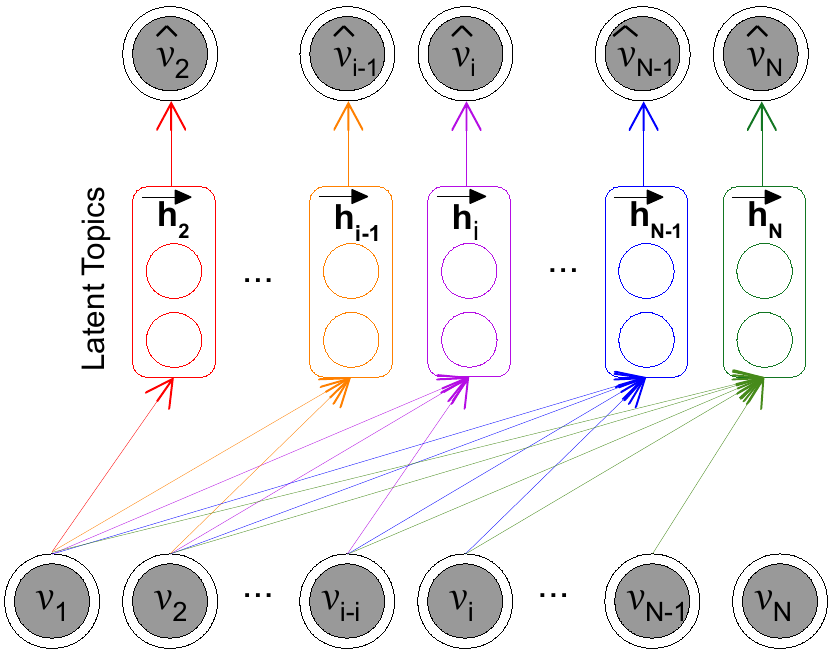}
\caption{DocNADE \cite{DBLP:Lauly_DocNADE} \enote{PG}{cite. show weight sharing. put in caption the significance of colors. why arrow on the hidden vector. it is uni-directional only. what is $\hat{v}$. Put here in caption} topic model with autoregressive connections between input softmax units and latent topic units. The arrow ($\rightarrow$) inside the hidden units ($\textbf{h}_i$) denotes that each hidden unit ($\textbf{h}_i$) takes into account the previous words only ($v_1, ..., v_{i-1}$) for regeneration probability $\hat{v}_i$ of the word $v_i$. Also, the connections between a particular softmax visible ($v_i$) and corresponding hidden units ($\textbf{h}_{i+1}, ..., \textbf{h}_N$) share the same weight parameter.}
\label{fig:docnade}
\end{figure}

To illustrate this, consider modeling a document of size $N$, DocNADE converts the document in a \textit{bag-of-words} vector $\textbf{v} = [v_1, v_2, ... , v_N]$ where each element $v_i \in \{1,2, ... , V\}$ is the index of $i$th word of the document in a vocabulary of size $V$. Then, DocNADE computes the joint probability of all the words, $p(\textbf{v})$, in the document $\textbf{v}$ according to the chain rule as follows: 

\begin{equation}
    p(\textbf{v}) = \prod_{i=1}^N p(v_i|\textbf{v}_{<i})
\end{equation}

where each conditional, $p(v_i|\textbf{v}_{<i})$, is computed in an autoregressive fashion using the preceeding observations $\textbf{v}_{<i} = [v_1, ... , v_{i-1}]$ in a feed-forward neural network for $i \in \{1, ... , N\}$. DocNADE computes the hidden layer, ($\textbf{h}_i^{DN}$), and probability conditional, $p(v_i|\textbf{v}_{<i})$, according to equations (~\ref{eq:ctx_docnade_dconade_hidden}) and (~\ref{eq:ctx_docnade_dconade_conditional}). Additionally, we ignore the computational efficiency of probabilistic binary tree, used in the original DocNADE model, and use a linear softmax over the vocabulary size $V$ to calculate the prediction probabilities of words at the output.

\begin{equation}
    \textbf{h}_i^{DN}(\textbf{v}_{<i}) = g(\textbf{b} + \sum_{k<i} \textbf{W}_{:,v_k})
    \label{eq:ctx_docnade_dconade_hidden}
\end{equation}

\begin{equation}
    p(v_i|\textbf{v}_{<i}) = \mbox{softmax}(\textbf{c} + \textbf{U} \textbf{h}_i^{DN}(\textbf{v}_{<i}))
    \label{eq:ctx_docnade_dconade_conditional}
\end{equation}


where, 

\begin{itemize}
    \item $g()$ is any activation function.
    \item $\textbf{W} \in \mathbb{R}^{H\times V}$ is a word representation matrix (encoding matrix) connecting the input layer to the hidden layer, where each column $\textbf{W}_{:,v_i}$ is a vector representation of word $v_i$ in the vocabulary $V$ and $H$ is the number of hidden units i.e., the number of topics.
    \item $\textbf{U} \in \mathbb{R}^{V\times H}$ is a weight matrix (decoding matrix) connecting the hidden layer to the softmax output layer.
    \item $\textbf{b} \in \mathbb{R}^H$ and $\textbf{c} \in \mathbb{R}^V$ are the input (encoding) and output (decoding) biases.
\end{itemize}


It is important to note here, in the example, that the past observations ($\textbf{v}_{<i}$), which we use in the conditionals, $p(v_i|\textbf{v}_{<i})$, may not be the actual words preceding the $i$th word in the document. For the same document, LSTM-LM takes into account the actual order of words in the document $\textbf{x} = [x_1, x_2, ... , x_N]$ for $N$ words in the document as seen in Fig ~\ref{fig:ctxdocnade} (right). Here, $x_i$ is represented by an embedding vector of dimension $H$ using the same word representation matrix $\textbf{W}$ as used in DocNADE. Let $\textbf{c}_i = [x_1, x_2, ... , x_{i-1}]$ be the preceeding context for each element $v_i \in \textbf{v}$. Therefore, LSTM-LM take into account the exact previous context $\textbf{c}_i$ to get the hidden layer representation, $\textbf{h}_i^{LM}(\textbf{c}_i)$, of language model for prediction of word $v_i$ as mentioned in eq. (~\ref{eq:ctx_docnade_lstm_hidden}).

\begin{equation}
    \textbf{h}_i^{LM}(\textbf{c}_i) = \mbox{LSTM-LM}(\textbf{c}_i, embedding = \textbf{W})
    \label{eq:ctx_docnade_lstm_hidden}
\end{equation}


\enote{PG}{use mbox within equation for text}

As we can see in fig. ~\ref{fig:ctxdocnade}, using  the hidden layer from DocNADE ($\textbf{h}_i^{DN}$), which represents the latent topic representation, and the hidden layer from LSTM-LM ($\textbf{h}_i^{LM}$), which represents the language structure and semantics, are combined in a complementary fashion, eq. (~\ref{eq:ctx_docnade_combined_hidden}), for prediction of the next word ($\hat{v}_i$), eq. (~\ref{eq:ctx_docnade_conditional}). We use an attention parameter ($\lambda$) over the hidden layer of LSTM-LM ($\textbf{h}_i^{LM}$) to control the inflow of language structure information into the DocNADE topic model.


\begin{equation}
    \textbf{h}_i(\textbf{v}_{<i}) = \textbf{h}_i^{DN}(\textbf{v}_{<i}) + \lambda  \textbf{h}_i^{LM}(\textbf{c}_i)
    \label{eq:ctx_docnade_combined_hidden}
\end{equation}

\begin{equation}
    p(v_i|\textbf{v}_{<i}) = \mbox{softmax}(\textbf{c} + \textbf{U} \textbf{h}_i(\textbf{v}_{<i}))
    \label{eq:ctx_docnade_conditional}
\end{equation}

\begin{equation}
    \log p(\textbf{v}) = \sum_{i=1}^N \log p(v_i|\textbf{v}_{<i})
    \label{eq:ctx_docnade_pseudo_LL}
\end{equation}


The ctx-DocNADE model is jointly optimized to maximize pseudo log-likelihood, eq. (~\ref{eq:ctx_docnade_pseudo_LL}), using stochastic gradient descent (SGD) algorithm. We say pseudo log-likelihood because the combined hidden representation $\textbf{h}_i(\textbf{v}_{<i})$ might incorporate some words from the forward context $\textbf{v}_{>i} = [v_i, ... , v_N]$ of the word $v_i$ using LSTM-LM hidden representation $\textbf{h}_i^{LM}(\textbf{c}_i)$. The ctx-DocNADE model learns better textual representations , which we have quantified via generalizability (i.e., perplexity), interpretability (i.e., topic coherence) and applicability (i.e., information retrieval and classification).



One thing to note here is that both DocNADE and LSTM-LM share the word representation matrix $\textbf{W}$ of DocNADE to learn better word representations, i.e., word embeddings, by accumulating the knowledge of global word co-occurences (from DocNADE) and local language structure \& semantics (from LSTM-LM), such that the better word representations would help in generating more coherent topics.



\section{Account for data sparsity using distributional compositional priors}

While incorporating language structure and word order helps in learning better latent representation (topics) for long texts and corpus of large number of documents, it is still a big challenge to learn from context for short text or corpus of small number of documents. The challenge comes from many reasons like: 

\begin{enumerate}
    \item \textit{Low frequency of word co-occurences} because of short length of text or less number of documents in the corpus.
    \item Significant \textit{word non-overlap} across the document corpus.
\end{enumerate}

However, distributed word representations i.e., word embeddings, learned on large corpus have been shown to learn the semantic relatedness between words. For example, consider these two sentences: 

\enote{PG}{why uppercase all?}
\enote{PG}{why red color? use italics?}
\enote{PG}{eqns. $-->$ equations}

\begin{itemize}
    \item \texttt{Brace for market share drops}
    \item \texttt{Deal with stock index falls}
\end{itemize}

Traditional topic models will not be able to learn the semantic relation between ``drops'' and ``falls'' due to the lack of context and word non-overlap. However, the relatedness can be shown in the Word2Vec \cite{DBLP:word2vec_mikolov} word embedding space where cosine similarity between ``drops'' and ``falls'' is 0.6816 and top 5 nearest neighbors of the word ``drops'' are \{``falls'', ``drop'', ``tumbles'', ``rises'', ''plummets''\}.


Related works such as Sahami \& Heilman (2006) \cite{DBLP:Sahami} employed web search results to improve the information in short texts and Petterson et al. (2010) \cite{DBLP:Petterson} introduced word similarity via thesauri and dictionaries into LDA. Das et al. (2015) \cite{DBLP:Das_gaussian_LDA} and Nguyen et al. (2015) \cite{DBLP:Nguyen_gloveDMM} integrated word embeddings into LDA and Dirichlet Multinomial Mixture (DMM) (Nigam et al., 2000) \cite{nigam} models respectively. However, these works are based on topic models (TMs) without considering language structure, e.g. word order. In addition, DocNADE (Larochelle \& Lauly, 2012) \cite{DBLP:Lauly_DocNADE} outperforms LDA and RSM topic models in terms of perplexity (PPL) and information retrieval (IR).


We incorporate pre-trained word embeddings in ctx-DocNADE model via LSTM-LM to supplement the topic model (DocNADE) in learning better latent representations ($\textbf{h}_i(\textbf{v}_{<i})$) of a small corpus of documents and/or a short text datasets as mentioned in eq. (~\ref{eq:ctx_docnade_lstm_hidden_embedding}) where, $\textbf{W}$ is the word representation matrix from DocNADE and \textbf{E} is the pre-trained distributional embedding matrix. The probability conditionals are calculated using eq. (~\ref{eq:ctx_docnade_conditional}). Then, the model is jointly optimized to maximize pseudo log-likelihood, eq. (~\ref{eq:ctx_docnade_pseudo_LL}), using stochastic gradient descent (SGD) algorithm.

\begin{equation}
    \textbf{h}_i^{LM}(\textbf{c}_i) = \mbox{LSTM-LM}(\textbf{c}_i, embedding = \textbf{W} + \textbf{E})
    \label{eq:ctx_docnade_lstm_hidden_embedding}
\end{equation}


Using the semantic relatedness via distributed word embeddings (\textbf{E}), from GloVe (Pennington et al., 2014) \cite{pennington_glove}, in ctx-DocNADE results in a much more coherent latent topic representation of the underlying document corpus. Therefore, we combine the advantages of complementary learning via LSTM-LM and external knowledge via distributed word embeddings (\textbf{E}), to model both short and long text documents in a unified neural autoregressive framework, named as ctx-DocNADEe. This model learns better latent representations of the short and long text documents, which we have quantified via generalizability (i.e., perplexity), interpretability (i.e., topic coherence) and applicability (i.e., information retrieval and classification).


In ctx-DocNADE model the word representation matrix $\textbf{W}$ of DocNADE is randomly initialized and shared with LSTM-LM as embedding layer. But, in ctx-DocNADEe the combination of randomly initialized word representation matrix $\textbf{W}$ of DocNADE and GloVe word embedding matrix $\textbf{E}$ is used as the embedding layer of LSTM-LM. Note that the $\textbf{W}$ is a trainable model parameter, while $\textbf{E}$ is a static prior. The trained models can be used to extract the combined textual representation $\textbf{h}(\textbf{v}^*)$ for a document $\textbf{v}^*$ of size $\textbf{N}^*$as per equations (~\ref{eq:ctx_docnade_representation_1}), (~\ref{eq:ctx_docnade_representation_2}) and (~\ref{eq:ctx_docnade_representation}).

\begin{equation}
    \textbf{h}^{DN}(\textbf{v}^*) = g(\textbf{b} + \sum_{k<=\textbf{N}^*} \textbf{W}_{:,v_k})
    \label{eq:ctx_docnade_representation_1}
\end{equation}

\begin{equation}
    \textbf{h}^{LM}(\textbf{c}_{N+1}^*) = \mbox{LSTM-LM}(\textbf{c}_{N+1}^*, embedding = \textbf{W}  \, \mbox{or} \,  (\textbf{W} + \textbf{E}))
    \label{eq:ctx_docnade_representation_2}
\end{equation}

\begin{equation}
    \textbf{h}(\textbf{v}^*) = \textbf{h}^{DN}(\textbf{v}^*) + \lambda \textbf{h}^{LM}(\textbf{c}_{N+1}^*)
    \label{eq:ctx_docnade_representation}
\end{equation}

As DocNADE and LSTM-LM both are based on feed forward neural network architecture, therefore by adding more number of hidden layers to ctx-DocNADE and ctx--DocNADEe, they both can be extended to their deep variants named as ctx-DeepDNE and ctx-DeepDNEe respectively. The more number of hidden layers allow for learning higher levels of abstraction of the global topic representation of DocNADE ($\textbf{h}^{DN}(\textbf{v}^*)$) and the local contextual representation of LSTM-LM ($\textbf{h}^{LM}(\textbf{c}_{N+1}^*)$). In the deep version, the first hidden layer of DocNADE and LSTM-LM ($\textbf{h}_{i,1}^{DN}(\textbf{v}_{<i})$ and $\textbf{h}_{i,1}^{LM}(\textbf{c}_i)$) are computed as per equations (~\ref{eq:ctx_docnade_dconade_hidden}) and (~\ref{eq:ctx_docnade_lstm_hidden}) respectively. Subsequent hidden layers are computed as per equations (~\ref{eq:ctx_docnade_deep_docnade}) and (~\ref{eq:ctx_docnade_deep_lstm}) respectively.

\begin{equation}
    \textbf{h}_{i,d}^{DN}(\textbf{v}_{<i}) = g(\textbf{b} + \textbf{W}_d \textbf{h}_{i,d-1}(\textbf{v}_{<i}))
    \label{eq:ctx_docnade_deep_docnade}
\end{equation}

\begin{equation}
    \textbf{h}_{i,d}^{LM}(\textbf{c}_i) = \mbox{deepLSTM-LM}(\textbf{c}_i, depth = d, embedding = \textbf{W} \, \mbox{or} \, (\textbf{W} + \textbf{E}))
    \label{eq:ctx_docnade_deep_lstm}
\end{equation}

for $d \in \{2, ... , n\}$, where n (total depth) is the total number of hidden layers. For $d=1$ the hidden vectors $\textbf{h}_{i,1}^{DN}(\textbf{v}_{<i})$ and $\textbf{h}_{i,1}^{LM}(\textbf{c}_i)$ correspond to equations (~\ref{eq:ctx_docnade_dconade_hidden}) and (~\ref{eq:ctx_docnade_lstm_hidden}) respectively. The final pseudo log-likelihood is computed using the last combined hidden layer ($\textbf{h}_{i,n} = \textbf{h}_{i,n}^{DN} + \lambda \textbf{h}_{i,n}^{LM}$) like eq. ~\ref{eq:ctx_docnade_representation}.

\begin{figure}[t]
\center
\includegraphics[scale=0.70]{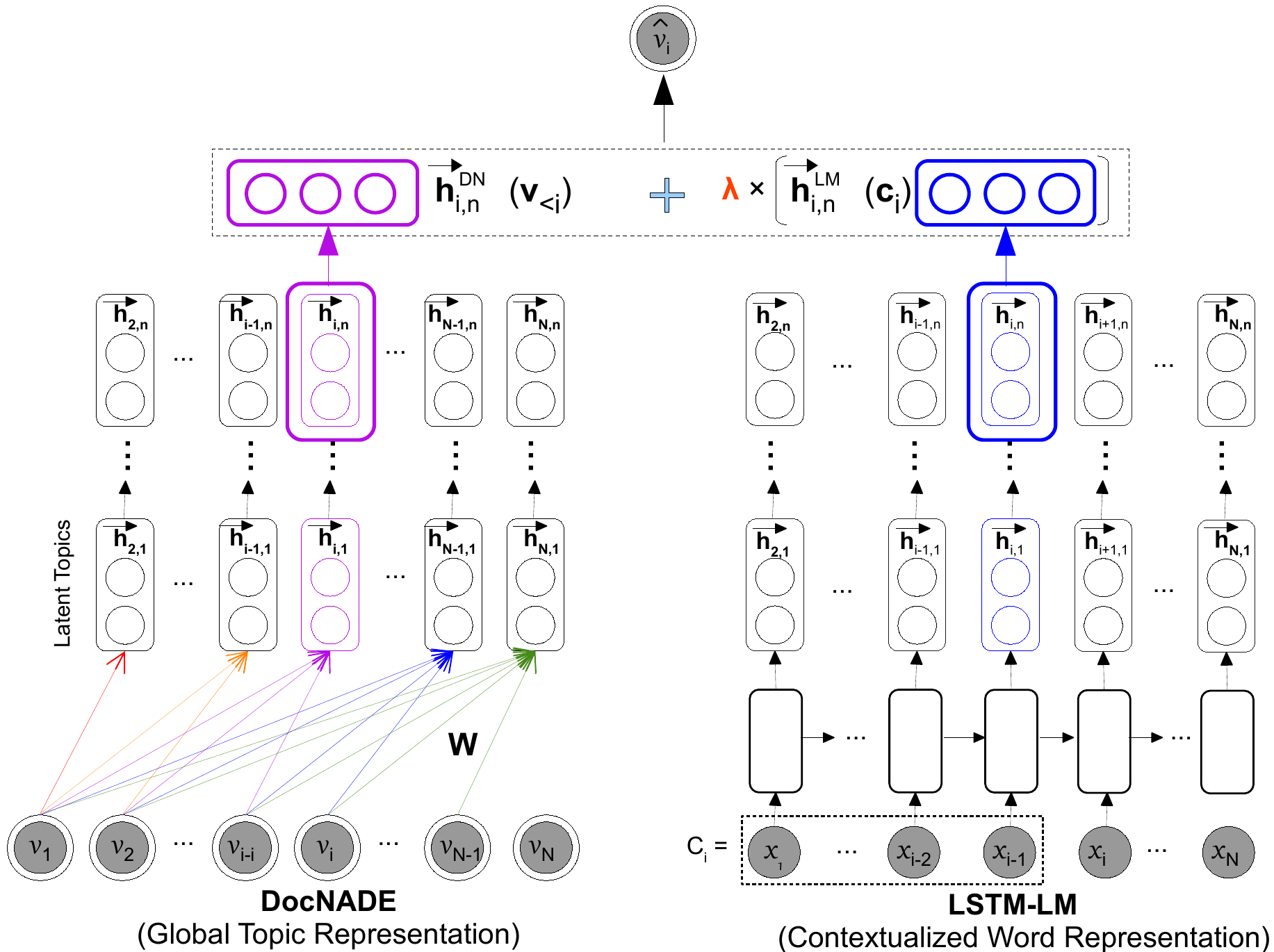}
\caption{contextualized deep DocNADE (ctx-DeepDNE) model. Similar architecture to ctx-DocNADE but with extra layers of hidden units.}
\label{fig:ctx_deep_docnade}
\end{figure}

\chapter{Lifelong learning topic model}
\label{chapter:lifelong_learning}

The motivation for \textit{lifelong learning} can be derived from a simple statement that ``\textit{you cannot learn everything at once}''. Even the most complicated artificial intelligence algorithms based on reinforcement learning, in recent years, work in a similar fashion and learn from different actions under different scenarios in thousands of simulations and boost their understanding while remembering the previous learning. Some of these algorithms have been able to defeat humans at some of the most complicated strategy games. Another example would be a student who is enrolled in ``introduction to machine learning'' course at TU Munich, would learn some basic mathematical concepts of machine learning, but when that student gets enrolled in ``Advanced deep learning'' in the next semester, he/she will learn more refined/complicated concepts based on the basic concepts while retaining the learning from previous course(s).


Therefore, we demonstrate the applicability of \textit{lifelong learning} in topic modeling using DocNADE as the base topic model. We perform topic modeling on a sequence of datasets to learn new information from each dataset in the sequence while retaining the previous learning or minimizing the \textit{catastrophic forgetting}. Let's say, we have three datasets (\textbf{D1}, \textbf{D2} and \textbf{D3}). So, we would start in a particular sequence ($\textbf{D1} \rightarrow \textbf{D2} \rightarrow \textbf{D3}$) with \textbf{D1} as current dataset and perform topic modeling to learn its latent topical representations. After that, we move to \textbf{D2} as current dataset to learn new latent representations while retaining latent representations of previous dataset \textbf{D1}. Similarly, we move to \textbf{D3} as current dataset to learn new latent representations with retention of learning from previous datasets i.e., \textbf{D2} \& \textbf{D1}. 


Previously, we discussed the problems associated with topic modeling of short text documents or small corpus of documents due to low word co-occurrence statistics. We also discussed to alleviate this problem using pre-trained distributional embeddings transfer from GloVe ~\cite{pennington_glove} which is trained on huge corpus, like Wikipedia, apart from the inclusion of language structure information from LSTM-LM. Similarly, the accumulated learning in a lifelong fashion might help in modeling the short text documents effectively by improving the word co-occurrence statistics. Even for the long text documents, the learning from previous datasets might help in generating more meaningful latent topic representations on the current dataset.


For future references of \textit{lifelong learning}, we use \textit{target dataset} for current dataset and \textit{source datasets(s)} for previous datasets(s). Also, we use $\theta_{new} \in \mathbb{R}^{H \times V^{'}}$ for the new parameter learned on target dataset and $\theta_{old} \in \mathbb{R}^{H \times V^{'}}$ for the parameters learned on source dataset(s), where $H$ is the size of hidden unit and $V^{'}$ is the common vocabulary of target and source datasets. $\textbf{T}$ is the total number of source dataset(s) and $t \in \{1, ... T\}$ is an indicator for source dataset. We use DocNADE as the base topic model for the inclusion \textit{lifelong learning}. To transfer the relevant important information from source dataset(s) to targe dataset and retention of past learning, we have subdivided the process of \textit{lifelong learning} into three different tasks as mentioned hereafter.

\section{Explicit knowledge transfer}

We transfer the word representations learned from the source dataset(s) into topic modeling of current dataset as static embedding prior(s) to help in building more coherent latent representations. We use an attention parameter $\lambda_{EmbTF}^t$ over pretrained embedding to control the flow of information from source dataset $t$ as seen in eq. (~\ref{eq:lifelong_TL}).

\begin{equation}
    EmbTF = \sum_{t=1}^T \lambda_{EmbTF}^t \theta_{old}^t
	\label{eq:lifelong_TL}
\end{equation}

As the encoding matrix \textbf{W} of DocNADE contains the word representations, we only use \textbf{W} matrix to transfer embedding prior(s). If the source and target datasets have an overlapping domains then the word representations from the source dataset(s) contains global semantics, i.e., \textit{coarse granularity}, similar to target domain. Hence, the value of $\lambda_{EmbTF}$ would be high otherwise it would be low. We call this task \textit{Embedding Transfer} (EmbTF).

\section{Implicit knowledge transfer}

We carefully select those documents from source dataset(s) which have a domain overlap with the target dataset. To analyze the domain overlap, we calculate a binary threshold parameter ($\beta_i^t$) (eq. ~\ref{eq:lifelong_sal_threshold_3}) by comparing the perplexity $PPL_{new,i}^t$ of document ($\textbf{v}_i^t$), containing $N_i^t$ number of words, in the source dataset $t$ (eq. ~\ref{eq:lifelong_sal_threshold_1}) with the old average perplexity value $PPL_{old}^t$ of all the documents ($D^t$) in the source dataset $t$ calculated using old parameter $\theta_{old}^t$ (eq. ~\ref{eq:lifelong_sal_threshold_2}). Then, the value of $\beta$ decides which document in the source dataset(s) is important for the target dataset. 

\enote{PG}{thresh...use a different symbol. Use $\beta$ or $\gamma$.}

\begin{equation}
    PPL_{old}^t = \exp (- \frac{1}{D^t} \sum_{i=1}^{D^t} \frac{1}{N_i^t} \log p(\textbf{v}_i^t|\theta_{old}^t))
    \label{eq:lifelong_sal_threshold_1}
\end{equation}

\begin{equation}
    PPL_{new,i}^t = \exp (- \frac{1}{N_i^t} \log p(\textbf{v}_i^t|\theta_{new}))
    \label{eq:lifelong_sal_threshold_2}
\end{equation}

\begin{equation}
    \begin{array}{l}
        \beta_i^t = 0, if (PPL_{new,i}^t > PPL_{old}^t) \\
        \beta_i^t = 1, if (PPL_{new,i}^t < PPL_{old}^t)
    \end{array}
    \label{eq:lifelong_sal_threshold_3}
\end{equation}

After this, we use the selected documents from all the source datasets for co-training with the target dataset to improve the word co-occurence statistics in the target dataset using eq. (~\ref{eq:lifelong_SAL}). We use attention parameter $\lambda_{SAL}^t$ over the selected documents to control the inflow of implicit information transfer. We call this task \textit{Selective Augmentation Learning} (SAL).

\begin{equation}
    SAL = - \sum_{t=1}^T \lambda_{SAL}^t \sum_{i=1}^{D^t} \beta_i^t \log p(\textbf{v}_i^t|\theta_{new})
	\label{eq:lifelong_SAL}
\end{equation}

Also, with the inclusion of the selected source documents in the training process, SAL enables the topic model to implicitly learn about part of the source datasets. Therefore, SAL task partly helps in the retention of past learnings, if not completely.

\section{Retaining previous knowledge}

To prevent \textit{catastrophic forgetting} of past learning, we enforce an L2 constraint on the difference of projected target model parameter $P^t\theta_{new}$ and source parameters $\theta_{old}^t$ as mentioned in eq. (~\ref{eq:lifelong_CF}). Here, $P^t \in \mathbb{R}^{H\times H}$ is a projection matrix which learns the transformation between target embedding space and source embedding space, where $H$ is the size of hidden units. We use an attention parameter $\lambda_{RK}^t$ to control the severity of the constraint. Large value of $\lambda_{RK}^t$ would enforce this constraint more strictly, while small value of $\lambda_{RK}^t$ would mellow the effect of the constraint. We call this task \textit{Retention of Knowledge} (RK).

\begin{equation}
    RK = \sum_{t=1}^T \lambda_{RK}^t {\lVert P^t \theta_{new} - \theta_{old}^t \lVert}_2^2
	\label{eq:lifelong_CF}
\end{equation}

\begin{equation}
    RK = \sum_{t=1}^T \lambda_{RK}^t {\lVert P_{W}^t \textbf{W}_{new} - \textbf{W}_{old}^t \lVert}_2^2 + \sum_{t=1}^T \lambda_{RK}^t {\lVert P_{U}^t \textbf{U}_{new} - \textbf{U}_{old}^t \lVert}_2^2
	\label{eq:lifelong_CF_explained}
\end{equation}

\enote{PG}{why bold in subscript}
\enote{PG}{change name of 'consolidated loss' in the equation, use some symbol name?}

\begin{figure}[h]
\center
\includegraphics[scale=0.65]{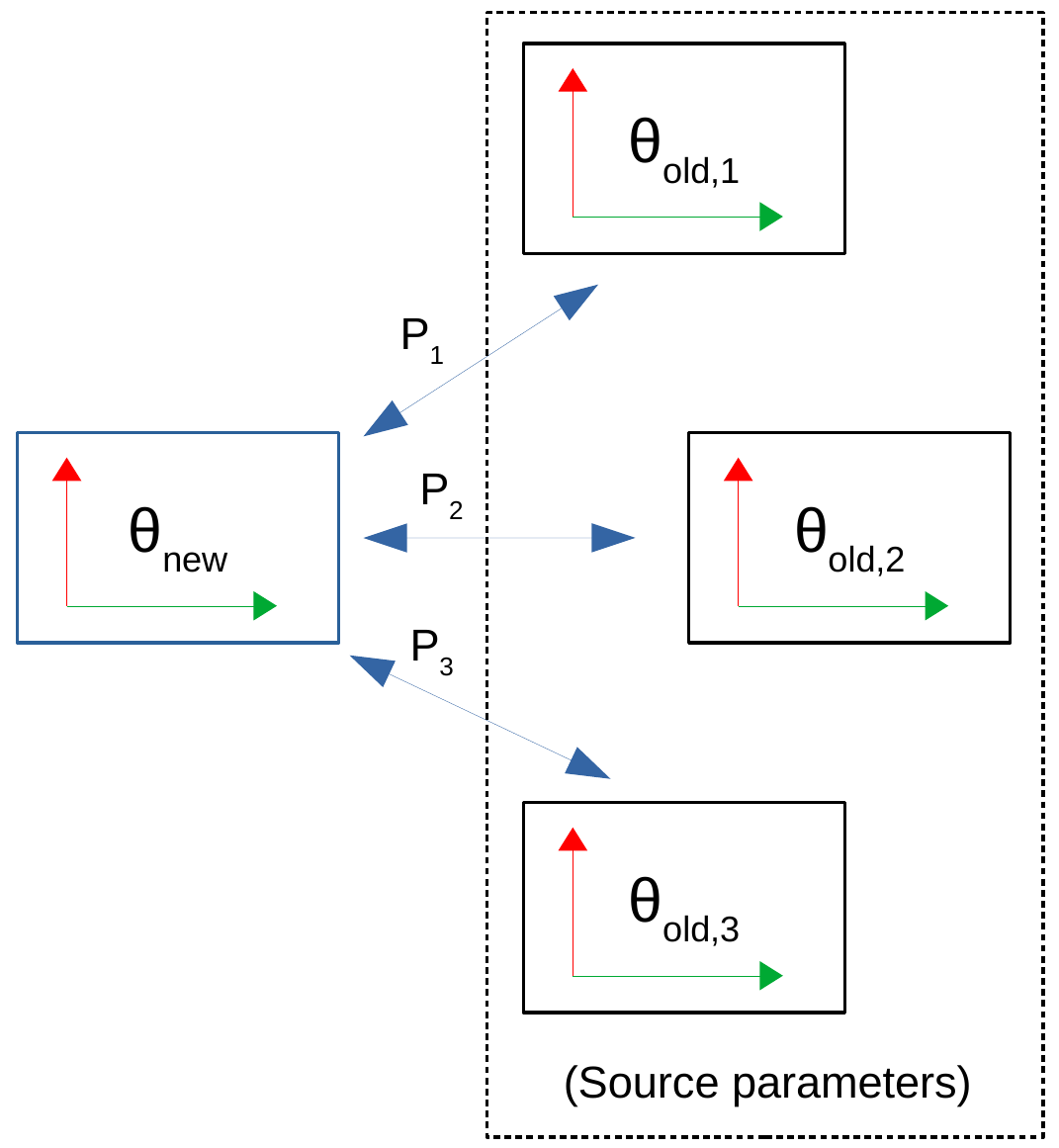}
\caption{Maximizing retention of knowledge (RK) by minimizing the distance between projected target embedding space and source embedding space(s). \enote{PG}{use source instead of old. Use some notations for source and target.}}
\label{fig:lifelong_cf}
\end{figure}

However, eq. (~\ref{eq:lifelong_CF}) describes the general form of the equation using $\theta_{new}$ and $\theta_{old}^t$, but eq. (~\ref{eq:lifelong_CF_explained}) explains the constraint in terms of DocNADE parameters $\textbf{W}$ and $\textbf{U}$. Also, fig. ~\ref{fig:lifelong_cf} illustrates the idea of minimizing the distance between source and target embedding spaces.


\section{Consolidation}

As discussed in chapter ~\ref{chapter:background}, DocNADE defines the negative log-likelihood loss $\mathcal{L}({\bf v})$ of document \textbf{v} of the target dataset as given in eq. (~\ref{eq:lifelong_target_loss}), where N is the number of words in the document \textbf{v}. Therefore, to train DocNADE in a complete \textit{lifelong learning} fashion, we need to add equations (~\ref{eq:lifelong_SAL}) and (~\ref{eq:lifelong_CF_explained}) in eq. (~\ref{eq:lifelong_target_loss}) to get a consolidated loss $\mathcal{L}({\bf v})_c$ as shown in eq. (~\ref{eq:lifelong_consolidated_loss}). Also, we need to use the words representations (EmbTF) from source datasets as static prior information, while training DocNADE using $\mathcal{L}({\bf v})_c$.

\begin{equation}
    \mathcal{L}({\bf v}) = - \sum_{i=1}^{N} \log p(\textbf{v}_{i}|\theta_{new}^{'})
	\label{eq:lifelong_target_loss}
\end{equation}

\begin{equation}
    \theta_{new}^{'} = \theta_{new} + EmbTF
	\label{eq:lifelong_target_loss_new_theta}
\end{equation}

\begin{equation}
    \mathcal{L}({\bf v})_c = \mathcal{L}({\bf v}) + SAL + RK
	\label{eq:lifelong_consolidated_loss}
\end{equation}

Equation (~\ref{eq:lifelong_consolidated_loss}) and algorithm (~\ref{algo:lifelong_learning}) describe the integration of \textit{lifelong learning} in DocNADE topic model by combining the different \textit{lifelong learning} tasks, i.e., EmbTF, SAL \& RK, as discussed before. It shows how we can use already existing knowledge from source dataset(s) to get the better latent topic representations on the target dataset while, at the same time, minimizing the \textit{catastrophic forgetting} of the past learning. In algorithm ~\ref{algo:lifelong_learning}, by setting the respective parameters EmbTF, SAK and RK as either True or False, these lifelong learning tasks can be included in the DocNADE topic modeling process. By setting all of them as True we get the consolidated loss $\mathcal{L}({\bf v})_c$ as the final loss as mentioned in eq. (~\ref{eq:lifelong_consolidated_loss}).

\begin{algorithm}[t]
\caption{{\small Computation of $\log p({\bf v})$ and Loss $\mathcal{L}({\bf v})$ for DocNADE topic model in \textit{lifelong learning} fashion}}\label{algo:lifelong_learning} 
{\small 
\begin{algorithmic}
\Statex \textbf{Input}: A target training document ${\bf v}$ with $N$ number of words and source training documents $\{\textbf{v}^1, ..., \textbf{v}^T\}$ with $\{N^1, N^2, ..., N^T\}$ number of words from \textbf{T} source datasets
\Statex \textbf{Input}: Knowledge Base of latent topic and word representations $\{{\bf W}^{1}, ..., {\bf W}^{T}\}$ from \textbf{T} source datasets; where ${\bf W}^{t}$ is the same as ${\bf W}_{old}^{t}$
\Statex \textbf{Input}: Knowledge Base of decoding matrices $\{{\bf U}^{1}, ..., {\bf U}^{T}\}$ from \textbf{T} source datasets; where ${\bf U}^{t}$ is the same as ${\bf U}_{old}^{t}$
\Statex \textbf{Input}: Knowledge Base of perplexity-per-word (PPL) of \textbf{T} source datasets $\{PPL^{1}, ..., PPL^{T}\}$; where $PPL^{t}$ is the same as $PPL_{old}^{t}$
\Statex \textbf{Parameters}: ${\bf \theta} = \{{\bf b}, {\bf c}, {\bf W}, {\bf U},  {\bf P}_W^{1}, ...,  {\bf P}_W^{T}, {\bf P}_U^{1}, ...,  {\bf P}_U^{T}\}$; where ${\bf W}$ is the same as ${\bf W}_{new}$ and ${\bf U}$ is the same as ${\bf U}_{new}$
\Statex \textbf{hyper-parameters}: $\{H, \lambda_{EmbTF}^{1}, ..., \lambda_{EmbTF}^{T}, \lambda_{SAL}^{1}, ..., \lambda_{SAL}^{T}, \lambda_{RK}^{1}, ..., \lambda_{RK}^{T}$\}
\item[]
\State Initialize ${\bf a} \gets {\bf c}$ and $ p({\bf v}) \gets 1$
\For{$i$ from $1$ to $N$}
    \State  ${\bf h}_{i} ({\bf v}_{<i})  \gets g({\bf a}$), where $g$ = \{sigmoid, tanh\}
    \State $p(v_{i}=w | {\bf v}_{<i}) \gets \frac{\exp (b_w + {\bf U}_{w,:} {\bf h}_i ({\bf v}_{<i}))}{\sum_{w'} \exp (b_{w'} + {\bf U}_{w',:} {\bf h}_i ({\bf v}_{<i}))}$
    \State $ p({\bf v}) \gets  p({\bf v}) p(v_{i} | {\bf v}_{<i})$
    \State compute pre-activation at step, $i$: ${\bf a} \gets {\bf a} + {\bf W}_{:, v_{i}}$  
    \If  {\texttt{EmbTF}} 
         \State get word embedding for $v_i$ from source domain(s) 
         \State ${\bf a} \gets {\bf a} + \sum_{t=1}^{T} \lambda_{EmbTF}^t \ {\bf W}_{:, v_{i}}^{t}$
    \EndIf
\EndFor

\State $\mathcal{L}({\bf v}) \gets  - \log p({\bf v})$

\item[]
\If {\texttt{SAL}}
    \For{$t$ from $1$ to $T$}
        \State Initialize ${\bf a}^t \gets {\bf c}$ and $ p({\bf v}^t) \gets 1$
        \For{$i$ from $1$ to $N^t$}
            \State  ${\bf h}^t_{i} ({\bf v}^t_{<i})  \gets g({\bf a}^t$), where $g$ = \{sigmoid, tanh\}
            \State $p(v_{i}^t=w | {\bf v}_{<i}^t) \gets \frac{\exp (b_w + {\bf U}_{w,:} {\bf h}_i^t ({\bf v}_{<i}^t))}{\sum_{w'} \exp (b_{w'} + {\bf U}_{w',:} {\bf h}_i^t ({\bf v}_{<i}^t))}$
    	    \State $ p({\bf v}^t) \gets  p({\bf v}^t) p(v_{i}^t | {\bf v}_{<i}^t)$
            \State compute pre-activation at step, $i$: ${\bf a}^t \gets {\bf a}^t + {\bf W}_{:, v_{i}^t}$
        \EndFor
        \State $\mathcal{L}({\bf v}^t) \gets  - \log p({\bf v}^t)$
        \State $PPL_{new}^t = \exp(\frac{\mathcal{L}({\bf v}^t)}{N^t})$
        \State $\beta^t = 0$
        \If {$PPL_{new}^t < PPL^t$}
            \State $\beta^t = 1$
        \EndIf
        \State $\mathcal{L}({\bf v}) \gets \mathcal{L}({\bf v}) + ( \lambda_{SAL}^t \times \beta^t \times \mathcal{L}({\bf v}^t)$)
    \EndFor
\EndIf

\item[]
\If  {\texttt{RK}} 
         \State $\mathcal{L}({\bf v}) \gets \mathcal{L}({\bf v}) + \sum_{t=1}^{T} \lambda_{RK}^t  \  || {\bf P}_{W}^{t} {\bf W} - {\bf W}^{t} ||_2^2 + \sum_{t=1}^{T} \lambda_{RK}^t  \  || {\bf P}_{U}^{t} {\bf U} - {\bf U}^{t} ||_2^2$
\EndIf

\end{algorithmic}}
\end{algorithm}



\chapter{Experiments: Contextualized topic model}
\label{chapter:results_ctx_docnade}

\section{Dataset Description}
\label{section:res_ctx_dataset}

In the previous chapters we have seen the motivations, problem statements, background theories and our proposed models and methodologies for tackling the aforementioned problem statements. Now, remains the most critical part of any research i.e., experiments and validation. For the same purpose, we have applied our proposed models (ctx-DocNADE, ctx-DocNADEe), for improving topic models, on 8 short-text and 7 long-text datasets of different sizes. Some of them are multi-labeled datasets from public as well as industrial corpus. The detailed statistics about the 15 datasets is mentioned in Table ~\ref{table:data_statistics} where 20NS and R21578 signify 20NewsGroups and Reuters21578 respectively.


A short description about each one of the 15 datasets is given below: 
\begin{enumerate}

\item \textit{20NSshort}: We take documents from 20NewsGroups data, with document size less (in terms of number of words) than 20.
\item \textit{TREC6}: a set of questions.
\item \textit{Reuters21578title}: a collection of new stories from nltk.corpus. We take titles of the documents.
\item \textit{Subjectivity}: sentiment analysis data.
\item \textit{Polarity}: a collection of positive and negative snippets acquired from Rotten Tomatoes.
\item \textit{TMNtitle}: Titles of the Tag My News (TMN) news dataset.
\item \textit{AGnewstitle}: Titles of the AGnews dataset.
\item \textit{Reuters8}: a collection of news stories, processed and released by Reuters.
\item \textit{Reuters21578}: a collection of new stories from nltk.corpus.
\item \textit{20NewsGroups}: a collection of news stories from nltk.corpus.
\item \textit{20NSsmall}: We sample 20 document for training from each class of the 20NS dataset. For validation and test, 10 document for each class.
\item \textit{TMN}: The Tag My News (TMN) news dataset.
\item \textit{BNC}: The British National Corpus (BNC) is a 100 million word collection of samples of written and spoken language from a wide range of sources.
\item \textit{AGnews} (long text): News articles from the AG’s corpus of news articles on the web containing the 4 largest classes.
\item \textit{Sixxx Requirement OBjects (SiROBs)}: a collection of paragraphs extracted from industrial tender documents (our industrial corpus).

\end{enumerate}

\begin{table}[h]
\centering
\resizebox{.8\textwidth}{!}{
\begin{tabular}{r||rrrrrrrr|}
\hline
\multicolumn{9}{c||}{\texttt{short-text}} \\ 
 \multicolumn{1}{c|}{\textbf{Data}} &  \multicolumn{1}{c}{\textbf{Train}} &  \multicolumn{1}{c}{\textbf{Val}} & \multicolumn{1}{c}{\textbf{Test}} &  \multicolumn{1}{c}{$|${\textbf{RV}}$|$}& \multicolumn{1}{c}{$|${\textbf{FV}}$|$}  & \multicolumn{1}{c}{\textbf{L}} & \multicolumn{1}{c}{\textbf{C}} & \multicolumn{1}{c||}{\textbf{Domain}} \\ \hline
20NSshort             & 1.3k & 0.1k & 0.5k &    1.4k       &   1.4k    &  13.5     &   20      &     News         \\
TREC6                  & 5.5k  &  0.5k &  0.5k  &      2k          &   2295 &   9.8          & 6        &   Q\&A      \\
R21578title$^\dagger$ &  7.3k &  0.5k & 3.0k     &   2k        &   2721       &   7.3   &     90  &   News   \\
Subjectivity           &  8.0k &  .05k &  2.0k  & 2k       &   7965   &   23.1       &  2     &   Senti      \\
Polarity                 & 8.5k &  .05k & 2.1k &    2k       &   7157    &   21.0     &   2              &  Senti        \\
TMNtitle                & 22.8k  &  2.0k & 7.8k  &       2k      &   6240     &   4.9    &     7    &   News    \\
TMN                 & 22.8k  &  2.0k & 7.8k  &      2k         &   12867  &    19          &       7   &  News             \\
AGnewstitle           & 118k &  2.0k &  7.6k  &        5k         &   17125   &  6.8      &    4      &  News     \\ 
\hline
\multicolumn{9}{c}{\texttt{long-text}} \\
\multicolumn{1}{c|}{\textbf{Data}} &  \multicolumn{1}{c}{\textbf{Train}} &  \multicolumn{1}{c}{\textbf{Val}} & \multicolumn{1}{c}{\textbf{Test}} &  \multicolumn{1}{c}{$|${\textbf{RV}}$|$} &  \multicolumn{1}{c}{$|${\textbf{FV}}$|$}  & \multicolumn{1}{c}{\textbf{L}} & \multicolumn{1}{c}{\textbf{C}} & \multicolumn{1}{c}{\textbf{Domain}}   \\ \hline
20NSsmall             & 0.4k &  0.2k & 0.2k  &      2k       &   4555   &   187.5     &    20      & News   \\
Reuters8             & 5.0k &  0.5k & 2.2k  &      2k    &   7654   &   102       &    8      & News   \\
20NS                  & 7.9k & 1.6k & 5.2k  &     2k       &   33770    &   107.5     &   20       &   News   \\
R21578$^\dagger$ &  7.3k&  0.5k & 3.0k  &      2k           &   11396    &   128       &   90      &    News   \\
BNC               &   15.0k& 1.0k & 1.0k &     9.7k       &  41370     &    1189     &   -     & News    \\
SiROBs$^\dagger$ &  27.0k &  1.0k & 10.5k  &      3k       &   9113  &  39       &  22    &  Indus   \\
AGNews              & 118k & 2.0k &   7.6k &        5k       &   34071   &  38       &     4    &   News   \\
\end{tabular}}
\caption{Data statistics: Short/long texts and/or small/large corpora from diverse domains. Symbols-  Avg: average, $L$: avg text length (\#words), $|RV|$ and $|FV|$: size of reduced (RV) and full vocabulary (FV), $C$: number of classes, Senti: Sentiment, Indus: Industrial, `k':thousand and $\dagger$: multi-label. For short-text,  $L$$<$$25$.
\enote{PG}{why double line at end of the table.}
}
\label{table:data_statistics}
\end{table}

The SiROBs is an industrial corpus of Siemens, extracted from industrial tender documents. The documents contain requirement specifications of an industrial project. There are 22 different types of requirements i.e. class labels (multi-class), where a requirement is a paragraph or collection of paragraphs within a document. We name the requirement as Requirement Objects (ROBs). Some of the requirement types are project management, testing, legal, risk analysis, technical requirement, etc. So, we analyze such documents to automate decision making, tender comparison, similar tender as well as ROB retrieval and assigning ROBs to a relevant department(s) to optimize/expedite tender analysis. See some examples of ROBs from SiROBs corpus in Table ~\ref{table:SiROBs_examples}.

\section{Experimental settings}

In chapter ~\ref{chapter:ctx_docnade}, we discussed two different drawbacks of topic models, namely:

\begin{enumerate}
    \item \textit{Missing language structure information in topic models}, and
    \item \textit{Data sparsity in short text documents}
\end{enumerate}

Therefore, to address the above-mentioned drawbacks, we conduct experiments with our proposed models (under different experimental settings) as mentioned below:

\begin{enumerate}
    \item \textit{ctx-DocNADE}: Our proposed model which combines the complementary information of DocNADE topic model and LSTM-LM language model, to address the \textit{missing language structure information} in DocNADE topic model.
    \item \textit{ctx-DocNADEe}: Our proposed model which incorporate semantically informative pre-trained distributional word embeddings as static prior knowledge source in ctx-DocNADE topic models, to address the missing semantic knowledge, i.e., \textit{data sparsity}, resulting from low word co-occurences in short text datasets.
\end{enumerate}

\section{Evaluation metrics}
\label{section:ctx_docnade_evaluation_metrics}
We have used four quantitative measures for evaluation of our proposed models (ctx-DocNADE and ctx-DocNADEe):

\begin{enumerate}
    \item \textit{Perplexity-per-word} (PPL): The generative performance of a model can be measured by the average negative log-likelihood (NLL) of the model over the test set. To evaluate the generative performance of the topic models we can do the same, but in the recent past, researchers have adopted a slightly different evaluation criterion called average perplexity-per-word (PPL). While NLL is calculated on document level as seen in eq. (~\ref{eq:nll}), PPL is calculated at word level, as seen in eq. (~\ref{eq:ppl}), which is more descriptive for a model generating word after word and not the whole sentence together. Also, PPL has an exponentiation term which helps in noticing the slight changes in word prediction probabilities.

    \begin{equation}
        NLL = - \frac{1}{D} \sum_{i=1}^{D} \log p(\textbf{v}_i)
        \label{eq:nll}
    \end{equation}
    
    \begin{equation}
        PPL = exp(- \frac{1}{D} \sum_{i=1}^{D} \frac{1}{N_i} \log p(\textbf{v}_i))
        \label{eq:ppl}
    \end{equation}
    
    \enote{PG}{missing reference to equation and figures in running text}
    
    where $D$ is total number of documents in the corpus and $N_i$ is the number of words in the document $\textbf{v}_i$.
    
    \item \textit{Topic coherence} (COH): Topic models help to understand, summarize and organize a large collection of documents by finding some latent features called \textit{topics}. Topics are collection of words based on co-occurence statistics in the document corpus. While it is important to evaluate topics models on the criterion of \textit{generalization} and \textit{applicability}, it is equally important to have a quantifiable evaluation of the these latent topics learned by the model to distinguish good topics from bad topics. 
    
    Therefore, we use topic coherence as the criterion to assess the meaningfulness of the underlying topics captured by the model. We use the coherence measure proposed by R\"oder et al. (2015) \cite{DBLP:Roder} , which identfies context features for each topic word using a sliding window over the reference corpus. The topics with high scores imply more coherency. We use the gensim module (radimrehurek.com/gensim/models/\-coherencemodel.html, coherence type = c\_v) to calculate topic coherence over top 10 and 20 words for each of the 200 topics generated by our proposed models and baseline models.
    
    \item \textit{Information retrieval} (IR): When it comes to practicality of topic modeling, \textit{document retrieval} is a critical evaluation. Given a particular query document, \textit{document retrieval} is defined as finding the most semantically related documents in a given document corpus. \textit{Document retrieval} is a particular form of \textit{information retrieval} where a higher level document representation i.e., latent vector representation, is used for retrieval task. For the topic models, this higher level representation of a document is, generally, a topic mixture representation i.e., a vector with mixture coefficients for all latent topics learned by the model. 
    
    Therefore, it is important to learn the vector representation of two most semantically related documents in such a way that the similarity distance between the vector representation of the two documents is very less as compared to other semantically unrelated documents. The similarity distance can be either \textit{cosine similarity} or \textit{euclidean distance}. Hence, it is very important for a topic model to learn all the different type of semantics present in a document corpus. For our proposed models we call these vector representations as \textit{contextualized} representations.
    
    \item \textit{Classification} (F1): To get a list of most related text documents to a given query is a very important task but, equally important is the classification of text documents into a predefined set of different categories i.e., text categorization. Text categorization does not require the presence of a query document but it is done on an absolute scale. It gives one or more tag(s) to each document based on it's semantic information which eventually put each document in different categories, hence reducing cluttering and facilitate in easy search and navigation of the user. For example, action, adventure, thriller, romantic etc are different tags that can be given to each movie plot (text) which will categorize them into different genres.

    
    In topic modeling, text categorization can be done in two ways. First, during training the label information can be leveraged to perform supervised classification along-with unsupervised regeneration of documents. Second, after learning latent document topic representations in an unsupervised fashion, use those representations as a static input data to perform supervised classification. We adopt the second method to perform text categorization using \textit{contextualized} representations of our proposed models and document representations of all other baselines models we have used.
\end{enumerate}

While, it is very important to perform an extensive evaluation of our proposed models, it is also equally important do to a fair comparison with all of the related baselines. Therefore, we have compared our proposed models with the following mentioned baselines:

\begin{enumerate}
\item \textit{Document representation using GloVe (Pennington et al., 2014)} \cite{pennington_glove}: A document representation is created using the sum of GloVe word embeddings of all the words in the document.
\item \textit{Document representation using doc2vec (Le \& Mikolov, 2014)} \cite{DBLP:LE_mikolov}: doc2vec gives a vector representation of a document.
\item \textit{ProdLDA (Srivastava and Sutton, 2017)} \cite{DBLP:ProdLDA} and \textit{SCHOLAR (Card et al., 2017)} \cite{DBLP:Card_SCHOLAR}: These both are LDA based \textit{bag-of-words} topic models, which give a topic mixture representation of every document. SCHOLAR focuses on incorporating the metadata i.e., date, author information, into topic models. If we remove the metadata information from SCHOLAR model then we get the ProdLDA model.
\item \textit{DocNADE (Lauly et al., 2017)} \cite{DBLP:Lauly_2017} and \textit{NTM (Cao et al., 2015)} \cite{DBLP:Cao}: Feed forward neural network based topic models, which give a topic mixture representation of every document.
\item \textit{GaussianLDA (Das et al., 2015)} \cite{DBLP:Das_gaussian_LDA} and \textit{glove-DMM \& glove-LDA (Nguyen et al., 2015)} \cite{DBLP:Nguyen_gloveDMM}: Topic models which uses pretrained word embeddings like GloVe and give a topic mixture representation of every document
\item \textit{TDLM (Lau et al., 2017)} \cite{DBLP:Lau_TDLM}, \textit{Topic-RNN (Dieng et al., 2016)} \cite{DBLP:Dieng_TopicRNN} and \textit{TCNLM (Wang et al., 2018)} \cite{DBLP:wang_TCNLM}: Topic models jointly trained with language models, they all are focused on improving language model using topical information from topic model.
\item \textit{DeepDNE}: A deeper version of DocNADE topic model with three hidden layers.
\end{enumerate}

Due to constraints from linear complexity of softmax layer, DocNADE generally use a small vocabulary, i.e., a reduced set of words which occurs most in a document corpus, after preprocessing of corpus which we call Reduced Vocabulary (RV). This reduced set generally ignores functional words like  determiners, conjunctions, prepositions, pronouns, auxiliary verbs etc. However, the language model LSTM-LM requires the whole sentence to understand its structure and semantics, so we cannot remove the functional words from dataset for LSTM-LM. Hence, for LSTM-LM, we do preprocessing of dataset keeping all the words in a document and generate a vocabulary which we call Full Vocabulary (FV). 

Therefore, we also investigate training DocNADE on FV setting and compute document representations to perform different evaluation tasks. It also enable us to do a fair comparison of ctx-DocNADE variants with DocNADE. We have used the GloVe embeddings of size 200 and have fixed the number of topics to 200 to maintain consistency among all baselines and proposed models (ctx-DocNADE, ctx-DocNADEe and ctx-DeepDNEe) while quantifying the quality of representation learned form these models. LSTM-LM and DocNADE components of the ctx-DocNADE model learn the underlying syntactic and semantic information in the dataset at different speeds. Therefore, it is a very important to choose a good initialize point for the shared $\textbf{W}$ matrix. For the same, we perform a pre-training with $\lambda$ set to zero for 10 epochs and after that we perform complementary learning of DocNADE and LSTM-LM i.e., training of ctx-DocNADE model. Apart from this, a mixture weight ($\lambda$) is used to control the flow of local contextual and syntactic information form LSTM-LM into DocNADE network. Therefore an ablation study over mixture weight parameter ($\lambda$) is necessary. See the Table ~\ref{table:ctx_docnade_hyperparameters} for hyperparameters selection and Tables ~\ref{table:ctx_docnade_lambdappl} and ~\ref{table:ctx_docnade_lambdIR} for ablation study of $\lambda$ over validation set for different datasets.

\section{Results}
\label{section:res_ctx_evaluation}

\subsection{Generalization: Perplexity}

Since DocNADE is \textit{bag-of-words} model and LSTM-LM maintains the original word order of the document, therefore at each autoregressive step, the previous context of a word in DocNADE and LSTM-LM could be different. Therefore, for ctx-DocNADE we set the mixture parameter $\lambda$ to zero, i.e., considering only DocNADE, while evaluating test dataset to calculate the exact PPL; however, $\lambda$ was non zero during training and we did optimization on the pseudo log-likelihood. The optimal value of $\lambda$ is selected based on the performance on validation set. See the Tables ~\ref{table:ctx_docnade_hyperparameters} and ~\ref{table:ctx_docnade_lambdappl} in \textit{appendix} for hyperparameters selection and for ablation study of $\lambda$ over validation set for different datasets respectively. We also noticed that the PPL performance of DocNADE is already proved to be better than LDA (See Larochelle and Lauly (2012) \cite{DBLP:Lauly_DocNADE}) and DocNADE also outperforms ProdLDA by a huge margin on 20NS dataset (665/1375 vs 1168/2097 respectively in RV/FV settings). Therefore, for these reasons, we have only compared our proposed models with DocNADE.


\begin{table}[t]
\small
\begin{tabular*}{\textwidth}{r||c|c|c|c|c|c}
\hline
Model        & 20NSshort & Subjectivity & AGnewstitle & TMNtitle & Reuters8 & 20NS \\
\hline\hline
DocNADE      & \textbf{646}       & 980          & 846         & 1437     & 283      & 1375 \\
ctx-DocNADE  & 656       & 968          & 822         & 1430     & 276      & \textbf{1358} \\
ctx-DocNADEe & 648       & \textbf{966}          & \textbf{820}         & \textbf{1427}     & \textbf{272}      & 1361 \\
\hline
\end{tabular*}
\caption{Generalization: PPL evaluation scores; best scores are shown in \textbf{bold} font.}
\label{table:ppl}
\end{table}

Table ~\ref{table:ppl} shows the quantitative PPL scores after complementary learning of ctx-Doc\-NADE and ctx-DocNADEe, with optimal $\lambda = 0.01$, compared with DocNADE baseline. ctx-DocNADE achieves lower perplexity (PPL) scores than the baseline DocNADE for short texts (822 vs 846) and long texts (1358 vs 1375) on \textit{AGnewstitle} and \textit{20NS} datasets respectively in full vocabulary (FV) setting.

\subsection{Interpretability: Topic Coherence}

Table ~\ref{table:ctx_docnade_topiccoherence} shows the average coherence score over 200 topics for top 10 and 20 words in each topic. It can be noted that ctx-DocNADE achieves higher average score than DocNADE (.772 vs .755; averaged over 11 datasets), which suggests that syntactic and word-order information help in generating more coherent topics. For ctx-DocNADEe, the introduction of distributed word embeddings further boosts topic coherence which can be seen by the gain of 4.6\% (.790 vs .755) achieved on average over 11 datasets. It is important to note that ctx-DocNADE and ctx-DocNADEe also outperforms the baseline models glove-DMM and glove-LDA which also uses pre-trained word embeddings from GloVe.  To further illustrate the importance of syntactic and word-order information, Table ~\ref{table:ctx_docnade_topiccoherenceexamples} shows the increasing score of topic coherence of a similar topic, related to \textit{computer}, generated from DocNADE, ctx-DocNADE and ctx-DocNADEe. Note that the word ``cars'' in the topic generated from DocNADE is an intruder word, which is removed in topic from ctx-DocNADE using the local contextual information from LSTM-LM and a new word ``terminal'' has been added which is more related to \textit{computer} than ``cars''.

\begin{table}[H]
\centering
\begin{tabular}{r||cccccccccc|}
\hline
\multirow{2}{*}{Dataset} & \multicolumn{2}{c}{glove-DMM} & \multicolumn{2}{c}{glove-LDA} & \multicolumn{2}{c}{DocNADE} & \multicolumn{2}{c}{ctx-DNE} & \multicolumn{2}{c}{ctx-DNEe} \\

& W10           & W20           & W10           & W20           & W10          & W20          & W10            & W20            & W10             & W20 \\
\hline
{\textit{20NSshort}}       & .512          &  .575      &  .616         &  .767     & .669           & .779       & .682         & .794    &   {\textbf{.696}}      &  {\textbf{.801}} \\  

{\textit{TREC6}}            & .410          &  .475      &  .551         &  .736     & .699           & .818       & {\textbf{.714}}         & {\textbf{.810}}    &   .713      &  .809    \\

{\textit{R21578title}}     & .364          &  .458      &  .478         &  .677     & .701           & .812       & .713         & .802    &   {\textbf{.723}}      &  {\textbf{.834}}    \\

{\textit{Polarity}}         & .637          &  .363      &  .375         &  .468     & .610           & .742       & .611         & .756    &   {\textbf{.650}}      &  {\textbf{.779}}  \\   

{\textit{TMNtitle}}        & .633          &  .778      &  .651         &  .798     & .712           & .822       & .716         & .831    &   {\textbf{.735}}      &  {\textbf{.845}}     \\
           
{\textit{TMN}}            & .705          &  .444      &  .550         &  .683     & .642           & .762       & .639         & .759    &   {\textbf{.709}}      &  {\textbf{.825}}      \\

{\textit{Subjectivity}}   & .538          &  .433    &  -  &  -     & .613           & .749       & .629         & .767    &   {\textbf{.634}}      &  {\textbf{.771}}   \\

{\textit{AGnewstitle}}    & .584          &  .678    &  -  &  -      & .731           & .757       & .739         & .858    &  {\textbf{.746}}      &  {\textbf{.865}}    \\

{\textit{20NSsmall}}       & {\textbf{.578}}         &  .548    &  -  &  -         &  .508         &  .628     & .546           & .667       & .565         & {\textbf{.692}}    \\

{\textit{Reuters8}}   & .372         &  .302    &  -  &  -           &  .583         &  .710     & .584           & .710       & {\textbf{.592}}         & {\textbf{.714}}    \\

{\textit{20NS}}        & .458         &  .374    &  -  &  -           &  .606         &  .729     & .615           & .746       & {\textbf{.631}}         & {\textbf{.759}}     \\

\hline

{ Avg (all)}    & .527         &  .452    &  -  &  -     &  .643         &  .755     & .654          & .772       & {\textbf{.672}}         &  {\textbf{.790}} \\

\hline

\end{tabular}
\caption{Average coherence for {\textit{short}} and {\textit{long}} texts over 200 topics in FV setting, where {\textit{DocNADE}} is written as {\textit{DNE}; best scores are shown in \textbf{bold} font.}
\enote{PG}{remove line in the column}
}
\label{table:ctx_docnade_topiccoherence}
\end{table}

\begin{table}[t]
\centering
\begin{tabular}{c|c|c}
\hline
{\textit{DocNADE}}   &     {\textit{ctx-DocNADE}}        & {\textit{ctx-DocNADEe}}           \\ \hline\hline
\textit{vga, screen, computer,}  &  \textit{computer, color, screen,}  &  \textit{svga, graphics, bar, macintosh,} \\
\textit{sell, color, powerbook,}  &  \textit{offer, vga, card,}  &  \textit{san, windows, utility,} \\
\textit{sold, cars, svga, offer}  &  \textit{terminal, forsale, gov, vesa}  &  \textit{monitor, computer,  processor} \\ \hline
.554  &  .624  &  \underline{.667} \\
\hline
\end{tabular}
\caption{A topic of 20NS dataset with coherence values; best score is underlined.}
\label{table:ctx_docnade_topiccoherenceexamples}
\end{table}


As discussed earlier, there exist some related works that combine topic modeling and language modeling paradigms into one composite model such as TDLM (Lau et al., 2017) \cite{DBLP:Lau_TDLM}, Topic-RNN (Dieng et al., 2016) \cite{DBLP:Dieng_TopicRNN} and TCNLM (Wang et al., 2018) \cite{DBLP:wang_TCNLM}. So, it becomes important to compare our proposed models with these models. However, it is important to note that the common motivation behind all these related works is ``to improve the language models using global latent topic information from topic models'' while our motivation is ``to improve the topic models using local syntactic and contextual information from language models''.

\begin{table}[H]
\center
\small
\begin{tabular}{l|c|c}
{\textbf{Topic}}   &  {\textbf{Model}}    &   {\textbf{Topic-words (increasing probabilities from left to right)}} \\ \hline 
                            &    TCNLM\#                &        courses, training, students, medau, education    \\ 
{\textit{education}}       &   ctx-DocNADE         &     teachers, curriculum, workshops, learning, medau      \\ 
                           &    ctx-DocNADEe      &      medau, pupils, teachers, schools, curriculum  \\ \hline
				&   TCNLM\#	   &    pollution, emissions, nuclear, waste, environmental \\ 
{\textit{environment}}	&    ctx-DocNADE  &    ozone, pollution, emissions, warming, waste   \\ 
                                & ctx-DocNADEe    &    pollution, emissions, dioxide, warming, environmental \\ \hline

                              &    TCNLM\#              &        album, band, guitar, music, film    \\ 
{\textit{art}}                    &    ctx-DocNADE         &   guitar, album, band, bass, tone      \\ 
                             &    ctx-DocNADEe     &     guitar, album, pop, guitars, song\\ \hline
                             
                               &    TCNLM\#              &       elections, economic, minister, political, democratic  \\ 
{\textit{politics}}               &    ctx-DocNADE       &    elections, democracy, votes, democratic, communist   \\ 
                               &    ctx-DocNADEe     &    democrat, candidate, voters, democrats, poll    \\  \hline

                            &    TCNLM\#                &        eye, looked, hair, lips, stared      \\ 
{\textit{expression}}    &   ctx-DocNADE         &     nodded, shook, looked, smiled, stared      \\ 
                           &    ctx-DocNADEe      &      charming, smiled, nodded, dressed, eyes    \\ \hline
                           
                            &     TCNLM\#               &        bedrooms, hotel, garden, situated, rooms     \\ 
{\textit{facilities}}         &    ctx-DocNADE         &    bedrooms, queen, hotel, situated, furnished     \\ 
                            &     ctx-DocNADEe      &    hotel, bedrooms, golf, resorts, relax     \\  \hline 

                            &    TCNLM\#                &       corp, turnover, unix, net, profits     \\ 
{\textit{business}}        &    ctx-DocNADE         &    shares, dividend, shareholders, stock, profits     \\ 
                            &    ctx-DocNADEe      &     profits, growing, net, earnings, turnover      \\ 
\end{tabular}

\caption{The top 5 words of seven topics from our proposed models and TCNLM. 
The hash (\#) indicates the topics taken from TCNLM
\enote{PG}{please shuffle the rows. looks cpoying from the paper. please od the same in other table anf figures whereever applies.}
 \cite{DBLP:wang_TCNLM}. }
\label{table:bnctopicstable}
\end{table}

\begin{table}[H]
\center
\small
\begin{tabular}{l|ccc}
\multirow{2}{*} {\textbf{Model}} &  \multicolumn{3}{c}{\textbf{Coherence (NPMI)}}\\ 
 & 50 & 100  &  150 \\   \hline

\multicolumn{4}{c}{(\textbf{Baseline models})}   \\ \hline
\multicolumn{4}{c}{({\textit{sliding window}}=20)}   \\ 

Topic-RNN(s)\#                        &     .102    &    .108    &     .102  \\
Topic-RNN(l)\#                           &   .100   &     .105   &      .097  \\
LDA\# 				       &     .106    &    .119     &    .119  \\
NTM\# 					&    .081    &    .070    &     .072  \\
TDLM(s)\# 				&    .102    &    .106	&      .100  \\
TDLM(l)\#    				&    .095	 &    .101    &    	.104  \\
TCNLM(s)\#                               &    .114   &     .111   &      .107   \\
TCNLM(l)\#                                &     .101  &    .104   &      .102   \\ \hline

\multicolumn{4}{c}{(\textbf{Proposed models})}   \\ \hline
\multicolumn{4}{c}{({\textit{sliding window}}=20)}   \\

DocNADE                                    &   .097     &   .095    &   .097   \\  \hline 
ctx-DocNADE*($\lambda$=0.2)         &   .102    &   .103     &   .102   \\
ctx-DocNADE*($\lambda$=0.8)        &    .106    &   .105     &   .104   \\
ctx-DocNADEe*($\lambda$=0.2)      &   .098    &   .101      &    -        \\
ctx-DocNADEe*($\lambda$=0.8)      &   .105   &    .104     &     -       \\  \hline

 \multicolumn{4}{c}{({\textit{sliding window}}=110)}   \\

DocNADE                                  &   .133    &   .131     &   .132     \\ \hline
ctx-DocNADE*($\lambda$=0.2)      &   .134     &   .141     &   .138       \\
ctx-DocNADE*($\lambda$=0.8)      &   .139    &   .142     &   .140   \\
ctx-DocNADEe*($\lambda$=0.2)    &   .133    &   .139     &     -       \\
ctx-DocNADEe*($\lambda$=0.8)    &   .135    &   .141     &     -     

\end{tabular}
\caption{Topic coherence (NMPI) scores of different models for 50, 100 and 150 topics on BNC dataset. 
The {\textit{sliding window}} is one of the hyper-parameters for computing topic coherence  \cite{DBLP:Roder, DBLP:wang_TCNLM}. 
A {\textit{sliding window}} of 20 is used in TCNLM; in addition we also present results for a window of size 110.  
$\lambda$ is the mixture weight of the LM component in the topic modeling process, and  
(s) and (l) indicate small and large model, respectively.  
The symbol '-' indicates no result, since word embeddings of 150 dimensions are not available from glove vectors.}
\label{table:ctx_docnade_topiccoherence_bnc}
\end{table}


We follow the experimental setup of the most recent work, TCNLM, to do quantitative comparison, in terms of topic coherence (NPMI), of our proposed models (ctx-DocNADE and ctx-DocNADEe) with TCNLM model on BNC dataset. It is important to note that the TCNLM paper compares itself with TDLM, TopicRNN, NTM and LDA. Table ~\ref{table:ctx_docnade_topiccoherence_bnc} shows the NPMI scores of different models on BNC dataset, where the importance of including language structure into the DocNADE topic model has been proved by the results of ctx-DocNADE and ctx-DocNADEe. Note that as the value of $\lambda$ increases (0.8 vs 0.2) the topic coherence also increases which further proves the importance of local language structure and contextual information (DocNADE corresponds to $\lambda = 0$). Similarly, the inclusion of distributed pre-trained word embeddings in ctx-DocNADEe further has a positive effect on the topic coherence score against DocNADE. While, the motivation behind ctx-DocNADEe is improvement in sparse data settings, it is important to note that BNC is neither a short-text nor a small corpus of documents. Therefore, there is no significant improvement in ctx-DocNADEe over ctx-DocNADE in terms of topic coherence. Also, we compare topic coherence (.24/.19) of topics learned by ProdLDA with topic coherence (.15/.12) of topics learned by DocNADE for (50/200) topics in reduced vocabulary setting on 20NS dataset, which suggest that ProdLDA is better topic model at capturing latent topics than DocNADE.


In Table \ref{table:bnctopicstable}, we further qualitatively show the top 5 words of the seven learned topics (names of the topics are summarized by Wang et al. (2018) \cite{DBLP:wang_TCNLM}) from our proposed models (i.e., ctx-DocNADE and ctx-DocNADEe) and TCNLM. Since the BNC dataset is unlabeled, we are here restricted to comparing model performance in terms of topic coherence (COH) only.



\subsection{Applicability: Text Retrieval}


To show the applicability of our proposed models, we perform a document retrieval task for short and long-text document using their label information. We follow the experimental setup of Lauly et al. (2017) \cite{DBLP:Lauly_2017}, where all test documents are treated as the query document to retrieve a fraction of the closest documents in the original training set using cosine similarity measure between their \textit{contextualized} representations. Let's say we have a document $D_1$ with a particular label $l_1$, now we perform similarity distance between \textit{contextualized} representations of $D_1$ and all the other documents in training dataset. After that we take a particular fraction $f_1$ (e.g., 0.0001, 0.005, 0.01, 0.02, 0.05, etc.) of most similar training documents, and then count the fraction of document with the same label as query i.e., $l_1$, We do this for all documents in test dataset and then take an average to calculate \textit{retrieval precision} at a particular fraction $f_1$. Since, Salakhutdinov \& Hinton (2009) \cite{DBLP:ruslan_RSM} and Lauly et al. (2017) \cite{DBLP:Lauly_2017} have shown that RSM and DocNADE strictly outperform LDA on this task, we solely compare DocNADE baseline with our proposed models. See Table ~\ref{table:ctx_docnade_lambdIR}, in \textit{appendix}, for ablation study of $\lambda$ for IR task on validation set of all training datasets.



Table ~\ref{table:ctx_docnade_IRscoresshorttext} and ~\ref{table:ctx_docnade_IRscoreslongtext} show the retrieval precision scores at a particular retrieval fraction of $0.02$ for short and long text datasets respectively. It is to be noted that for short-text documents, introduction of both pre-trained distributional embeddings (from GloVe) and language structure information (from LSTM-LM) leads to improvement in the IR task. For the full vocabulary (FV) setting, we noticed that the DocNADE(FV) and glove(FV) achieve an improvement in IR precision score over DocNADE(RV) or glove(RV) respectively with reduced vocabulary setting. For this reason, we opt for FV setting in our proposed extension also. On an average over the 8 short-text and 6 long-text datasets, ctx-DocNADEe reports a gain of 7.1\% (.630 vs .588) and 6.0\% (.601 vs .567) (Table 4), respectively in precision compared to DocNADE(RV). We also compared our proposed models with TDLM and found out that ctx-DocNADE and ctx-DocNADEe outperform TDLM by a noticeable margin, on an average, for all short-text datasets with a gain of 14.5\% (.630 vs .550; with ctx-DocNADE and TDLM models respectively) in IR precision. In addition, the deep variant ($d=3$) with embeddings, i.e., ctx-DeepDNEe shows competitive performance on TREC6 and Subjectivity datasets.

\begin{table}[H]
\center
\small
\begin{tabular}{r||c|c|c|c|c|c|c|c||c}
\textbf{Model}        & \textbf{ST-1} & \textbf{ST-2} & \textbf{ST-3} & \textbf{ST-4} & \textbf{ST-5} & \textbf{ST-6} & \textbf{ST-7} & \textbf{ST-8} & \textbf{Average} \\

\hline
\multicolumn{10}{c}{\textbf{Baseline models}}                                        \\
\hline

\textit{GaussianLDA}  & 0.080              & 0.325          & \textbf{0.713}        & 0.558                 & 0.505             & 0.408             & 0.367                & 0.516                & 0.434            \\
\textit{glove-DMM}    & 0.183              & 0.370          & 0.551        & 0.738                 & 0.515             & 0.445             & 0.273                & 0.540                & 0.451            \\
\textit{glove-LDA}    & 0.160              & 0.300          & 0.428        & 0.610                 & 0.517             & 0.260             & 0.387                & 0.547                & 0.401            \\
\textit{glove(RV)}    & 0.236              & 0.480          & 0.638        & 0.754                 & 0.543             & 0.513             & 0.587                & 0.588                & 0.542            \\
\textit{glove(FV)}    & 0.236              & 0.480          & 0.643        & 0.775                 & 0.553             & 0.545             & 0.595                & 0.612                & 0.554            \\
\textit{doc2vec}      & 0.090              & 0.260          & 0.220        & 0.571                 & 0.510             & 0.190             & 0.518                & 0.265                & 0.328            \\
\textit{TDLM}         & 0.219              & 0.521          & 0.672        & 0.839                 & 0.520             & 0.535             & 0.563                & 0.534                & 0.550            \\
\textit{DocNADE(RV)}  & 0.290              & 0.550          & 0.652        & 0.820                 & 0.560             & 0.524             & \textbf{0.657}                & 0.656                & 0.588            \\
\textit{DocNADE(FV)}  & 0.290              & 0.546          & 0.687        & 0.848                 & 0.576             & 0.525             & 0.654                & 0.678                & 0.600            \\
\textit{DeepDNE}      & 0.100              & 0.479          & 0.671        & 0.865                 & 0.503             & 0.536             & 0.630                & 0.682                & 0.558            \\

\hline
\multicolumn{10}{c}{\textbf{Proposed models}}                                        \\
\hline

\textit{ctx-DocNADE}  & 0.296              & 0.595          & 0.692        & 0.874                 & 0.591             & 0.560             & 0.641                & 0.691                & 0.617            \\
\textit{ctx-DocNADEe} & \textbf{0.306}              & 0.599          & 0.698        & 0.874                 & \textbf{0.605}             & \textbf{0.595}             & 0.656                & \textbf{0.703}                & \textbf{0.630}            \\
\textit{ctx-DeepDNEe} & 0.278              & \textbf{0.606}          & 0.687        & \textbf{0.878}                 & 0.591             & 0.576             & 0.647                & 0.689                & 0.620           
\end{tabular}
\caption{State-of-the-art comparison of IR-precision (at 0.02 fraction) for {\textit{short}} texts, where $Average$: average over the row values, best scores are shown in \textbf{bold} font. (ST-1: \textit{20NSshort}; ST-2: \textit{TREC6}; ST-3: \textit{TMN}; ST-4: \textit{Subjectivity}; ST-5: \textit{Polarity}; ST-6: \textit{TMNtitle}, ST-7: \textit{R21578title}; ST-8: \textit{AGnewstitle})}
\label{table:ctx_docnade_IRscoresshorttext}
\end{table}

\begin{table}[H]
\center
\small
\begin{tabular}{r||c|c|c|c|c|c||c}
\textbf{Model}        & \textbf{20NSsmall} & \textbf{Reuters8} & \textbf{20NS} & \textbf{R21578} & \textbf{SiROBs} & \textbf{AGnews} & \textbf{Average} \\

\hline
\multicolumn{8}{c}{\textbf{Baseline models}}     \\
\hline

\textit{GaussianLDA}  & 0.090              & 0.712             & 0.142         & 0.539           & 0.232           & 0.456           & 0.361            \\
\textit{glove-DMM}    & 0.060              & 0.623             & 0.092         & 0.501           & 0.226           & -               & -                \\
\textit{glove(RV)}    & 0.214              & 0.845             & 0.200         & 0.644           & 0.273           & 0.725           & 0.483            \\
\textit{glove(FV)}    & 0.238              & 0.837             & 0.253         & 0.659           & 0.285           & 0.737           & 0.501            \\
\textit{doc2vec}      & 0.200              & 0.586             & 0.216         & 0.524           & 0.282           & 0.387           & 0.365            \\
\textit{DocNADE(RV)}  & 0.270              & \textbf{0.884}             & 0.366         & \textbf{0.723}           & 0.374           & 0.787           & 0.567            \\
\textit{DocNADE(FV)}  & 0.299              & 0.879             & 0.427         & 0.715           & 0.382           & 0.794           & 0.582            \\

\hline
\multicolumn{8}{c}{\textbf{Proposed models}}     \\
\hline

\textit{ctx-DocNADE}  & 0.313              & 0.880             & 0.472         & 0.714           & 0.386           & 0.791           & 0.592            \\
\textit{ctx-DocNADEe} & \textbf{0.327}              & 0.883             & \textbf{0.486}         & 0.721           & \textbf{0.390}           & \textbf{0.796}           & \textbf{0.601}           
\end{tabular}
\caption{State-of-the-art comparison of IR-precision (at 0.02 fraction) for {\textit{long}} texts, where $Average$: average over the row values, best scores are shown in \textbf{bold} font.}
\label{table:ctx_docnade_IRscoreslongtext}
\end{table}

\begin{figure*}[t]
\centering

\begin{subfigure}{0.48\textwidth}
\centering
\begin{tikzpicture}[scale=0.8][baseline]
\small
\begin{axis}[
    x label style={at={(axis description cs:0.5,-0.05)},anchor=north},
    xlabel={\textbf{Fraction of Retrieved Documents (Recall)}},
    ylabel={\textbf{Precision (\%)}},
    xmin=0, xmax=10,
    ymin=0.09, ymax=0.51,
   /pgfplots/ytick={.10,.14,...,.51},
    xtick={0,1,2,3,4,5,6,7,8,9, 10,11,12},
    xticklabels={0.0005, 0.001, 0.002, 0.005, 0.01, 0.02, 0.05, 0.1, 0.2, 0.3, 0.5},
    x tick label style={rotate=45,anchor=east},
    legend pos=north east,
    ymajorgrids=true, xmajorgrids=true,
    grid style=dashed,
]

\addplot[
	color=blue,
	mark=square,
	]
	plot coordinates {
    (0, 0.472)
    (1, 0.472)
    (2, 0.449)
    (3, 0.394)
    (4, 0.356)
    (5, 0.306)
    (6, 0.256)
    (7, 0.212)
    (8, 0.162)
    (9, 0.134)
    (10, 0.102)
    (11, 0.077)
    (12, 0.066)
	};
\addlegendentry{ctx-DocNADEe}

\addplot[
	color=red,
	mark=square,
	]
	plot coordinates {
    (0, 0.463)
    (1, 0.463)
    (2, 0.429)
    (3, 0.379)
    (4, 0.338)
    (5, 0.296)
    (6, 0.274)
    (7, 0.204)
    (8, 0.155)
    (9, 0.128)
    (10, 0.10)
    (11, 0.076)
    (12, 0.066)
	};
\addlegendentry{ctx-DocNADE}

\addplot[
	color=cyan,
	mark=triangle,
	]
	plot coordinates {
    (0, 0.470)
    (1, 0.470)
    (2, 0.418)
    (3, 0.375)
    (4, 0.335)
    (5, 0.290)
    (6, 0.242)
    (7, 0.199)
    (8, 0.153)
    (9, 0.127)
    (10, 0.099)
    (11, 0.076)
    (12, 0.066)
    
	};
\addlegendentry{DocNADE}

\addplot[
	color=green,
	mark=*,
	]
	plot coordinates {
    (0, 0.388)
    (1, 0.388)
    (2, 0.378)
    (3, 0.327)
    (4, 0.281)
    (5, 0.236)
    (6, 0.188)
    (7, 0.154)
    (8, 0.123)
    (9, 0.107)
    (10, 0.088)
    (11, 0.072)
    (12, 0.066)
	};
\addlegendentry{glove}

\end{axis}
\end{tikzpicture}%
\caption{{\textbf{IR:}} 20NSshort} \label{fig:IR20NSshort}
\end{subfigure}\hspace*{\fill}
\begin{subfigure}{0.48\textwidth}
\centering
\begin{tikzpicture}[scale=0.8][baseline]
\small
\begin{axis}[
    x label style={at={(axis description cs:0.5,-0.05)},anchor=north},
    xlabel={\textbf{Fraction of Retrieved Documents (Recall)}},
    xmin=0, xmax=13,
    ymin=0.50, ymax=0.91,
    /pgfplots/ytick={.50,.55,...,.91},
    xtick={0,1,2,3,4,5,6,7,8,9,10, 11,12,13},
    xticklabels={0.0001, 0.0005, 0.001, 0.002, 0.005, 0.01, 0.02, 0.05, 0.1, 0.2, 0.3, 0.5, 0.8, 1.0},
    x tick label style={rotate=48,anchor=east},
    legend pos=south west,
    ymajorgrids=true, xmajorgrids=true,
    grid style=dashed,
]

\addplot[
	color=blue,
	mark=square,
	]
	plot coordinates {
    (0, 0.896)
    (1, 0.893)
    (2, 0.889)
    (3, 0.885)
    (4, 0.882)
    (5, 0.878)
    (6, 0.874)
    (7, 0.865)
    (8, 0.854)
    (9, 0.833)
    (10, 0.810)
    (11, 0.739)
    (12, 0.583)
    (13, 0.499)
	};
\addlegendentry{ctx-DocNADEe}

\addplot[
	color=red,
	mark=square,
	]
	plot coordinates {
    (0, 0.888)
    (1, 0.880)
    (2, 0.880)
    (3, 0.882)
    (4, 0.879)
    (5, 0.877)
    (6, 0.874)
    (7, 0.869)
    (8, 0.861)
    (9, 0.848)
    (10, 0.829)
    (11, 0.763)
    (12, 0.586)
    (13, 0.499)
	};
\addlegendentry{ctx-DocNADE}

\addplot[
	color=cyan,
	mark=triangle,
	]
	plot coordinates {
    (0, 0.889)
    (1, 0.872)
    (2, 0.868)
    (3, 0.864)
    (4, 0.858)
    (5, 0.854)
    (6, 0.848)
    (7, 0.838)
    (8, 0.826)
    (9, 0.804)
    (10, 0.780)
    (11, 0.713)
    (12, 0.578)
    (13, 0.499)
    
	};
\addlegendentry{DocNADE(FV)}

\addplot[
	color=black,
	mark=*,
	]
	plot coordinates {
    (0, 0.837)
    (1, 0.840)
    (2, 0.837)
    (3, 0.834)
    (4, 0.828)
    (5, 0.824)
    (6, 0.820)
    (7, 0.811)
    (8, 0.801)
    (9, 0.782)
    (10, 0.760)
    (11, 0.701)
    (12, 0.576)
     (13, 0.499)
	};
\addlegendentry{DocNADE(RV)}

\addplot[
	color=green,
	mark=*,
	]
	plot coordinates {
    (0, 0.846)
    (1, 0.831)
    (2, 0.826)
    (3, 0.820)
    (4, 0.805)
    (5, 0.791)
    (6, 0.775)
    (7, 0.744)
    (8, 0.713)
    (9, 0.668)
    (10, 0.634)
    (11, 0.581)
    (12, 0.525)
    (13, 0.499)
	};
\addlegendentry{glove(FV)}

\addplot[
	color=orange,
	mark=*,
	]
	plot coordinates {
    (0, 0.813)
    (1, 0.807)
    (2, 0.800)
    (3, 0.794)
    (4, 0.782)
    (5, 0.770)
    (6, 0.753)
    (7, 0.725)
    (8, 0.695)
    (9, 0.653)
    (10, 0.620)
    (11, 0.571)
    (12, 0.521)
    (13, 0.499)
	};
\addlegendentry{glove(RV)}

\end{axis}
\end{tikzpicture}%
\caption{{\textbf{IR:}} Subjectivity} \label{fig:IRsubjectivity}
\end{subfigure}
~
\begin{subfigure}{0.48\textwidth}
\centering
\begin{tikzpicture}[scale=0.8][baseline]
\small
\begin{axis}[
    x label style={at={(axis description cs:0.5,-0.05)},anchor=north},
    xlabel={\textbf{Fraction of Retrieved Documents (Recall)}},
    ylabel={\textbf{Precision (\%)}},
    xmin=0, xmax=13,
    ymin=0.50, ymax=0.67,
   /pgfplots/ytick={.50,.52,...,.67},
    xtick={0,1,2,3,4,5,6,7,8,9,10,11,12,13},
    xticklabels={0.0001, 0.0005, 0.001, 0.002, 0.005, 0.01, 0.02, 0.05, 0.1, 0.2, 0.3, 0.5, 0.8, 1.0},
    x tick label style={rotate=45,anchor=east},
    legend pos=north east,
    ymajorgrids=true, xmajorgrids=true,
    grid style=dashed,
]

\addplot[
	color=blue,
	mark=square,
	]
	plot coordinates {
    (0, 0.660)
    (1, 0.650)
    (2, 0.642)
    (3, 0.634)
    (4, 0.624)
    (5, 0.614)
    (6, 0.603)
    (7, 0.587)
    (8, 0.575)
    (9, 0.559)
    (10, 0.549)
    (11, 0.533)
    (12, 0.513)
   (13, 0.499)
	};
\addlegendentry{ctx-DocNADEe}

\addplot[
	color=red,
	mark=square,
	]
	plot coordinates {
    (0, 0.650)
    (1, 0.636)
    (2, 0.625)
    (3, 0.619)
    (4, 0.607)
    (5, 0.599)
    (6, 0.591)
    (7, 0.578)
    (8, 0.567)
    (9, 0.553)
    (10, 0.544)
    (11, 0.530)
    (12, 0.512)
   (13, 0.499)
	};
\addlegendentry{ctx-DocNADE}

\addplot[
	color=cyan,
	mark=triangle,
	]
	plot coordinates {
    (0, 0.647)
    (1, 0.632)
    (2, 0.617)
    (3, 0.605)
    (4, 0.595)
    (5, 0.585)
    (6, 0.576)
    (7, 0.562)
    (8, 0.552)
    (9, 0.542)
    (10, 0.534)
    (11, 0.523)
    (12, 0.509)
    (13, 0.499)
	};
\addlegendentry{DocNADE(FV)}

\addplot[
	color=black,
	mark=*,
	]
	plot coordinates {
    (0, 0.643)
    (1, 0.623)
    (2, 0.610)
    (3, 0.596)
    (4, 0.581)
    (5, 0.570)
    (6, 0.560)
    (7, 0.546)
    (8, 0.537)
    (9, 0.528)
    (10, 0.523)
    (11, 0.515)
    (12, 0.506)
    (13, 0.499)
    
	};
\addlegendentry{DocNADE(RV)}

\addplot[
	color=green,
	mark=*,
	]
	plot coordinates {
    (0, 0.606)
    (1, 0.591)
    (2, 0.588)
    (3, 0.577)
    (4, 0.566)
    (5, 0.559)
    (6, 0.553)
    (7, 0.541)
    (8, 0.532)
    (9, 0.523)
    (10, 0.518)
    (11, 0.511)
    (12, 0.503)
  (13, 0.499)
	};
\addlegendentry{glove(FV)}

\addplot[
	color=orange,
	mark=*,
	]
	plot coordinates {
    (0, 0.591)
    (1, 0.585)
    (2, 0.579)
    (3, 0.568)
    (4, 0.557)
    (5, 0.550)
    (6, 0.542)
    (7, 0.534)
    (8, 0.526)
    (9, 0.519)
    (10, 0.515)
    (11, 0.509)
    (12, 0.503)
  (13, 0.499)
	};
\addlegendentry{glove(RV)}

\end{axis}
\end{tikzpicture}%
\caption{{\textbf{IR:}} Polarity} \label{fig:IRPolarity}
\end{subfigure}\hspace*{\fill}
\begin{subfigure}{0.48\textwidth}
\centering
\begin{tikzpicture}[scale=0.8][baseline]
\small
\begin{axis}[
    x label style={at={(axis description cs:0.5,-0.05)},anchor=north},
    xlabel={\textbf{Fraction of Retrieved Documents  (Recall)}},
    xmin=0, xmax=11,
    ymin=0.22, ymax=0.76,
    /pgfplots/ytick={.25,.30,...,.76},
    xtick={0,1,2,3,4,5,6,7,8,9, 10,11,12,13},
    xticklabels={0.0001, 0.0005, 0.001, 0.002, 0.005, 0.01, 0.02, 0.05, 0.1, 0.2, 0.3, 0.5},
    x tick label style={rotate=48,anchor=east},
    legend pos=south west,
    ymajorgrids=true, xmajorgrids=true,
    grid style=dashed,
]

\addplot[
	color=blue,
	mark=square,
	]
	plot coordinates {
    (0, 0.739)
    (1, 0.698)
    (2, 0.681)
    (3, 0.665)
    (4, 0.641)
    (5, 0.620)
    (6, 0.595)
    (7, 0.549)
    (8, 0.496)
    (9, 0.417)
    (10, 0.357)
    (11, 0.273)
    (12, 0.200)
    (13, 0.169)
	};
\addlegendentry{ctx-DocNADEe}

\addplot[
	color=red,
	mark=square,
	]
	plot coordinates {
    (0, 0.719)
    (1, 0.679)
    (2, 0.660)
    (3, 0.640)
    (4, 0.612)
    (5, 0.588)
    (6, 0.560)
    (7, 0.512)
    (8, 0.463)
    (9, 0.394)
    (10, 0.341)
    (11, 0.266)
    (12, 0.199)
    (13, 0.169)
	};
\addlegendentry{ctx-DocNADE}

\addplot[
	color=cyan,
	mark=triangle,
	]
	plot coordinates {
    (0, 0.712)
    (1, 0.676)
    (2, 0.656)
    (3, 0.632)
    (4, 0.600)
    (5, 0.574)
    (6, 0.576)
    (7, 0.501)
    (8, 0.455)
    (9, 0.387)
    (10, 0.335)
    (11, 0.263)
    (12, 0.198)
    (13, 0.169)
	};
\addlegendentry{DocNADE(FV)}

\addplot[
	color=black,
	mark=*,
	]
	plot coordinates {
    (0, 0.665)
    (1, 0.634)
    (2, 0.620)
    (3, 0.600)
    (4, 0.573)
    (5, 0.549)
    (6, 0.524)
    (7, 0.480)
    (8, 0.435)
    (9, 0.372)
    (10, 0.324)
    (11, 0.259)
    (12, 0.198)
    (13, 0.169)
	};
\addlegendentry{DocNADE(RV)}

\addplot[
	color=green,
	mark=*,
	]
	plot coordinates {
    (0, 0.746)
    (1, 0.700)
    (2, 0.675)
    (3, 0.649)
    (4, 0.612)
    (5, 0.581)
    (6, 0.525)
    (7, 0.489)
    (8, 0.434)
    (9, 0.359)
    (10, 0.309)
    (11, 0.246)
    (12, 0.192)
    (13, 0.169)
	};
\addlegendentry{glove(FV)}

\addplot[
	color=orange,
	mark=*,
	]
	plot coordinates {
    (0, 0.694)
    (1, 0.653)
    (2, 0.633)
    (3, 0.610)
    (4, 0.574)
    (5, 0.544)
    (6, 0.512)
    (7, 0.461)
    (8, 0.411)
    (9, 0.343)
    (10, 0.296)
    (11, 0.239)
    (12, 0.190)
    (13, 0.169)
	};
\addlegendentry{glove(RV)}

\end{axis}
\end{tikzpicture}%
\caption{{\textbf{IR:}} TMNtitle} \label{fig:IRTMNtitle}
\end{subfigure}
~
\begin{subfigure}{0.48\textwidth}
\centering
\begin{tikzpicture}[scale=0.8][baseline]
\small
\begin{axis}[
    x label style={at={(axis description cs:0.5,-0.05)},anchor=north},
    xlabel={\textbf{Fraction of Retrieved Documents (Recall)}},
    ylabel={\textbf{Precision (\%)}},
    xmin=0, xmax=11,
    ymin=0.30, ymax=0.81,
    /pgfplots/ytick={.30,.35,...,.81},
    xtick={0,1,2,3,4,5,6,7,8,9, 10,11,12,13},
    xticklabels={0.0001, 0.0005, 0.001, 0.002, 0.005, 0.01, 0.02, 0.05, 0.1, 0.2, 0.3, 0.5}, 
    x tick label style={rotate=48,anchor=east},
    legend pos=south west,
    ymajorgrids=true, xmajorgrids=true,
    grid style=dashed,
]

\addplot[
	color=blue,
	mark=square,
	]
	plot coordinates {
    (0, 0.792)
    (1, 0.768)
    (2, 0.757)
    (3, 0.747)
    (4, 0.732)
    (5, 0.718)
    (6, 0.701)
    (7, 0.667)
    (8, 0.623)
    (9, 0.544)
    (10, 0.477)
    (11, 0.380)
    (12, 0.291)
    (13, 0.250)
	};
\addlegendentry{ctx-DocNADEe}

\addplot[
	color=red,
	mark=square,
	]
	plot coordinates {
    (0, 0.785)
    (1, 0.760)
    (2, 0.749)
    (3, 0.738)
    (4, 0.722)
    (5, 0.708)
    (6, 0.691)
    (7, 0.657)
    (8, 0.611)
    (9, 0.532)
    (10, 0.467)
    (11, 0.375)
    (12, 0.290)
    (13, 0.250)
	};
\addlegendentry{ctx-DocNADE}

\addplot[
	color=cyan,
	mark=triangle,
	]
	plot coordinates {
    (0, 0.789)
    (1, 0.759)
    (2, 0.746)
    (3, 0.732)
    (4, 0.713)
    (5, 0.697)
    (6, 0.678)
    (7, 0.643)
    (8, 0.598)
    (9, 0.519)
    (10, 0.457)
    (11, 0.369)
    (12, 0.288)
    (13, 0.250)
	};
\addlegendentry{DocNADE(FV)}

\addplot[
	color=black,
	mark=*,
	]
	plot coordinates {
    (0, 0.758)
    (1, 0.734)
    (2, 0.722)
    (3, 0.708)
    (4, 0.690)
    (5, 0.674)
    (6, 0.656)
    (7, 0.621)
    (8, 0.577)
    (9, 0.503)
    (10, 0.445)
    (11, 0.363)
    (12, 0.287)
    (13, 0.250)
	};
\addlegendentry{DocNADE(RV)}

\addplot[
	color=green,
	mark=*,
	]
	plot coordinates {
    (0, 0.777)
    (1, 0.738)
    (2, 0.718)
    (3, 0.698)
    (4, 0.669)
    (5, 0.643)
    (6, 0.612)
    (7, 0.558)
    (8, 0.502)
    (9, 0.429)
    (10, 0.383)
    (11, 0.324)
    (12, 0.273)
    (13, 0.250)
	};
\addlegendentry{glove(FV)}

\addplot[
	color=orange,
	mark=*,
	]
	plot coordinates {
    (0, 0.750)
    (1, 0.710)
    (2, 0.691)
    (3, 0.671)
    (4, 0.642)
    (5, 0.617)
    (6, 0.587)
    (7, 0.536)
    (8, 0.483)
    (9, 0.416)
    (10, 0.374)
    (11, 0.320)
    (12, 0.272)
    (13, 0.250)
	};
\addlegendentry{glove(RV)}

\end{axis}
\end{tikzpicture}%
\caption{{\textbf{IR:}} AGnewstitle} \label{fig:IRAGnewstitle}
\end{subfigure}\hspace*{\fill}
\begin{subfigure}{0.48\textwidth}
\centering
\begin{tikzpicture}[scale=0.8][baseline]
\small
\begin{axis}[
    x label style={at={(axis description cs:0.5,-0.05)},anchor=north},
    xlabel={\textbf{Fraction of Retrieved Documents (Recall)}},
    xmin=0, xmax=11,
    ymin=0.10, ymax=0.71,
   /pgfplots/ytick={.10,.20,...,.71},
    xtick={0,1,2,3,4,5,6,7,8,9, 10,11,12,13},
    xticklabels={0.0001, 0.0005, 0.001, 0.002, 0.005, 0.01, 0.02, 0.05, 0.1, 0.2, 0.3, 0.5},
    x tick label style={rotate=45,anchor=east},
    legend pos=north east,
    ymajorgrids=true, xmajorgrids=true,
    grid style=dashed,
]

\addplot[
	color=blue,
	mark=square,
	]
	plot coordinates {
    (0, 0.687)
    (1, 0.657)
    (2, 0.659)
    (3, 0.602)
    (4, 0.565)
    (5, 0.532)
    (6, 0.484)
    (7, 0.382)
    (8, 0.282)
    (9, 0.189)
    (10, 0.144)
    (11, 0.097)
    (12, 0.065)
    (13, 0.052)
	};
\addlegendentry{ctx-DocNADEe}

\addplot[
	color=red,
	mark=square,
	]
	plot coordinates {
    (0, 0.669)
    (1, 0.642)
    (2, 0.615)
    (3, 0.587)
    (4, 0.552)
    (5, 0.518)
    (6, 0.472)
    (7, 0.372)
    (8, 0.276)
    (9, 0.186)
    (10, 0.142)
    (11, 0.097)
    (12, 0.064)
    (13, 0.052)
	};
\addlegendentry{ctx-DocNADE}

\addplot[
	color=cyan,
	mark=triangle,
	]
	plot coordinates {
    (0, 0.657)
    (1, 0.618)
    (2, 0.588)
    (3, 0.557)
    (4, 0.517)
    (5, 0.478)
    (6, 0.428)
    (7, 0.336)
    (8, 0.251)
    (9, 0.174)
    (10, 0.137)
    (11, 0.096)
    (12, 0.064)
    (13, 0.052)
	};
\addlegendentry{DocNADE(FV)}

\addplot[
	color=black,
	mark=*,
	]
	plot coordinates {
    (0, 0.582)
    (1, 0.538)
    (2, 0.509)
    (3, 0.477)
    (4, 0.437)
    (5, 0.406)
    (6, 0.366)
    (7, 0.294)
    (8, 0.225)
    (9, 0.159)
    (10, 0.126)
    (11, 0.090)
    (12, 0.063)
    (13, 0.052)
	};
\addlegendentry{DocNADE(RV)}

\addplot[
	color=green,
	mark=*,
	]
	plot coordinates {
    (0, 0.524)
    (1, 0.466)
    (2, 0.428)
    (3, 0.387)
    (4, 0.335)
    (5, 0.295)
    (6, 0.253)
    (7, 0.196)
    (8, 0.155)
    (9, 0.117)
    (10, 0.098)
    (11, 0.076)
    (12, 0.059)
     (13, 0.052)
	};
\addlegendentry{glove(FV)}

\addplot[
	color=orange,
	mark=*,
	]
	plot coordinates {
    (0, 0.434)
    (1, 0.381)
    (2, 0.347)
    (3, 0.308)
    (4, 0.263)
    (5, 0.231)
    (6, 0.199)
    (7, 0.158)
    (8, 0.128)
    (9, 0.101)
    (10, 0.087)
    (11, 0.070)
    (12, 0.057)
     (13, 0.052)
	};
\addlegendentry{glove(RV)}

\end{axis}
\end{tikzpicture}%
\caption{{\textbf{IR:}} 20NS} \label{fig:IR20NS}
\end{subfigure}
\caption{
Retrieval performance (IR-precision) on 
6 datasets at different fractions
}
\label{fig:docretrieval}
\end{figure*}
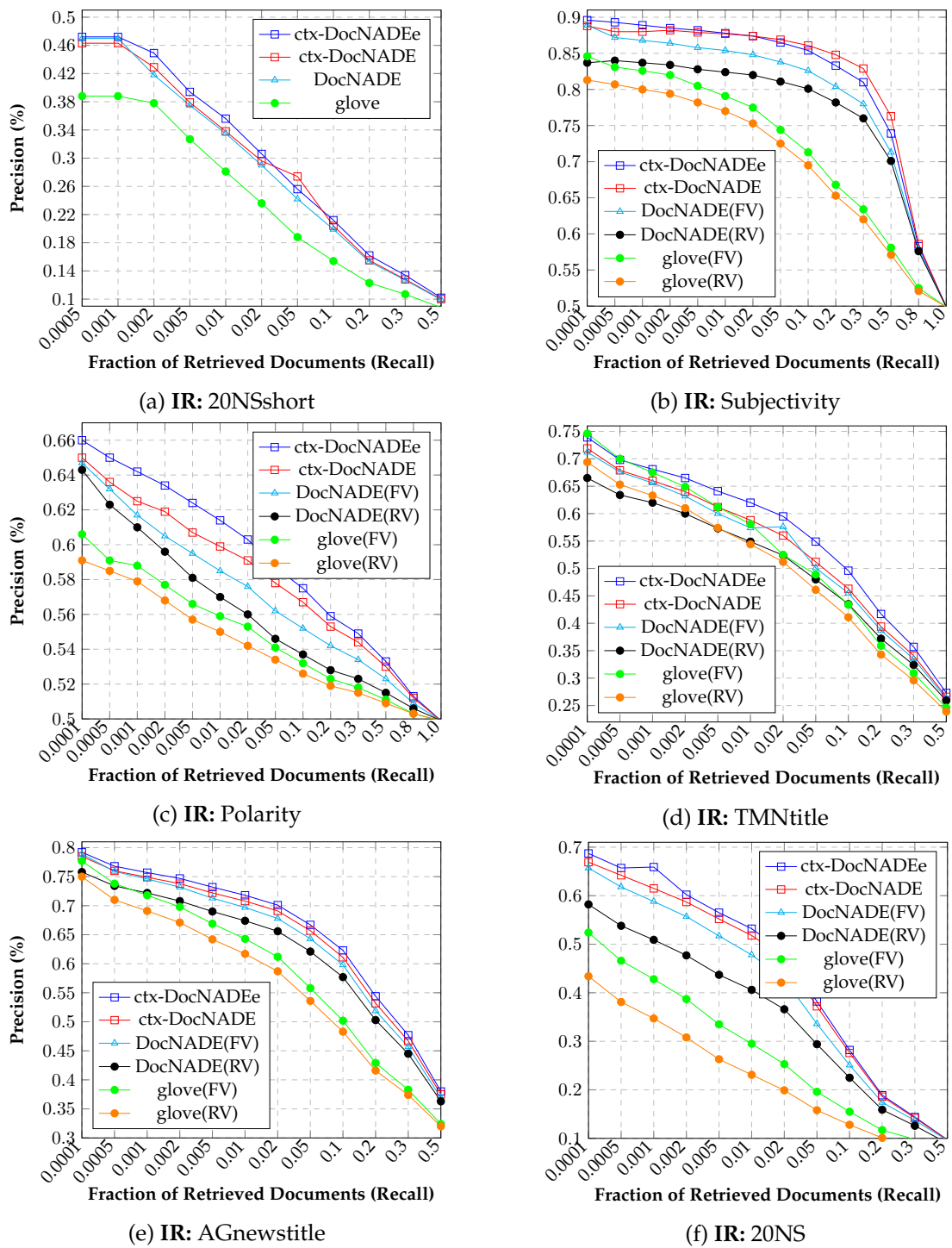

Figures (~\ref{fig:IR20NSshort}, ~\ref{fig:IRsubjectivity}, ~\ref{fig:IRPolarity}, ~\ref{fig:IRTMNtitle}, ~\ref{fig:IRAGnewstitle} and ~\ref{fig:IR20NS}) illustrate the average precision for the retrieval task on 6 datasets over a list of different fractions. Observe that the ctx-DocNADEe outperforms DocNADE(RV) at all the fractions and demonstrates a gain of 6.5\% (.615 vs .577 respectively) in IR precision at fraction 0.02, averaged over 14 datasets. Additionally, we evaluate IR precision task over topic mixture representation learned by ProdLDA model for 20NS dataset and it shows that our proposed models outperform both ProdLDA and TDLM by noticeable margins. However, we also evaluated IR performance of ProdLDA model on short-text datasets at retrieval fraction $0.02$ and achieved these precision scores: 20NSshort (.08), \textit{TREC6} (.24), \textit{R21578title} (.31), \textit{Subjectivity} (.63) and \textit{Polarity} (.51). Therefore, DocNADE, ctx-DocNADE and ctx-DocNADEe outperform ProdLDA in both the settings: short-text datasets (i.e., low word co-occurences) and long-text datasets (i.e., high word co-occurrences).

\subsection{Applicability: Categorization}

We use the same experimental setup as the \textit{information retrieval} task with 200 dimensional \textit{contextualized} representations. We use logistic regression classifier with L2 regularization. To do a fair comparison of ctx-DocNADEe and ctx-DeepDNEe, which use distributional pretrained embeddings prior, we compare them with the topic models baselines which uses embeddings prior.

\begin{table}[H]
\center
\small
\begin{tabular}{r||c|c|c|c|c|c|c|c||c}
\textbf{Model}        & \textbf{ST-1} & \textbf{ST-2} & \textbf{ST-3} & \textbf{ST-4} & \textbf{ST-5} & \textbf{ST-6} & \textbf{ST-7} & \textbf{ST-8} & \textbf{Average} \\

\hline
\multicolumn{10}{c}{\textbf{Baseline models}}                                                        \\
\hline

\textit{GaussianLDA}  & 0.118              & 0.202          & 0.692        & 0.676                 & 0.511             & 0.472             & 0.012                & 0.752                & 0.429            \\
\textit{glove-DMM}    & 0.213              & 0.454          & 0.666        & 0.834                 & 0.585             & 0.590             & 0.011                & 0.652                & 0.500            \\
\textit{glove-LDA}    & 0.320              & 0.600          & 0.627        & 0.805                 & 0.607             & 0.412             & 0.052                & 0.687                & 0.513            \\
\textit{glove(RV)}    & \textbf{0.493}              & 0.798          & 0.736        & 0.882                 & 0.715             & 0.693             & \textbf{0.356}                & 0.814                & 0.685            \\
\textit{glove(FV)}    & 0.488              & 0.785          & \textbf{0.813}        & 0.901                 & 0.728             & \textbf{0.736}             & \textbf{0.356}                & \textbf{0.830}                & 0.704            \\
\textit{doc2vec}      & 0.413              & 0.400          & 0.720        & 0.763                 & 0.624             & 0.582             & 0.176                & 0.600                & 0.534            \\
\textit{TDLM}         & 0.308              & 0.671          & 0.767        & 0.885                 & 0.599             & 0.657             & 0.174                & 0.722                & 0.586            \\
\textit{DocNADE(RV)}  & 0.440              & 0.804          & 0.759        & 0.889                 & 0.699             & 0.664             & 0.313                & 0.819                & 0.673            \\
\textit{DocNADE(FV)}  & 0.440              & 0.791          & 0.796        & 0.907                 & 0.724             & 0.688             & 0.302                & 0.821                & 0.683            \\
\textit{DeepDNE}      & 0.080              & 0.629          & 0.783        & 0.909                 & 0.531             & 0.661             & 0.221                & 0.825                & 0.560            \\

\hline
\multicolumn{10}{c}{\textbf{Proposed models}}                                                        \\
\hline

\textit{ctx-DocNADE}  & 0.440              & 0.817          & 0.793        & 0.910                 & 0.725             & 0.687             & 0.300                & 0.826                & 0.688            \\
\textit{ctx-DocNADEe} & 0.490              & \textbf{0.824}          & 0.806        & 0.917                 & \textbf{0.740}             & 0.726             & 0.308                & 0.828                & \textbf{0.705}            \\
\textit{ctx-DeepDNEe} & 0.406              & 0.804          & 0.796        & \textbf{0.920}                 & 0.723             & 0.694             & 0.244                & 0.826                & 0.688           
\end{tabular}
\caption{State-of-the-art comparison of F1 (classification) scores for \textit{short} texts, where $Average$: average over the row values, best scores are shown in \textbf{bold} font. (ST-1: \textit{20NSshort}; ST-2: \textit{TREC6}; ST-3: \textit{TMN}; ST-4: \textit{Subjectivity}; ST-5: \textit{Polarity}; ST-6: \textit{TMNtitle}, ST-7: \textit{R21578title}; ST-8: \textit{AGnewstitle})}
\label{table:F1scoresshorttext}
\end{table}

Table ~\ref{table:F1scoresshorttext} shows that \textit{glove} outperforms DocNADE for short-text datasets suggesting the need for distributional pretrained priors in DocNADE. However, Table ~\ref{table:F1scoreslongtext} and Table ~\ref{table:F1scoresshorttext} shows that ctx-DocNADEe achieves a gain of 4.8\% (.705 vs .673), on an average, in F1 accuracy for short and long-text datasets respectively over DocNADE(RV) model. For 20NS dataset, the F1 classification scores for other baselines models are as follows: DocNADE (0.734), NTM (0.72), SCHOLAR (0.71) against ctx-DocNADE (0.744) and ctx-DocNADEe (0.751). NTM and SCHOLAR results are taken from (Cao et al. 2015) \cite{DBLP:Cao} and (Card et al. 2017) \cite{DBLP:Card_SCHOLAR} respectively.


\begin{table}[t]
\center
\small
\begin{tabular}{r||c|c|c|c|c|c||c}
\textbf{Model}        & \textbf{20NSsmall} & \textbf{Reuters8} & \textbf{20NS} & \textbf{R21578} & \textbf{SiROBs} & \textbf{AGnews} & \textbf{Average} \\

\hline
\multicolumn{8}{c}{\textbf{Baseline models}}     \\
\hline

\textit{GaussianLDA}  & 0.080              & 0.557             & 0.340         & 0.114           & 0.070           & 0.818           & 0.329            \\
\textit{glove-DMM}    & 0.134              & 0.453             & 0.187         & 0.023           & 0.050           & -               & -                \\
\textit{glove(RV)}    & 0.442              & 0.830             & 0.608         & 0.316           & 0.202           & 0.870           & 0.544            \\
\textit{glove(FV)}    & 0.494              & 0.880             & 0.632         & \textbf{0.340}           & 0.217           & 0.890           & 0.575            \\
\textit{doc2vec}      & 0.450              & 0.852             & 0.691         & 0.215           & 0.226           & 0.713           & 0.524            \\
\textit{DocNADE(RV)}  & \textbf{0.530}              & 0.890             & 0.644         & 0.336           & 0.298           & 0.882           & 0.596            \\
\textit{DocNADE(FV)}  & 0.509              & \textbf{0.907}             & 0.727         & \textbf{0.340}           & 0.308           & 0.888           & 0.613            \\

\hline
\multicolumn{8}{c}{\textbf{Proposed models}}     \\
\hline

\textit{ctx-DocNADE}  & 0.526              & 0.898             & 0.732         & 0.315           & 0.309           & 0.890           & 0.611            \\
\textit{ctx-DocNADEe} & 0.524              & 0.900             & \textbf{0.745}         & 0.332           & \textbf{0.311}           & \textbf{0.894}           & \textbf{0.618}           
\end{tabular}
\caption{State-of-the-art comparison of F1 (classification) scores for \textit{long} texts, where $Average$: average over the row values, best scores are shown in \textbf{bold} font.}
\label{table:F1scoreslongtext}
\end{table}

\chapter{Experiments: Lifelong learning topic model}
\label{chapter:results_lifelong_learning}

\section{Dataset description}

For experiments and validation of \textit{lifelong learning} in topic modeling, we have selected a mix of 3 short text, 4 long-text datasets and 1 small corpus of documents. A brief description about each dataset is mentioned below (please refer to section ~\ref{section:res_ctx_dataset} for detailed statistics about these datasets):

\begin{enumerate}
    \item \textit{20NSshort} (short text): We take documents from 20NewsGroups data, with document size less (in terms of number of words) than 20.
    \item \textit{Reuters21578title} (short text): a collection of new stories from nltk.corpus. We take titles of the documents.
    \item \textit{TMNtitle} (short text): Titles of the Tag My News (TMN) news dataset.
    \item \textit{20NSsmall} (small corpus): We sample 20 document for training from each class of the 20NS dataset. For validation and test, 10 document for each class.
    \item \textit{Reuters21578} (long text): a collection of new stories from nltk.corpus.
    \item \textit{20NewsGroups} (long text): a collection of news stories from nltk.corpus.
    \item \textit{AGnews} (long text): News articles from the AG’s corpus of news articles on the web containing the 4 largest classes.
    \item \textit{TMN} (long text): The Tag My News (TMN) news dataset.
\end{enumerate}

\section{Experimental settings}

In the \textit{lifelong learning} chapter ~\ref{chapter:lifelong_learning}, we discussed three different tasks, namely:

\begin{enumerate}
    \item \textit{Explicit knowledge transfer} (EmbTF): It deals with the transfer of pre-trained word representations, i.e., word embeddings, from source dataset(s). This embedding knowledge transfer is controlled via $\lambda_{EmbTF}$ parameter to prevent the negative transfer, i.e., out of domain embedding transfer.
    \item \textit{Implicit knowledge transfer} (SAL): It deals with the smart selection of those documents from source datasets which have an overlapping domain with the target dataset. The inclusion of selected documents into target dataset is controlled via $\lambda_{SAL}$ parameter.
    \item \textit{Retaining previous knowledge} (RK): It deals with retention of previous learning of the topic model by penalizing the loss term if new model parameters deviates from previous model parameters. The degree of penalization is controlled via $\lambda_{RK}$ parameter.
\end{enumerate}

So, we conduct \textit{lifelong learning} experiments in 4 different combinations of these tasks with DocNADE topic model, to quantitatively identify the contribution due to each task and check the performance gain, due to interplay of these tasks, on DocNADE topic model. These are the four different task combinations we use in our experiments:

\begin{itemize}
    \item \textbf{E1}: DocNADE + EmbTF
    \item \textbf{E2}: DocNADE + RK
    \item \textbf{E3}: DocNADE + EmbTF + RK
    \item \textbf{E4}: DocNADE + EmbTF + RK + SAL
\end{itemize}

It is to be noted here that the \textbf{E1} combination is similar to ctx-Doc\-NADEe model with a difference that instead of transferring GloVe word embeddings we are transferring word embeddings from source dataset(s). Also, \textbf{E1} experiment combination itself does not focus on the retention of previous learning because of the absence of retention task (RK) and SAL task which are designed in a way to minimize forgetting and maximize retention. For baseline, we compare each one of these tasks combination with DocNADE topic model.

\section{Evaluation metrics}

To quantify the improvement in topic modeling process due to inclusion of above-mentioned tasks, we use three evaluation metrics as mentioned below (Refer to section ~\ref{section:ctx_docnade_evaluation_metrics} to get more details about the following evaluation metrics):

\begin{enumerate}
    \item \textit{Perplexity-per-word} (PPL): PPL is an indicator of the word regeneration capability of the model. For a corpus of $D$ number of documents, PPL can be calculated as follows:
    
    \begin{equation}
        PPL = exp(- \frac{1}{D} \sum_{i=1}^{D} \frac{1}{N_i} \log p(\textbf{v}_i))
        \label{eq:ppl_2}
    \end{equation}
    \enote{PG}{what is z}
    
    where, $N_i$ is the total number of words in a document. As evident from eq. (~\ref{eq:ppl_2}) PPL depends on the negative log-likelihood of the documents. Therefore, low value of PPL indicates better regeneration capability of the model.
    
    \item \textit{Information retrieval} (IR): IR is calculated as the average precision of all the test query documents, where precision is defined as the fraction of documents, of all the retrieved documents based on a similarity metric, with the same label as the query document. Therefore, high value of IR indicates more descriptive document representations of the model.
    
    \item \textit{Topic coherence} (COH): Topic coherence is used to quantify the meaningfulness of the latent topics, of a dataset, generated by the topic models. We use target dataset itself as the reference corpus against which topic coherence is calculated using the underlying word co-occurrence statistics. 
\end{enumerate}

\section{Results}

We perform topic modeling on multiple sequences of five datasets in a \textit{lifelong learning} fashion to evaluate the performance gain we get on target datasets. The particular sequences of datasets we use are mentioned below (with $\rightarrow$ arrows denote a source dataset at its tail and a target dataset at its head):

\begin{itemize}
    \item \textbf{S1}: AGnews $\rightarrow$ TMN  $\rightarrow$  R21578  $\rightarrow$  20NS $\rightarrow$ 20NSshort  
    \item \textbf{S2}: AGnews $\rightarrow$ TMN  $\rightarrow$  R21578  $\rightarrow$  20NS $\rightarrow$ TMNtitle   
    \item \textbf{S3}: AGnews $\rightarrow$ TMN  $\rightarrow$  R21578  $\rightarrow$  20NS $\rightarrow$ R21578title
\end{itemize}

It can be noted that the first four datasets in these sequences are fixed and the final dataset is varying to keep the experimental setup simple. The sequences are designed in a way that the first four datasets are long text and final dataset is a short text dataset to, particularly, study the effect of \textit{lifelong learning} on short text datasets. However, we evaluate the \textit{catastrophic forgetting} on the intermediate long text datasets, i.e., source datasets, which is the central to the idea of \textit{lifelong learning}. Refer to Table ~\ref{table:lifelong_hyperparameters} in \textit{appendix} to know about the hyperparameter settings used in the experiments.



Table ~\ref{table:lifelong_20NSshort}, ~\ref{table:lifelong_TMNtitle} and ~\ref{table:lifelong_R21578title} shows the evaluation results of \textit{lifelong learning} on 20NSshort, TMNtitle and R21578title datasets respectively, under four different experimental settings mentioned before using four different source datasets listed in \textbf{S1}, \textbf{S2} and \textbf{S3} respectively. The findings from the experimental results are mentioned below: 

\begin{enumerate}
    \item Transfer of global contextual knowledge via pre-trained embeddings (EmbTF) from source datasets helps in improving the IR precision score (at 0.02 fraction) by 6.9\% (0.31 vs 0.29), 4.5\% (0.687 vs 0.657) and 5.2\% (0.548 vs 0.521), against DocNADE baseline, in 20NSshort, R21578title and TMNtitle target datasets respectively.
    
    \item As row vectors in $\textbf{W}$ matrix of DocNADE represents latent topic representations, an L2 constraint on projected $\textbf{W}$ matrix, in RK task, facilitates the gradual transfer of latent topic knowledge from source datasets into new model parameters, while minimizing the \textit{catastrophic forgetting} at the same time. This helps in an improvement in topic coherence (COH) score by 7.8\% (0.719 vs 0.667), 4.0\% (0.742 vs 0.713) and 4.8\% (0.743 vs 0.709), against DocNADE baseline, for 20NSshort, R21578title and TMNtitle target datasets respectively. 
    
    \begin{figure}[t]
    \center
    \includegraphics[width=0.85\textwidth]{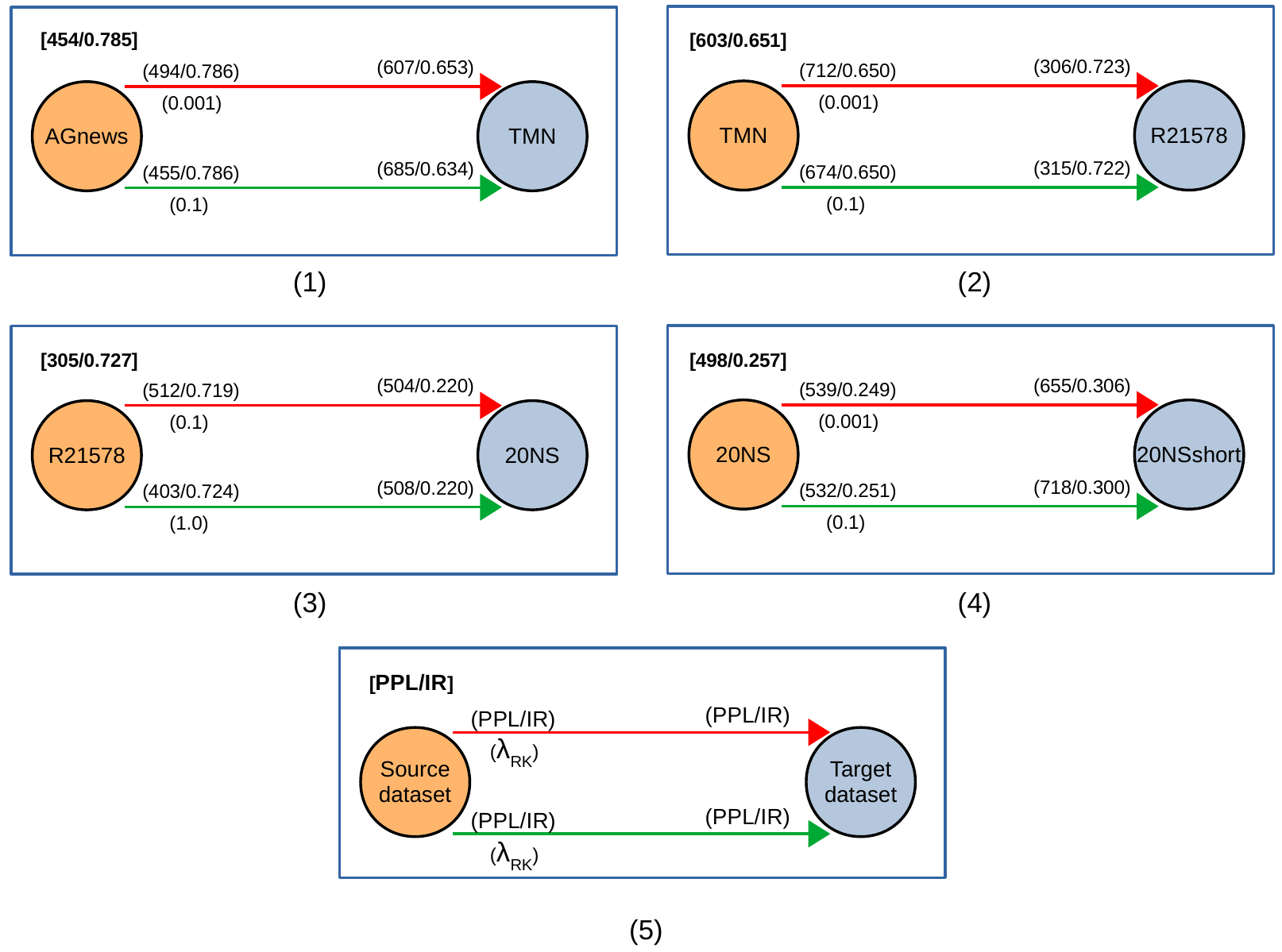}
    \caption{Catastrophic forgetting: sub-figure (5) indicates the format of the displayed results. Red arrow indicates the setting of \textit{high forgetting} (high deviation in source PPL and IR) and green arrow indicates the setting of \textit{high retention} (low deviation in source PPL and IR). Evaluation (PPL/IR) of source dataset (left) using $\theta_{new}$ is shown on the left of each arrow, alongwith its respective $\lambda_{RK}$ parameter, and evaluation of target dataset (right) using $\theta_{new}$ is shown on the right side of every arrow. For comparison, evaluation (PPL/IR) of every source dataset with its respective $\theta_{old}$ is given in \textbf{bold} above each source dataset.
    \enote{PG}{put what is source and target inthe figure. Above circle.}}
    \label{fig:lifelong_pairwise_forgetting}
    \end{figure}
    
    Also, the column vectors of $\textbf{W}$ matrix of DocNADE represent the latent word representations. Therefore, the L2 constraint also helps in transfer of global word context knowledge from source datasets into word representations of target dataset. This results in an improvement in IR precision score (at 0.02 fraction) by 5.5\% (0.306 vs 0.290) and 6.3\% (0.554 vs 0.521), against DocNADE baseline, for 20NSshort and TMNtitle target datasets respectively .
    
    The importance of the $\lambda_{RK}$ factor in controlling the retention of past learning is clearly illustrated in fig ~\ref{fig:lifelong_pairwise_forgetting}, where the sequence \textbf{S1} is split into four pairs of datasets in the order shown in \textbf{S1}. Here, the red arrow indicates the setting of high catastrophic forgetting or low retention of source dataset (left) and the green arrow indicates the setting of low catastrophic forgetting or high retention. The figure, clearly, shows that for high values of $\lambda_{RK}$, the L2 constraint is applied more strictly, thus leading to low performance on target datasets and vice versa.
    
    \begin{table}[h]
    \center
    \small
    \begin{tabular}{p{3.5cm}||r||c|c|c|c|c}
    \multirow{2}{*}{\textbf{Model}}  & \multirow{2}{*}{\textbf{Evaluation}} &  \textit{(Target dataset)} & \multicolumn{4}{c}{\textit{(Source datasets)}} \\
                        &     & \textbf{20NSshort}   & \textbf{20NS}        & \textbf{R21578}      & \textbf{TMN}         & \textbf{AGnews}          \\
    \hline
    \hline
    \multicolumn{7}{c}{\textit{Baseline model}}                                                                                                                                       \\
    \hline
    \multirow{3}{*}{\textit{DocNADE}} & PPL & 646                     & 470                     & 311                                            & 584                     & 454                   \\
                                      & COH & 0.667                     &  -                    & -          &  -                    &  -                  \\
                                      & IR  &  0.290                    & 0.268                     & 0.726          & 0.651                     & 0.785                   \\
    \hline
    \multicolumn{7}{c}{\textit{Proposed models}}                                                                                                                                       \\
    \hline
    \multirow{3}{*}{\parbox{3.5cm}{\textit{DocNADE + EmbTF}}}       & PPL & 647                     & -                     & -                                            & -                     & -                    \\
                                      & COH &  0.670                    & -                     & -          & -                     &  -                   \\
                                      & IR  & 0.310                     & -                     & -          & -                     & -                    \\
    \hline
    \multirow{3}{*}{\parbox{3.5cm}{\textit{DocNADE + RK}}}       & PPL & 655                     & 538                     & 382                                            & 690                     & 466                    \\
                                      & COH & 0.719                     &  -                    &  -         &   -                   & -                    \\
                                      & IR  & 0.306                     & 0.249                     & 0.724          & 0.649                     & 0.786                    \\
    \hline
    \multirow{3}{*}{\parbox{3.5cm}{\textit{DocNADE + EmbTF \\ + RK}}}       & PPL & 672                     & 541                     & 382                                            & 698                     & 469                    \\
                                      & COH & 0.728                     & -                     &  -         & -                     &  -                   \\
                                      & IR  & 0.297                     & 0.247                     & 0.724          & 0.649                     & 0.786                    \\
    \hline
    \multirow{3}{*}{\parbox{3.5cm}{\textit{DocNADE + EmbTF \\ + RK + SAL}}}       & PPL & \textbf{641}                     & 541                     & 380                                            & 696                     & 471                    \\
                                      & COH & \textbf{0.735}                     & -                     & -          & -                     & -                    \\
                                      & IR  & \textbf{0.324}                     & 0.248                     & 0.724          & 0.647                     & 0.786                    \\
    \hline
    \end{tabular}
    \caption{Lifelong learning results on \textit{20NSshort} dataset via four different long text datasets in the order shown in \textbf{S1}, best scores are shown in \textbf{bold} font.}
    \label{table:lifelong_20NSshort}
    \end{table}
    
    \item By combining the efforts of (1) global contextual knowledge transfer via EmbTF, (2) latent topical information transfer via RK and (3) implicit knowledge transfer via SAL tasks in (\textit{DocNADE+EmbTF+RK+SAL}), we achieve an even further improvement in topic coherence (COH) score by 10.0\% (0.735 vs 0.667), 4.8\% (0.747 vs 0.713) and 5.0\% (0.745 vs 0.709), against DocNADE baseline, for 20NSshort, R21578title and TMNtitle datasets respectively.
    
    Similarly, \textit{DocNADE+EmbTF+RK+SAL} setting of \textit{lifelong learning} achieves best IR precision scores on all the three target datasets. The improvement in IR precision scores at different fractions is demonstrated in figs ~\ref{fig:lifelong_IR20NSshort}, ~\ref{fig:lifelong_IRR21578title} and ~\ref{fig:lifelong_IRTMNtitle}. It can be noticed that \textit{DocNADE + EmbTF + RK + SAL} setting achieves the highest improvement in IR precision scores for all the three target datasets.
\end{enumerate}

\begin{figure}[h]
\includegraphics[width=0.90\textwidth]{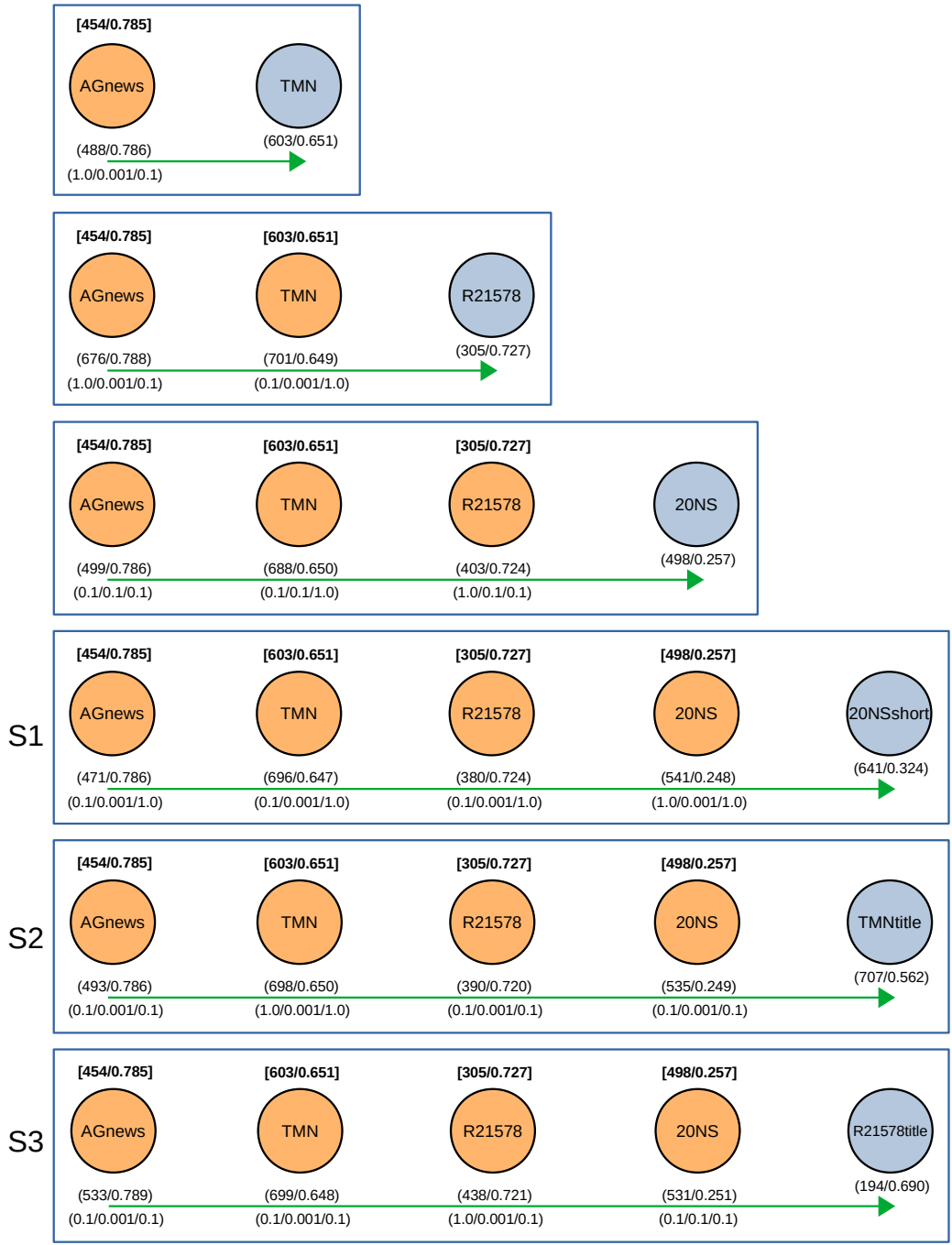}
\centering
\caption{Detailed step-by-step evaluation of target (blue color) and source (orange color) datasets in \textit{lifelong learning} on part and full sequences. Evaluation is shown in (PPL/IR) format, above the green arrow, of every dataset with the respective parameter setting in format ($\lambda_{EmbTF}$/$\lambda_{RK}$/$\lambda_{SAL}$), below the green arrow. At the tip of the arrow the evaluation of target dataset is shown in (PPL/IR) format.
\enote{PG}{menton the meaning of color in circle. source or target data}}
\label{fig:lifelong_results}
\end{figure}

To corroborate the high topic coherence scores (COH) on target datasets with \textit{lifelong learning} against DocNADE baseline model, table ~\ref{table:lifelong_topics} shows that the topics generated from models with \textit{lifelong learning} are more meaningful and coherent, which is supported by the topic coherence scores (COH) mentioned. However, to get an overall picture of \textit{lifelong learning} in DocNADE topic model, fig. ~\ref{fig:lifelong_results} demonstrates the step-by-step evaluations of the \textit{DocNADE+EmbTF+RK+SAL} model of all the three sequences \textbf{S1}, \textbf{S2} and \textbf{S3}.

\begin{table}[h]
\center
\small
\begin{tabular}{r||r|l}
\hline
\hline
\textbf{Model}  & \textbf{COH} & \textbf{Topics}  \\
\hline
\hline
\multirow{2}{*}{\textbf{DocNADE}}                                                         & 0.49                    & \textbf{T1}: \textit{nuclear, break, jobs, afghanistan, ipad}                \\
                                                                                 & 0.52                    & \textbf{T2}: \textit{gulf, bruins, japanese, michigan, radiation}            \\
\hline
\multirow{2}{*}{\textbf{DocNADE+RK}}                                                      & 0.52                    & \textbf{T1}: \textit{arts, android, iphone, tablet, ipad}                    \\
                                                                                 & 0.55                    & \textbf{T2}: \textit{rail, medicare, wildfire, radioactive, recession}       \\
\hline
\multirow{2}{*}{\textbf{\begin{tabular}[c]{@{}r@{}}DocNADE+EmbTF\\ +RK+SAL\end{tabular}}} & 0.53                    & \textbf{T1}: \textit{linkedin, android, tablet, ipad, iphone}                \\
                                                                                 & 0.55                    & \textbf{T2}: \textit{tornadoes, fukushima, radioactive, radiation, medicare} \\
\hline
\end{tabular}
\caption{Example topics from \textit{TMNtitle} dataset from 3 different experimental settings. It shows that training in \textit{lifelong fashion} increases the meaningfulness of the generated topics. \textbf{T1} is related to \textit{Apple} (company) and \textbf{T2} is related to \textit{Disaster}.
\enote{PG}{remove quotes. Put italics}}
\label{table:lifelong_topics}
\end{table}

\begin{table}[h]
\center
\small
\begin{tabular}{p{3.5cm}||r||c|c|c|c|c}
\multirow{2}{*}{\textbf{Model}}  & \multirow{2}{*}{\textbf{Evaluation}} &  \textit{(Target dataset)} & \multicolumn{4}{c}{\textit{(Source datasets)}} \\
                    &     & \textbf{R21578title}   & \textbf{20NS}        & \textbf{R21578}      & \textbf{TMN}         & \textbf{AGnews}          \\
\hline
\hline
\multicolumn{7}{c}{\textit{Baseline model}}                                                                                                                                       \\
\hline
\multirow{3}{*}{\textit{DocNADE}} & PPL & 192                     & 470                     & 311                                            & 584                     & 454                   \\
                                  & COH & 0.713                     &  -                    & -          &  -                    &  -                  \\
                                  & IR  &  0.657                    & 0.268                     & 0.726          & 0.651                     & 0.785                   \\
\hline
\multicolumn{7}{c}{\textit{Proposed models}}                                                                                                                                       \\
\hline
\multirow{3}{*}{\parbox{3.5cm}{\textit{DocNADE + EmbTF}}}       & PPL & \textbf{183}                     & -                     & -                                            & -                     & -                    \\
                                  & COH &  0.709                    & -                     & -          & -                     &  -                   \\
                                  & IR  & 0.678                     & -                     & -          & -                     & -                    \\
\hline
\multirow{3}{*}{\parbox{3.5cm}{\textit{DocNADE + RK}}}       & PPL & 208                     & 532                     & 451                                            & 704                     & 571                    \\
                                  & COH & 0.742                     &  -                    &  -         &   -                   & -                    \\
                                  & IR  & 0.668                     & 0.251                     & 0.722          & 0.649                     & 0.787                    \\
\hline
\multirow{3}{*}{\parbox{3.5cm}{\textit{DocNADE + EmbTF \\ + RK}}}       & PPL & 203                     & 532                     & 444                                            & 703                     & 550                    \\
                                  & COH & \textbf{0.752}                     & -                     &  -         & -                     &  -                   \\
                                  & IR  & 0.676                     & 0.251                     & 0.721          & 0.650                     & 0.788                    \\
\hline
\multirow{3}{*}{\parbox{3.5cm}{\textit{DocNADE + EmbTF \\ + RK + SAL}}}       & PPL & 194                     & 531                     & 438                                            & 699                     & 533                    \\
                                  & COH &     0.747                 & -                     & -          & -                     & -                    \\
                                  & IR  & \textbf{0.690}                     & 0.251                     & 0.721          & 0.648                     & 0.789                    \\
\hline
\end{tabular}
\caption{Lifelong learning results on \textit{R21578title} dataset via four different long text datasets in the order shown in \textbf{S2}, best scores are shown in \textbf{bold} font.}
\label{table:lifelong_R21578title}
\end{table}


\begin{table}[h]
\small
\center
\begin{tabular}{p{3.5cm}||r||c|c|c|c|c}
\multirow{2}{*}{\textbf{Model}}  & \multirow{2}{*}{\textbf{Evaluation}} &  \textit{(Target dataset)} & \multicolumn{4}{c}{\textit{(Source datasets)}} \\
                    &     & \textbf{TMNtitle}   & \textbf{20NS}        & \textbf{R21578}      & \textbf{TMN}         & \textbf{AGnews}          \\
\hline
\hline
\multicolumn{7}{c}{\textit{Baseline model}}                                                                                                                                       \\
\hline
\multirow{3}{*}{\textit{DocNADE}} & PPL & 706                     & 470                     & 311                                            & 584                     & 454                   \\
                                  & COH & 0.709                     &  -                    & -          &  -                    &  -                  \\
                                  & IR  &  0.521                    & 0.268                     & 0.726          & 0.651                     & 0.785                   \\
\hline
\multicolumn{7}{c}{\textit{Proposed models}}                                                                                                                                       \\
\hline
\multirow{3}{*}{\parbox{3.5cm}{\textit{DocNADE + EmbTF}}}       & PPL & \textbf{666}                     & -                     & -                                            & -                     & -                    \\
                                  & COH &  0.726                    & -                     & -          & -                     &  -                   \\
                                  & IR  & 0.548                     & -                     & -          & -                     & -                    \\
\hline
\multirow{3}{*}{\parbox{3.5cm}{\textit{DocNADE + RK}}}       & PPL & 736                     & 534                     & 389                                            & 700                     & 483                    \\
                                  & COH & 0.743                     &  -                    &  -         &   -                   & -                    \\
                                  & IR  & 0.554                     & 0.249                     & 0.725          & 0.650                     & 0.784                    \\
\hline
\multirow{3}{*}{\parbox{3.5cm}{\textit{DocNADE + EmbTF \\ + RK}}}       & PPL & 723                     & 534                     & 390                                            & 701                     & 484                    \\
                                  & COH & \textbf{0.750}                     & -                     &  -         & -                     &  -                   \\
                                  & IR  & 0.556                     & 0.249                     & 0.72          & 0.650                     & 0.786                    \\
\hline
\multirow{3}{*}{\parbox{3.5cm}{\textit{DocNADE + EmbTF \\ + RK + SAL}}}       & PPL & 707                     & 535                     & 390                                            & 698                     & 493                    \\
                                  & COH &   0.745                   & -                     & -          & -                     & -                    \\
                                  & IR  & \textbf{0.562}                     & 0.249                     & 0.720          & 0.650                     & 0.786                    \\
\hline
\end{tabular}
\caption{Lifelong learning results on \textit{TMNtitle} dataset via four different long text datasets in the order shown in \textbf{S3}, best scores are shown in \textbf{bold} font.}
\label{table:lifelong_TMNtitle}
\end{table}

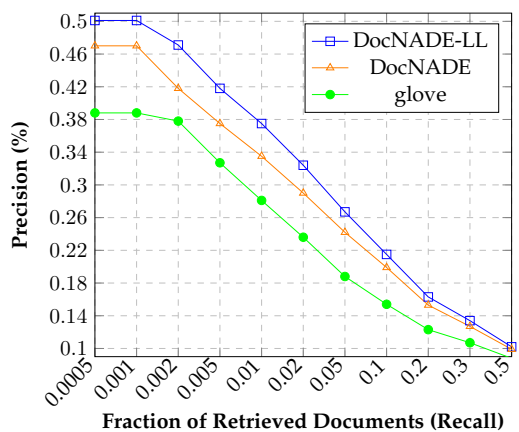
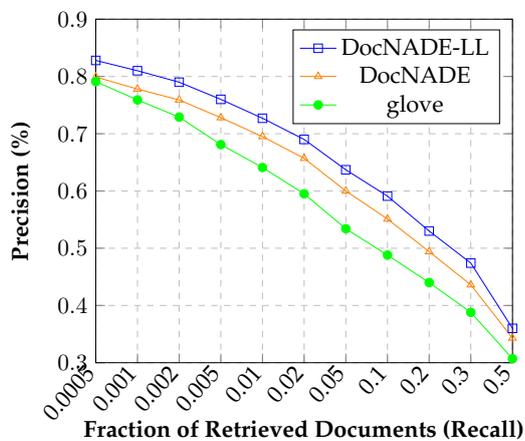
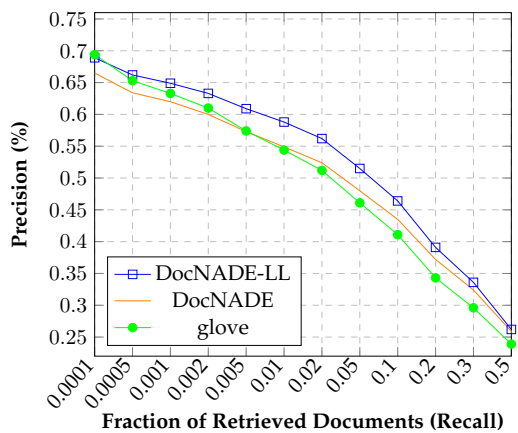
\begin{figure*}[t]
\centering

\begin{subfigure}{0.48\textwidth}
\centering
\begin{tikzpicture}[scale=0.8][baseline]
\small
\begin{axis}[
    x label style={at={(axis description cs:0.5,-0.05)},anchor=north},
    xlabel={\textbf{Fraction of Retrieved Documents (Recall)}},
    ylabel={\textbf{Precision (\%)}},
    xmin=0, xmax=10,
    ymin=0.09, ymax=0.51,
   /pgfplots/ytick={.10,.14,...,.51},
    xtick={0,1,2,3,4,5,6,7,8,9, 10,11,12},
    xticklabels={0.0005, 0.001, 0.002, 0.005, 0.01, 0.02, 0.05, 0.1, 0.2, 0.3, 0.5},
    x tick label style={rotate=45,anchor=east},
    legend pos=north east,
    ymajorgrids=true, xmajorgrids=true,
    grid style=dashed,
]

\addplot[
	color=blue,
	mark=square,
	]
	plot coordinates {
    (0, 0.501)
    (1, 0.501)
    (2, 0.471)
    (3, 0.418)
    (4, 0.375)
    (5, 0.324)
    (6, 0.267)
    (7, 0.215)
    (8, 0.163)
    (9, 0.134)
    (10, 0.102)
    (11, 0.077)
    (12, 0.066)
	};
\addlegendentry{DocNADE-LL}

\addplot[
	color=orange,
	mark=triangle,
	]
	plot coordinates {
    (0, 0.470)
    (1, 0.470)
    (2, 0.418)
    (3, 0.375)
    (4, 0.335)
    (5, 0.290)
    (6, 0.242)
    (7, 0.199)
    (8, 0.153)
    (9, 0.127)
    (10, 0.099)
    (11, 0.076)
    (12, 0.066)
    
	};
\addlegendentry{DocNADE}

\addplot[
	color=green,
	mark=*,
	]
	plot coordinates {
    (0, 0.388)
    (1, 0.388)
    (2, 0.378)
    (3, 0.327)
    (4, 0.281)
    (5, 0.236)
    (6, 0.188)
    (7, 0.154)
    (8, 0.123)
    (9, 0.107)
    (10, 0.088)
    (11, 0.072)
    (12, 0.066)
	};
\addlegendentry{glove}

\end{axis}
\end{tikzpicture}%
\caption{{\textbf{IR:}} 20NSshort} \label{fig:lifelong_IR20NSshort}
\end{subfigure}
\\
\begin{subfigure}{0.48\textwidth}
\centering
\begin{tikzpicture}[scale=0.8][baseline]
\begin{axis}[
    x label style={at={(axis description cs:0.5,-0.05)},anchor=north},
    xlabel={\textbf{Fraction of Retrieved Documents (Recall)}},
    ylabel={\textbf{Precision (\%)}},
    xmin=0, xmax=10,
    ymin=0.30, ymax=0.90,
    /pgfplots/ytick={.30,.40,...,.90},
    xtick={0,1,2,3,4,5,6,7,8, 9,10,11,12},
    xticklabels={0.0005, 0.001, 0.002, 0.005, 0.01, 0.02, 0.05, 0.1, 0.2, 0.3, 0.5},
    x tick label style={rotate=48,anchor=east},
    legend pos=north east,
    ymajorgrids=true, xmajorgrids=true,
    grid style=dashed,
]

\addplot[
	color=blue,
	mark=square,
	]
	plot coordinates {
    (0, 0.828)
    (1, 0.810)
    (2, 0.790)
    (3, 0.760)
    (4, 0.727)
    (5, 0.690)
    (6, 0.637)
    (7, 0.591)
    (8, 0.530)
    (9, 0.474)
    (10, 0.360)
    (11, 0.246)
    (12, 0.199)
	};
\addlegendentry{DocNADE-LL}

\addplot[
	color=orange,
	mark=triangle,
	]
	plot coordinates {
    (0, 0.799)
    (1, 0.778)
    (2, 0.759)
    (3, 0.728)
    (4, 0.695)
    (5, 0.657)
    (6, 0.600)
    (7, 0.551)
    (8, 0.494)
    (9, 0.436)
    (10, 0.343)
    (11, 0.244)
    (12, 0.199)
	};
\addlegendentry{DocNADE}

\addplot[
	color=green,
	mark=*,
	]
	plot coordinates {
    (0, 0.791)
    (1, 0.759)
    (2, 0.729)
    (3, 0.681)
    (4, 0.641)
    (5, 0.595)
    (6, 0.534)
    (7, 0.488)
    (8, 0.440)
    (9, 0.388)
    (10, 0.307)
    (11, 0.233)
    (12, 0.199)
	};
\addlegendentry{glove}


\end{axis}
\end{tikzpicture}%
\caption{{\textbf{IR:}} R21578title} \label{fig:lifelong_IRR21578title}
\end{subfigure}
\\
\begin{subfigure}{0.48\textwidth}
\centering
\begin{tikzpicture}[scale=0.8][baseline]
\small
\begin{axis}[
    x label style={at={(axis description cs:0.5,-0.05)},anchor=north},
    xlabel={\textbf{Fraction of Retrieved Documents  (Recall)}},
    ylabel={\textbf{Precision (\%)}},
    xmin=0, xmax=11,
    ymin=0.22, ymax=0.76,
    /pgfplots/ytick={.25,.30,...,.76},
    xtick={0,1,2,3,4,5,6,7,8,9, 10,11,12,13},
    xticklabels={0.0001, 0.0005, 0.001, 0.002, 0.005, 0.01, 0.02, 0.05, 0.1, 0.2, 0.3, 0.5},
    x tick label style={rotate=48,anchor=east},
    legend pos=south west,
    ymajorgrids=true, xmajorgrids=true,
    grid style=dashed,
]

\addplot[
	color=blue,
	mark=square,
	]
	plot coordinates {
    (0, 0.689)
    (1, 0.662)
    (2, 0.649)
    (3, 0.633)
    (4, 0.609)
    (5, 0.588)
    (6, 0.562)
    (7, 0.515)
    (8, 0.464)
    (9, 0.391)
    (10, 0.336)
    (11, 0.262)
    (12, 0.198)
    (13, 0.169)
	};
\addlegendentry{DocNADE-LL}

\addplot[
	color=orange,
	mark=traingle,
	]
	plot coordinates {
    (0, 0.665)
    (1, 0.634)
    (2, 0.620)
    (3, 0.600)
    (4, 0.573)
    (5, 0.549)
    (6, 0.524)
    (7, 0.480)
    (8, 0.435)
    (9, 0.372)
    (10, 0.324)
    (11, 0.259)
    (12, 0.198)
    (13, 0.169)
	};
\addlegendentry{DocNADE}

\addplot[
	color=green,
	mark=*,
	]
	plot coordinates {
    (0, 0.694)
    (1, 0.653)
    (2, 0.633)
    (3, 0.610)
    (4, 0.574)
    (5, 0.544)
    (6, 0.512)
    (7, 0.461)
    (8, 0.411)
    (9, 0.343)
    (10, 0.296)
    (11, 0.239)
    (12, 0.190)
    (13, 0.169)
	};
\addlegendentry{glove}

\end{axis}
\end{tikzpicture}%
\caption{{\textbf{IR:}} TMNtitle} \label{fig:lifelong_IRTMNtitle}
\end{subfigure}

\caption{Retrieval performance (IR-precision) on 
3 target datasets at different fractions. DocNADE-LL indicates the \textit{DocNADE + EmbTF + RK + SAL} setting of \textit{lifelong learning}. 
\enote{PG}{rearrange. 1 plot in one row.i.e., 3 plot one after another in new lines}}
\label{fig:lifelong_docretrieval}
\end{figure*}

\chapter{Conclusion}
\label{chapter:conclusion}

In this thesis work, we followed two different motivations for topic models. First motivation is to include language structure information into topic models to model documents with \textit{fine granularity} alongwith the \textit{coarse granularity} from topic models. Second motivation is the inclusion of \textit{lifelong learning} process in topic models to learn the new information from currently available data while retaining the past learning.


In the first motivation, we discussed the two shortcomings of topic models, namely, absence of word order information due to \textit{bag-of-words} approach and poor modeling of short text documents due to limited word co-occurrences. In chapter ~\ref{chapter:ctx_docnade}, we investigated a joint topic and language model called ctx-DocNADE on a variety of short and long text datasets to study the effect to inclusion of langauge structure information into the DocNADE topic model. Topic modeling of short text documents, generally, results in incoherent latent topic representations due to limited word co-occurrences. Therefore, a special focus is given to the topic modeling of the short text documents or small corpus of documents with the inclusion of pre-trained distributional word embeddings, trained on a large dataset like Wikipedia, to provide additional semantic information apart from the syntactic information provided by the LSTM-LM language model in ctx-DocNADE. We called this model ctx-DocNADEe. Chapter ~\ref{chapter:results_ctx_docnade} described that the language structure information of documents is important during topic modeling process as it provides details of local contextual information in \textit{fine granularity}. It is observed that the controlled inclusion of language structure information into the DocNADE topic model results in better topic coherence (COH), information retrieval (IR) and classification (F1) scores for the long and short text datasets. However, the short text datasets benefited much more from this new information because of the limited capacity of DocNADE, to model short text, resulting from the limited word co-occurrences.


In the second motivation, in chapter ~\ref{chapter:lifelong_learning}, we discussed the importance of \textit{lifelong learning} in topic modeling and its two key aspects. First, the controlled transfer of previous knowledge (EmbTF \& SAL) to efficiently model the target dataset. Second, the retention of previous knowledge (RK) while modeling the target dataset. We investigated four different modeling approaches to understand the effect of knowledge transfer and knowledge retention. We also investigated the interplay of forces between modeling of target dataset and retention of previous knowledge. In chapter ~\ref{chapter:results_lifelong_learning}, we found that there is an inverse correlation between the two. Hence, a strong retention of previous knowledge results in the poor modeling of target dataset and vice versa. It is also discovered that the retention of previous knowledge results in the improvement of latent topic and word representations of the target dataset. Therefore, it can be said that the large source dataset of long text documents would most likely result in learning better latent topic and word representations on a target dataset in a \textit{lifelong learning} topic model.


In conclusion, we can say that (1) the language structure information helps in learning better latent topic and word representations by providing local syntactic and semantic knowledge in \textit{fine granularity} to complement the global semantic knowledge
of topic models; (2) the introduction of semantic information, in word representations, via distributional word embeddings from a large corpus of documents like Wikipedia, results in even better performance for short text datasets by filling the gap of semantic understanding of words arising from limited word co-occurrences; and (3) the topic modeling in a \textit{lifelong learning} fashion results in learning better latent topic and word representations using large datasets of long text documents as the source of the transfer of more semantically informative word representations and restriction to source datasets' more coherent topic and word representations. Our results for \textit{contextualized topic model} (ctx-DocNADE, ctx-DocNADEe) have been accepted at the \textit{Seventh International Conference on Learning Representations} (ICLR-2019), New Orleans.

%
%
\part*{Appendix}
\addcontentsline{toc}{part}{Appendix}

\appendix 

\chapter{Experimental Setup and Hyperparameters}
\label{chapter:ExperimentalSetup}\label{appendix}
This appendix contains examples topics from Siemens SiROBs dataset and search space for different hyperparameters used in DocNADE, ctx-DocNADE and \textit{lifelong learning} topic model. This appendix also contains the tables of perplexity (PPL) scores and information retrieval (IR) scores on the validation sets of different datasets for selection of the optimal hyperparameter values for ctx-DocNADE.

\begin{table*}[t]
\centering
\resizebox{0.99\textwidth}{!}{
\begin{tabular}{|c|}
\hline 
{\textbf{Label}:}   {\textit{maintenance}}    \\ \hline
 The Contractor shall provide experienced staff for 24 hours per Day, 
7 Days per week, throughout the Year, \\
for call out to carry out On-call  Maintenance for the Signalling System.  \\ \hline \hline

{\textbf{Label}:}   {\textit{operations, interlocking}} \\ \hline
It shall be possible to switch any station Interlocking capable of reversing 
the service into  \\
``Auto-Turnaround Operation". This facility once selected shall automatically 
route Trains into and out  of \\
these stations, independently of the ATS system. At stations where multiple
 platforms can be used  to reverse\\
 the service it shall be possible to select one or both platforms for the service reversal. \\ \hline \hline

{\textbf{Label}:} {\textit{training}}              \\ \hline
Instructors shall have tertiary education and experience in the 
operation and maintenance \\
of the equipment or sub-system of Plant. They shall be proficient 
in the use of the English language both written \\
and oral. They shall be able to deliver instructions clearly and
 systematically. The curriculum vitae \\
of the instructors shall be submitted for acceptance by the 
Engineer at least 8 weeks before \\
the commencement of any training.  \\ \hline \hline
 
 {\textbf{Label}:}   {\textit{cables}} \\ \hline
Unless otherwise specified, this standard is applicable to all cables 
which include single and multi-core cables \\ 
and wires, Local Area Network  (LAN) cables and Fibre Optic (FO) cables.  \\ \hline \hline
 
 {\textbf{Label}:}   {\textit{installation}} \\ \hline
The Contractor shall provide and permanently install the asset labels onto 
all equipment supplied \\
under this Contract. The Contractor shall liaise and  co-ordinate with 
the Engineer for the format \\
and the content of the labels. The Contractor shall submit the final format
 and size of the labels as well \\ 
as the installation layout of the labels on the respective equipment, to 
the Engineer for acceptance.\\ \hline \hline 

\end{tabular}}
\caption{SiROBs data: Example Documents (Requirement Objects) with their types (label).}
\label{table:SiROBs_examples}
\end{table*}

\begin{table}[t]
      \centering
        \begin{tabular}{c|c}
         \hline 
         {\textbf{Hyperparameter}}               & {\textbf{Search Space}} \\ \hline
           learning rate        &    [0.001]                   \\
           hidden units        &       [200]                      \\
           iterations        &      [2000]      \\
           activation function      &    [sigmoid, tanh]            \\ 
          $\lambda$        &      [1.0, 0.8, 0.5, 0.3, 0.1, 0.01, 0.001]  \\ \hline      
       \end{tabular}
         \caption{Different values of hyperparameter settings in the DocNADE and ctx-DocNADE variants for 200 topics; \textit{sigmoid} is used as activation function in case of PPL/COH, while \textit{tanh} is used as the activation function in case of IR}
         \label{table:ctx_docnade_hyperparameters}
 \end{table}%

\begin{figure*}[t]
\centering

\begin{subfigure}{0.48\textwidth}
\centering
\begin{tikzpicture}[scale=0.8][baseline]
\begin{axis}[
    x label style={at={(axis description cs:0.5,-0.05)},anchor=north},
    xlabel={\textbf{Fraction of Retrieved Documents (Recall)}},
    ylabel={\textbf{Precision (\%)}},
    xmin=0, xmax=13,
    ymin=0.20, ymax=0.87,
    /pgfplots/ytick={.20,.30,...,.87},
    xtick={0,1,2,3,4,5,6,7,8,9, 10,11,12,13},
    xticklabels={0.0001, 0.0005, 0.001, 0.002, 0.005, 0.01, 0.02, 0.05, 0.1, 0.2, 0.3, 0.5, 0.8, 1.0},
    x tick label style={rotate=48,anchor=east},
    legend pos=south west,
    ymajorgrids=true, xmajorgrids=true,
    grid style=dashed,
]

\addplot[
	color=orange,
	mark=*,
	]
	plot coordinates {
    (0, 0.00)
    (1, 0.00)
    (2, 0.00)
    (3, 0.00)
    (4, 0.00)
    (5, 0.00)
    (6, 0.00)
    (7, 0.00)
    (8, 0.00)
    (9, 0.00)
	};
\addlegendentry{DeepDNE}

\addplot[
	color=blue,
	mark=square,
	]
	plot coordinates {
    (0, 0.856)
    (1, 0.791)
    (2, 0.766)
    (3, 0.754)
    (4, 0.694)
    (5, 0.646)
    (6, 0.599)
    (7, 0.504)
    (8, 0.418)
    (9, 0.343)
    (10, 0.300)
    (11, 0.251)
    (12, 0.211)
    (13, 0.193)
	};
\addlegendentry{ctx-DocNADEe}

\addplot[
	color=red,
	mark=square,
	]
	plot coordinates {
    (0, 0.858)
    (1, 0.785)
    (2, 0.759)
    (3, 0.732)
    (4, 0.688)
    (5, 0.647)
    (6, 0.595)
    (7, 0.507)
    (8, 0.426)
    (9, 0.342)
    (10, 0.299)
    (11, 0.250)
    (12, 0.211)
    (13, 0.193)
	};
\addlegendentry{ctx-DocNADE}

\addplot[
	color=violet,
	mark=square,
	]
	plot coordinates {
    (0, 0.00)
    (1, 0.00)
    (2, 0.00)
    (3, 0.00)
    (4, 0.00)
    (5, 0.00)
    (6, 0.00)
    (7, 0.00)
    (8, 0.00)
    (9, 0.00)
	};
\addlegendentry{DocNADE-LM}

\addplot[
	color=cyan,
	mark=triangle,
	]
	plot coordinates {
    (0, 0.792)
    (1, 0.722)
    (2, 0.722)
    (3, 0.703)
    (4, 0.640)
    (5, 0.589)
    (6, 0.546)
    (7, 0.460)
    (8, 0.390)
    (9, 0.326)
    (10, 0.288)
    (11, 0.245)
    (12, 0.215)
    (13, 0.193)
	};
\addlegendentry{DocNADE(FV)}

\addplot[
	color=black,
	mark=*,
	]
	plot coordinates {
    (0, 0.79)
    (1, 0.776)
    (2, 0.738)
    (3, 0.710)
    (4, 0.641)
    (5, 0.591)
    (6, 0.550)
    (7, 0.468)
    (8, 0.398)
    (9, 0.327)
    (10, 0.287)
    (11, 0.244)
    (12, 0.211)
    (13, 0.193)
	};
\addlegendentry{DocNADE(RV)}

\addplot[
	color=green,
	mark=*,
	]
	plot coordinates {
    (0, 0.712)
    (1, 0.667)
    (2, 0.670)
    (3, 0.628)
    (4, 0.564)
    (5, 0.509)
    (6, 0.479)
    (7, 0.504)
    (8, 0.358)
    (9, 0.308)
    (10, 0.281)
    (11, 0.246)
    (12, 0.211)
    (13, 0.193)
	};
\addlegendentry{glove}

\end{axis}
\end{tikzpicture}%
\caption{{\textbf{IR:}} TREC6} \label{IRTREC6}
\end{subfigure}\hspace*{\fill}
\begin{subfigure}{0.48\textwidth}
\centering
\begin{tikzpicture}[scale=0.8][baseline]
\begin{axis}[
    x label style={at={(axis description cs:0.5,-0.05)},anchor=north},
    xlabel={\textbf{Fraction of Retrieved Documents (Recall)}},
    xmin=0, xmax=12,
    ymin=0.11, ymax=0.80,
    /pgfplots/ytick={.20,.30,...,.80},
    xtick={0,1,2,3,4,5,6,7,8, 9,10,11,12},
    xticklabels={0.0005, 0.001, 0.002, 0.005, 0.01, 0.02, 0.05, 0.1, 0.2, 0.3, 0.5, 0.8, 1.0},
    x tick label style={rotate=48,anchor=east},
    legend pos=north east,
    ymajorgrids=true, xmajorgrids=true,
    grid style=dashed,
]

\addplot[
	color=orange,
	mark=*,
	]
	plot coordinates {
    (0, 0.00)
    (1, 0.00)
    (2, 0.00)
    (3, 0.00)
    (4, 0.00)
    (5, 0.00)
    (6, 0.00)
    (7, 0.00)
    (8, 0.00)
    (9, 0.00)
	};
\addlegendentry{DeepDNE}

\addplot[
	color=blue,
	mark=square,
	]
	plot coordinates {
    (0, 0.786)
    (1, 0.766)
    (2, 0.747)
    (3, 0.719)
    (4, 0.691)
    (5, 0.656)
    (6, 0.606)
    (7, 0.562)
    (8, 0.506)
    (9, 0.451)
    (10, 0.351)
    (11, 0.245)
    (12, 0.199)
	};
\addlegendentry{ctx-DocNADEe}

\addplot[
	color=red,
	mark=square,
	]
	plot coordinates {
    (0, 0.773)
    (1, 0.751)
    (2, 0.733)
    (3, 0.704)
    (4, 0.675)
    (5, 0.641)
    (6, 0.592)
    (7, 0.550)
    (8, 0.496)
    (9, 0.441)
    (10, 0.346)
    (11, 0.245)
    (12, 0.199)
	};
\addlegendentry{ctx-DocNADE}

\addplot[
	color=violet,
	mark=square,
	]
	plot coordinates {
    (0, 0.00)
    (1, 0.00)
    (2, 0.00)
    (3, 0.00)
    (4, 0.00)
    (5, 0.00)
    (6, 0.00)
    (7, 0.00)
    (8, 0.00)
    (9, 0.00)
	};
\addlegendentry{DocNADE-LM}

\addplot[
	color=cyan,
	mark=triangle,
	]
	plot coordinates {
    (0, 0.789)
    (1, 0.770)
    (2, 0.752)
    (3, 0.721)
    (4, 0.690)
    (5, 0.654)
    (6, 0.600)
    (7, 0.554)
    (8, 0.498)
    (9, 0.441)
    (10, 0.346)
    (11, 0.244)
    (12, 0.199)
	};
\addlegendentry{DocNADE(FV)}

\addplot[
	color=black,
	mark=*,
	]
	plot coordinates {
    (0, 0.00)
    (1, 0.00)
    (2, 0.00)
    (3, 0.00)
    (4, 0.00)
    (5, 0.00)
    (6, 0.00)
    (7, 0.00)
    (8, 0.00)
    (9, 0.00)
	};
\addlegendentry{DocNADE(RV)}

\addplot[
	color=green,
	mark=*,
	]
	plot coordinates {
    (0, 0.791)
    (1, 0.759)
    (2, 0.729)
    (3, 0.681)
    (4, 0.641)
    (5, 0.595)
    (6, 0.534)
    (7, 0.488)
    (8, 0.440)
    (9, 0.388)
    (10, 0.307)
    (11, 0.233)
    (12, 0.199)
	};
\addlegendentry{glove}


\end{axis}
\end{tikzpicture}%
\caption{{\textbf{IR:}} R21578title} \label{IRR21578title}
\end{subfigure}
~
\begin{subfigure}{0.48\textwidth}
\centering
\begin{tikzpicture}[scale=0.8][baseline]
\begin{axis}[
    x label style={at={(axis description cs:0.5,-0.05)},anchor=north},
    xlabel={\textbf{Fraction of Retrieved Documents (Recall)}},
    ylabel={\textbf{Precision (\%)}},
    xmin=0, xmax=13,
    ymin=0.15, ymax=0.85,
   /pgfplots/ytick={.15,.35,...,.85},
    xtick={0,1,2,3,4,5,6,7,8,9, 10,11,12,13},
    xticklabels={0.0001, 0.0005, 0.001, 0.002, 0.005, 0.01, 0.02, 0.05, 0.1, 0.2, 0.3, 0.5, 0.8, 1.0},
    x tick label style={rotate=45,anchor=east},
    legend pos=south west,
    ymajorgrids=true, xmajorgrids=true,
    grid style=dashed,
]

\addplot[
	color=blue,
	mark=square,
	]
	plot coordinates {
    (0, 0.822)
    (1, 0.792)
    (2, 0.780)
    (3, 0.766)
    (4, 0.745)
    (5, 0.724)
    (6, 0.698)
    (7, 0.645)
    (8, 0.581)
    (9, 0.480)
    (10, 0.399)
    (11, 0.291)
    (12, 0.204)
    (13, 0.169)
	};
\addlegendentry{ctx-DocNADEe}

\addplot[
	color=red,
	mark=square,
	]
	plot coordinates {
    (0, 0.807)
    (1, 0.780)
    (2, 0.769)
    (3, 0.757)
    (4, 0.736)
    (5, 0.717)
    (6, 0.692)
    (7, 0.641)
    (8, 0.578)
    (9, 0.479)
    (10, 0.397)
    (11, 0.289)
    (12, 0.204)
    (13, 0.169)
	};
\addlegendentry{ctx-DocNADE}

\addplot[
	color=violet,
	mark=square,
	]
	plot coordinates {
    (0, 0.00)
    (1, 0.00)
    (2, 0.00)
    (3, 0.00)
    (4, 0.00)
    (5, 0.00)
    (6, 0.00)
    (7, 0.00)
    (8, 0.00)
    (9, 0.00)
	};
\addlegendentry{DocNADE-LM}

\addplot[
	color=cyan,
	mark=triangle,
	]
	plot coordinates {
    (0, 0.812)
    (1, 0.788)
    (2, 0.774)
    (3, 0.760)
    (4, 0.737)
    (5, 0.715)
    (6, 0.687)
    (7, 0.632)
    (8, 0.571)
    (9, 0.476)
    (10, 0.397)
    (11, 0.291)
    (12, 0.204)
    (13, 0.169)
    
	};
\addlegendentry{DocNADE(FV)}

\addplot[
	color=black,
	mark=*,
	]
	plot coordinates {
    (0, 0.775)
    (1, 0.750)
    (2, 0.737)
    (3, 0.724)
    (4, 0.702)
    (5, 0.680)
    (6, 0.653)
    (7, 0.598)
    (8, 0.539)
    (9, 0.452)
    (10, 0.382)
    (11, 0.285)
    (12, 0.203)
    (13, 0.169)
	};
\addlegendentry{DocNADE(RV)}

\addplot[
	color=green,
	mark=*,
	]
	plot coordinates {
    (0, 0.811)
    (1, 0.774)
    (2, 0.755)
    (3, 0.735)
    (4, 0.705)
    (5, 0.676)
    (6, 0.642)
    (7, 0.582)
    (8, 0.519)
    (9, 0.430)
    (10, 0.362)
    (11, 0.272)
    (12, 0.199)
    (13, 0.169)
	};
\addlegendentry{glove}

\end{axis}
\end{tikzpicture}%
\caption{{\textbf{IR:}} TMN} \label{IRTMN}
\end{subfigure}\hspace*{\fill}
\begin{subfigure}{0.48\textwidth}
\centering
\begin{tikzpicture}[scale=0.8][baseline]
\begin{axis}[
    x label style={at={(axis description cs:0.5,-0.05)},anchor=north},
    xlabel={\textbf{Fraction of Retrieved Documents (Recall)}},
    xmin=0, xmax=10,
    ymin=0.05, ymax=0.50,
    /pgfplots/ytick={.05,.16,...,.50},
    xtick={0,1,2,3,4,5,6,7,8, 9,10},
    xticklabels={0.002, 0.005, 0.01, 0.02, 0.05, 0.1, 0.2, 0.3, 0.5, 0.8, 1.0},
    x tick label style={rotate=48,anchor=east},
    legend pos=north east,
    ymajorgrids=true, xmajorgrids=true,
    grid style=dashed,
]
\addplot[
	color=blue,
	mark=square,
	]
	plot coordinates {
    (0, 0.44)
    (1, 0.432)
    (2, 0.378)
    (3, 0.327)
    (4, 0.237)
    (5, 0.185)
    (6, 0.136)
    (7, 0.109)
    (8, 0.082)
    (9, 0.059)
    (10, 0.050)
	};
\addlegendentry{ctx-DocNADEe}

\addplot[
	color=red,
	mark=square,
	]
	plot coordinates {
    (0, 0.445)
    (1, 0.42)
    (2, 0.358)
    (3, 0.313)
    (4, 0.226)
    (5, 0.176)
    (6, 0.129)
    (7, 0.107)
    (8, 0.081)
    (9, 0.059)
    (10, 0.050)
	};
\addlegendentry{ctx-DocNADE}

\addplot[
	color=violet,
	mark=square,
	]
	plot coordinates {
    (0, 0.00)
    (1, 0.00)
    (2, 0.00)
    (3, 0.00)
    (4, 0.00)
    (5, 0.00)
    (6, 0.00)
    (7, 0.00)
    (8, 0.00)
    (9, 0.00)
	};
\addlegendentry{DocNADE-LM}

\addplot[
	color=cyan,
	mark=triangle,
	]
	plot coordinates {
    (0, 0.415)
    (1, 0.397)
    (2, 0.348)
    (3, 0.299)
    (4, 0.222)
    (5, 0.169)
    (6, 0.125)
    (7, 0.105)
    (8, 0.079)
    (9, 0.059)
    (10, 0.050)
	};
\addlegendentry{DocNADE(FV)}

\addplot[
	color=black,
	mark=*,
	]
	plot coordinates {
    (0, 0.405)
    (1, 0.367)
    (2, 0.303)
    (3, 0.270)
    (4, 0.209)
    (5, 0.165)
    (6, 0.124)
    (7, 0.103)
    (8, 0.079)
    (9, 0.059)
    (10, 0.050)
	};
\addlegendentry{DocNADE(RV)}

\addplot[
	color=green,
	mark=*,
	]
	plot coordinates {
    (0, 0.340)
    (1, 0.330)
    (2, 0.277)
    (3, 0.238)
    (4, 0.181)
    (5, 0.143)
    (6, 0.108)
    (7, 0.089)
    (8, 0.070)
    (9, 0.056)
    (10, 0.050)
	};
\addlegendentry{glove}


\end{axis}
\end{tikzpicture}%
\caption{{\textbf{IR:}} 20NSsmall} \label{IR20NSsmall}
\end{subfigure}
~
\begin{subfigure}{0.48\textwidth}
\centering
\begin{tikzpicture}[scale=0.8][baseline]
\begin{axis}[
    x label style={at={(axis description cs:0.5,-0.05)},anchor=north},
    xlabel={\textbf{Fraction of Retrieved Documents (Recall)}},
    ylabel={\textbf{Precision (\%)}},
    xmin=0, xmax=12,
    ymin=0.35, ymax=0.98,
    /pgfplots/ytick={.35,.40,...,.98},
    xtick={0,1,2,3,4,5,6,7,8,9,10, 11,12},
    xticklabels={0.0005, 0.001, 0.002, 0.005, 0.01, 0.02, 0.05, 0.1, 0.2, 0.3, 0.5, 0.5, 0.8, 1.0},
    x tick label style={rotate=48,anchor=east},
    legend pos=south west,
    ymajorgrids=true, xmajorgrids=true,
    grid style=dashed,
]

\addplot[
	color=blue,
	mark=square,
	]
	plot coordinates {
    (0, 0.948)
    (1, 0.942)
    (2, 0.934)
    (3, 0.921)
    (4, 0.905)
    (5, 0.883)
    (6, 0.844)
    (7, 0.803)
    (8, 0.750)
    (9, 0.708)
    (10, 0.600)
    (11, 0.434)
    (12, 0.356)
	};
\addlegendentry{ctx-DocNADEe}

\addplot[
	color=red,
	mark=square,
	]
	plot coordinates {
    (0, 0.948)
    (1, 0.939)
    (2, 0.933)
    (3, 0.919)
    (4, 0.903)
    (5, 0.880)
    (6, 0.840)
    (7, 0.798)
    (8, 0.744)
    (9, 0.703)
    (10, 0.598)
    (11, 0.432)
    (12, 0.356)
	};
\addlegendentry{ctx-DocNADE}

\addplot[
	color=violet,
	mark=square,
	]
	plot coordinates {
    (0, 0.00)
    (1, 0.00)
    (2, 0.00)
    (3, 0.00)
    (4, 0.00)
    (5, 0.00)
    (6, 0.00)
    (7, 0.00)
    (8, 0.00)
    (9, 0.00)
	};
\addlegendentry{DocNADE-LM}

\addplot[
	color=cyan,
	mark=triangle,
	]
	plot coordinates {
    (0, 0.948)
    (1, 0.939)
    (2, 0.928)
    (3, 0.918)
    (4, 0.900)
    (5, 0.879)
    (6, 0.838)
    (7, 0.796)
    (8, 0.746)
    (9, 0.707)
    (10, 0.599)
    (11, 0.433)
    (12, 0.356)
    
	};
\addlegendentry{DocNADE(FV)}

\addplot[
	color=black,
	mark=*,
	]
	plot coordinates {
    (0, 0.947)
    (1, 0.939)
    (2, 0.932)
    (3, 0.922)
    (4, 0.906)
    (5, 0.884)
    (6, 0.843)
    (7, 0.802)
    (8, 0.752)
    (9, 0.714)
    (10, 0.609)
    (11, 0.434)
    (12, 0.356)
	};
\addlegendentry{DocNADE(RV)}

\addplot[
	color=green,
	mark=*,
	]
	plot coordinates {
    (0, 0.926)
    (1, 0.917)
    (2, 0.910)
    (3, 0.891)
    (4, 0.869)
    (5, 0.837)
    (6, 0.789)
    (7, 0.747)
    (8, 0.694)
    (9, 0.647)
    (10, 0.539)
    (11, 0.408)
    (12, 0.356)
	};
\addlegendentry{glove}

\end{axis}
\end{tikzpicture}%
\caption{{\textbf{IR:}} Reuters8} \label{IRReuters8}
\end{subfigure}\hspace*{\fill}
\begin{subfigure}{0.48\textwidth}
\centering
\begin{tikzpicture}[scale=0.8][baseline]
\begin{axis}[
    x label style={at={(axis description cs:0.5,-0.05)},anchor=north},
    xlabel={\textbf{Fraction of Retrieved Documents  (Recall)}},
    xmin=0, xmax=12,
    ymin=0.20, ymax=0.87,
    /pgfplots/ytick={.20,.30,...,.87},
    xtick={0,1,2,3,4,5,6,7,8,9, 10,11,12},
    xticklabels={0.0005, 0.001, 0.002, 0.005, 0.01, 0.02, 0.05, 0.1, 0.2, 0.3, 0.5, 0.8, 1.0},
    x tick label style={rotate=48,anchor=east},
    legend pos=south west,
    ymajorgrids=true, xmajorgrids=true,
    grid style=dashed,
]

\addplot[
	color=blue,
	mark=square,
	]
	plot coordinates {
    (0, 0.854)
    (1, 0.836)
    (2, 0.819)
    (3, 0.787)
    (4, 0.755)
    (5, 0.721)
    (6, 0.666)
    (7, 0.615)
    (8, 0.551)
    (9, 0.486)
    (10, 0.355)
    (11, 0.244)
    (12, 0.199)
	};
\addlegendentry{ctx-DocNADEe}

\addplot[
	color=red,
	mark=square,
	]
	plot coordinates {
    (0, 0.850)
    (1, 0.830)
    (2, 0.814)
    (3, 0.785)
    (4, 0.752)
    (5, 0.714)
    (6, 0.664)
    (7, 0.616)
    (8, 0.551)
    (9, 0.487)
    (10, 0.354)
    (11, 0.243)
    (12, 0.199)
	};
\addlegendentry{ctx-DocNADE}

\addplot[
	color=violet,
	mark=square,
	]
	plot coordinates {
    (0, 0.00)
    (1, 0.00)
    (2, 0.00)
    (3, 0.00)
    (4, 0.00)
    (5, 0.00)
    (6, 0.00)
    (7, 0.00)
    (8, 0.00)
    (9, 0.00)
	};
\addlegendentry{DocNADE-LM}

\addplot[
	color=cyan,
	mark=triangle,
	]
	plot coordinates {
    (0, 0.854)
    (1, 0.834)
    (2, 0.816)
    (3, 0.785)
    (4, 0.752)
    (5, 0.715)
    (6, 0.661)
    (7, 0.612)
    (8, 0.550)
    (9, 0.490)
    (10, 0.357)
    (11, 0.244)
    (12, 0.199)
    
	};
\addlegendentry{DocNADE(FV)}

\addplot[
	color=black,
	mark=*,
	]
	plot coordinates {
    (0, 0.861)
    (1, 0.842)
    (2, 0.825)
    (3, 0.794)
    (4, 0.761)
    (5, 0.723)
    (6, 0.668)
    (7, 0.618)
    (8, 0.558)
    (9, 0.501)
    (10, 0.360)
    (11, 0.244)
    (12, 0.199)
	};
\addlegendentry{DocNADE(RV)}

\addplot[
	color=green,
	mark=*,
	]
	plot coordinates {
    (0, 0.823)
    (1, 0.803)
    (2, 0.781)
    (3, 0.737)
    (4, 0.701)
    (5, 0.659)
    (6, 0.599)
    (7, 0.549)
    (8, 0.477)
    (9, 0.410)
    (10, 0.316)
    (11, 0.234)
    (12, 0.199)
	};
\addlegendentry{glove}

\end{axis}
\end{tikzpicture}%
\caption{{\textbf{IR:}} R21578} \label{IRR21578}
\end{subfigure}
\caption{
Retrieval performance (IR-precision) on 
short-text and long-text datasets at different fractions
}
\label{fig:docretrieval}
\end{figure*}

\begin{figure*}[t]
\centering

\begin{subfigure}{0.48\textwidth}
\centering
\begin{tikzpicture}[scale=0.8][baseline]
\begin{axis}[
    x label style={at={(axis description cs:0.5,-0.05)},anchor=north},
    xlabel={\textbf{Fraction of Retrieved Documents  (Recall)}},
    ylabel={\textbf{Precision (\%)}},
    xmin=0, xmax=13,
    ymin=0.20, ymax=0.65,
    /pgfplots/ytick={.20,.30,...,.65},
    xtick={0,1,2,3,4,5,6,7,8,9, 10,11,12,13},
    xticklabels={0.0001, 0.0005, 0.001, 0.002, 0.005, 0.01, 0.02, 0.05, 0.1, 0.2, 0.3, 0.5, 0.8, 1.0},
    x tick label style={rotate=48,anchor=east},
    legend pos=north east,
    ymajorgrids=true, xmajorgrids=true,
    grid style=dashed,
]

\addplot[
	color=blue,
	mark=square,
	]
	plot coordinates {
    (0, 0.643)
    (1, 0.565)
    (2, 0.535)
    (3, 0.503)
    (4, 0.459)
    (5, 0.424)
    (6, 0.388)
    (7, 0.337)
    (8, 0.298)
    (9, 0.262)
    (10, 0.245)
    (11, 0.224)
    (12, 0.209)
    (13, 0.203)
	};
\addlegendentry{ctx-DocNADEe}

\addplot[
	color=red,
	mark=square,
	]
	plot coordinates {
    (0, 0.641)
    (1, 0.560)
    (2, 0.529)
    (3, 0.498)
    (4, 0.454)
    (5, 0.421)
    (6, 0.386)
    (7, 0.336)
    (8, 0.298)
    (9, 0.262)
    (10, 0.245)
    (11, 0.225)
    (12, 0.208)
    (13, 0.208)
	};
\addlegendentry{ctx-DocNADE}

\addplot[
	color=violet,
	mark=square,
	]
	plot coordinates {
    (0, 0.00)
    (1, 0.00)
    (2, 0.00)
    (3, 0.00)
    (4, 0.00)
    (5, 0.00)
    (6, 0.00)
    (7, 0.00)
    (8, 0.00)
    (9, 0.00)
	};
\addlegendentry{DocNADE-LM}

\addplot[
	color=cyan,
	mark=triangle,
	]
	plot coordinates {
    (0, 0.640)
    (1, 0.557)
    (2, 0.526)
    (3, 0.494)
    (4, 0.451)
    (5, 0.417)
    (6, 0.382)
    (7, 0.334)
    (8, 0.296)
    (9, 0.261)
    (10, 0.244)
    (11, 0.225)
    (12, 0.209)
    (13, 0.202)
    
	};
\addlegendentry{DocNADE(FV)}

\addplot[
	color=black,
	mark=*,
	]
	plot coordinates {
    (0, 0.624)
    (1, 0.544)
    (2, 0.515)
    (3, 0.484)
    (4, 0.441)
    (5, 0.408)
    (6, 0.374)
    (7, 0.327)
    (8, 0.291)
    (9, 0.257)
    (10, 0.242)
    (11, 0.224)
    (12, 0.209)
    (13, 0.202)
	};
\addlegendentry{DocNADE(RV)}

\addplot[
	color=green,
	mark=*,
	]
	plot coordinates {
    (0, 0.527)
    (1, 0.420)
    (2, 0.385)
    (3, 0.353)
    (4, 0.321)
    (5, 0.301)
    (6, 0.284)
    (7, 0.263)
    (8, 0.246)
    (9, 0.227)
    (10, 0.218)
    (11, 0.209)
    (12, 0.203)
    (13, 0.202)
	};
\addlegendentry{glove}

\end{axis}
\end{tikzpicture}%
\caption{{\textbf{IR:}} SiROBs} \label{IRSiROBs}
\end{subfigure}\hspace*{\fill}
\begin{subfigure}{0.48\textwidth}
\centering
\begin{tikzpicture}[scale=0.8][baseline]
\begin{axis}[
    x label style={at={(axis description cs:0.5,-0.05)},anchor=north},
    xlabel={\textbf{Fraction of Retrieved Documents  (Recall)}},
    xmin=0, xmax=13,
    ymin=0.25, ymax=0.87,
    /pgfplots/ytick={.25,.30,...,.86},
    xtick={0,1,2,3,4,5,6,7,8,9, 10,11,12,13},
    xticklabels={0.0001, 0.0005, 0.001, 0.002, 0.005, 0.01, 0.02, 0.05, 0.1, 0.2, 0.3, 0.5, 0.8, 1.0},
    x tick label style={rotate=48,anchor=east},
    legend pos=south west,
    ymajorgrids=true, xmajorgrids=true,
    grid style=dashed,
]

\addplot[
	color=blue,
	mark=square,
	]
	plot coordinates {
     (0, 0.875)
    (1, 0.860)
    (2, 0.851)
    (3, 0.842)
    (4, 0.828)
    (5, 0.814)
    (6, 0.796)
    (7, 0.757)
    (8, 0.709)
    (9, 0.618)
    (10, 0.532)
    (11, 0.407)
    (12, 0.297)
    (13, 0.249)
	};
\addlegendentry{ctx-DocNADEe}

\addplot[
	color=red,
	mark=square,
	]
	plot coordinates {
    (0, 0.873)
    (1, 0.858)
    (2, 0.850)
    (3, 0.841)
    (4, 0.826)
    (5, 0.812)
    (6, 0.791)
    (7, 0.756)
    (8, 0.708)
    (9, 0.617)
    (10, 0.531)
    (11, 0.406)
    (12, 0.297)
    (13, 0.249)
	};
\addlegendentry{ctx-DocNADE}

\addplot[
	color=violet,
	mark=square,
	]
	plot coordinates {
    (0, 0.00)
    (1, 0.00)
    (2, 0.00)
    (3, 0.00)
    (4, 0.00)
    (5, 0.00)
    (6, 0.00)
    (7, 0.00)
    (8, 0.00)
    (9, 0.00)
	};
\addlegendentry{DocNADE-LM}

\addplot[
	color=cyan,
	mark=triangle,
	]
	plot coordinates {
    (0, 0.871)
    (1, 0.856)
    (2, 0.849)
    (3, 0.839)
    (4, 0.825)
    (5, 0.811)
    (6, 0.794)
    (7, 0.760)
    (8, 0.718)
    (9, 0.633)
    (10, 0.547)
    (11, 0.415)
    (12, 0.299)
    (13, 0.249)
    
	};
\addlegendentry{DocNADE(FV)}

\addplot[
	color=black,
	mark=*,
	]
	plot coordinates {
    (0, 0.863)
    (1, 0.848)
    (2, 0.840)
    (3, 0.831)
    (4, 0.816)
    (5, 0.803)
    (6, 0.787)
    (7, 0.756)
    (8, 0.718)
    (9, 0.639)
    (10, 0.552)
    (11, 0.415)
    (12, 0.299)
    (13, 0.249)
	};
\addlegendentry{DocNADE(RV)}

\addplot[
	color=green,
	mark=*,
	]
	plot coordinates {
    (0, 0.851)
    (1, 0.824)
    (2, 0.812)
    (3, 0.798)
    (4, 0.778)
    (5, 0.760)
    (6, 0.737)
    (7, 0.695)
    (8, 0.642)
    (9, 0.550)
    (10, 0.476)
    (11, 0.376)
    (12, 0.289)
    (13, 0.249)
	};
\addlegendentry{glove}

\end{axis}
\end{tikzpicture}%
\caption{{\textbf{IR:}} AGnews} \label{IRAGnews}
\end{subfigure}\hspace*{\fill}%
\caption{
Retrieval performance (IR-precision) on 
short-text and long-text datasets at different fractions
}
\label{fig:docretrieval}
\end{figure*}

\begin{table*}[t]
\center
\renewcommand*{\arraystretch}{1.2}
\resizebox{.6\textwidth}{!}{
\begin{tabular}{c|c||ccc}
\hline
\multicolumn{1}{c}{Dataset}   & \multicolumn{1}{c}{Model} & \multicolumn{3}{c}{\textbf{$\lambda$}}        \\
                     &    & 1.0    & 0.1    & 0.01      \\ \hline

\multirow{2}{*}{TMNtitle}     & ctx-DocNADE               & 1898.1 & 1482.7 & 1487.1        \\
                              & ctx-DocNADEe              & 1877.7 & 1480.2 & 1484.7        \\  \hline
\multirow{2}{*}{AGnewstitle}  & ctx-DocNADE               & 1296.1 &  861.1     & 865           \\
                              & ctx-DocNADEe              & 1279.2 & 853.3      & 862.9       \\ \hline
\multirow{2}{*}{20NSshort}    & ctx-DocNADE               & 899.04 & 829.5  &  842.1            \\
                              & ctx-DocNADEe              & 890.3  & 828.8  & 832.4          \\  \hline

\multirow{2}{*}{Subjectivity} & ctx-DocNADE               & 982.8      & 977.8  & 966.5     \\
                              & ctx-DocNADEe              & 977.1      & 975.0 & 964.2        \\  \hline
\multirow{2}{*}{Reuters-8}    & ctx-DocNADE               & 336.1  & 313.2    & 311.9        \\
                              & ctx-DocNADEe              & 323.3  & 312.0     & 310.2        \\ \hline
\multirow{2}{*}{20NS}         & ctx-DocNADE               & 1282.1 & 1209.3 & 1207.2      \\
                              & ctx-DocNADEe              & 1247.1 & 1211.6      & 1206.1       

\end{tabular}}
\caption{Evaluation of PPL scores on validation set of different datasets for selection of optimal setting of $\lambda$ hyperparameter over hidden unit of LSTM-LM}
\label{table:ctx_docnade_lambdappl}
\end{table*}

\begin{table}[t]
\center
\renewcommand*{\arraystretch}{1.2}
\resizebox{.60\textwidth}{!}{
\begin{tabular}{l|l|llll}
\hline
\multicolumn{1}{c}{Dataset}   & \multicolumn{1}{c}{Model} & \multicolumn{4}{c}{\textbf{$\lambda$}}     \\ \hline
                              &                           & 1.0   & 0.8   & 0.5   & 0.3   \\
\multirow{2}{*}{Polarity}     & ctx-DocNADE               & 0.587 & 0.588 & 0.591 & 0.587 \\
                              & ctx-DocNADEe              & 0.602 & 0.603 & 0.601 & 0.599 \\ \hline
\multirow{2}{*}{TMNtitle}     & ctx-DocNADE               & 0.556 & 0.557 & 0.559 & 0.568 \\
                              & ctx-DocNADEe              & 0.604 & 0.604 & 0.6   & 0.6   \\ \hline
\multirow{2}{*}{20NSshort}    & ctx-DocNADE               & 0.264 & 0.265 & 0.265 & 0.265 \\
                              & ctx-DocNADEe              & 0.277 & 0.277 & 0.278 & 0.276 \\ \hline
\multirow{2}{*}{Subjectivity} & ctx-DocNADE               & 0.874 & 0.874 & 0.873 & 0.874 \\
                              & ctx-DocNADEe              & 0.868 & 0.868 & 0.874 & 0.87  \\ \hline
\multirow{2}{*}{TMN}          & ctx-DocNADE               & 0.683 & 0.689 & 0.692 & 0.694 \\  
                              & ctx-DocNADEe              & 0.696 & 0.698 & 0.698 & 0.7   \\ \hline
\multirow{2}{*}{AGnewstitle}  & ctx-DocNADE               & 0.665 & 0.668 & 0.678 & 0.689 \\
                              & ctx-DocNADEe              & 0.686 & 0.688 & 0.695 & 0.696 \\  \hline
\multirow{2}{*}{20NSsmall}    & ctx-DocNADE               & 0.352 & 0.356 & 0.366 & 0.37  \\
                              & ctx-DocNADEe              & 0.381 & 0.381 & 0.375 & 0.353 \\  \hline
\multirow{2}{*}{R21578}       & ctx-DocNADE               & 0.714 & 0.714 & 0.714 & 0.714 \\
                              & ctx-DocNADEe              & 0.715 & 0.715 & 0.715 & 0.714 \\   \hline
\multirow{2}{*}{SiROBs}       & ctx-DocNADE               & 0.409 & 0.409 & 0.408 & 0.408 \\
                              & ctx-DocNADEe              & 0.41  & 0.411 & 0.411 & 0.409 \\  \hline
\multirow{2}{*}{Reuters-8}    & ctx-DocNADE               &    0.863   &  0.866     & 0.87  & 0.87  \\
                              & ctx-DocNADEe              & 0.875      & 0.872 & 0.873 &   0.872    \\  \hline
\multirow{2}{*}{20NS}         & ctx-DocNADE               & 0.503 & 0.506 & 0.513 & 0.512 \\
                              & ctx-DocNADEe              & 0.524 & 0.521 & 0.518 & 0.511 \\  \hline
\multirow{2}{*}{AGnews}       & ctx-DocNADE               & 0.786 & 0.789 & 0.792 & 0.797 \\
                              & ctx-DocNADEe              & 0.795 & 0.796 & 0.8   & 0.799
\end{tabular}}
\caption{Evaluation of IR precision scores (at fraction 0.02) on validation set of different datasets for selection of optimal setting of $\lambda$ hyperparameter over hidden unit of LSTM-LM}
\label{table:ctx_docnade_lambdIR}
\end{table}

\begin{table}[t]
      \centering
        \begin{tabular}{c|c}
         \hline 
         {\textbf{Hyperparameter}}               & {\textbf{Search Space}} \\ \hline
           learning rate        &    [0.001]                   \\
           hidden units        &       [200]                      \\
           iterations        &      [2000]      \\
           activation function      &    [sigmoid, tanh]            \\ 
          $\lambda_{EmbTF}$        &      [1.0, 0.5, 0.1]  \\ 
          $\lambda_{CF}$        &      [1.0, 0.5, 0.1]  \\ 
          $\lambda_{SAL}$        &      [1.0, 0.1, 0.01, 0.001]  \\ 
          \hline      
       \end{tabular}
         \caption{Different values of hyperparameter settings in \textit{lifelong learning} task; \textit{sigmoid} is used as activation function in case of PPL/COH, while \textit{tanh} is used as the activation function in case of IR}
         \label{table:lifelong_hyperparameters}
 \end{table}

 \clearemptydoublepage

 \printglossaries

 \addcontentsline{toc}{chapter}{Bibliography}

\end{document}